\documentclass[aps,prd,reprint,preprintnumbers,showpacs,floatfix,nofootinbib,superscript address,longbibliography]{revtex4-2}

\usepackage{stix}
\usepackage{textcomp}
\usepackage{amsmath,amsfonts,amssymb}
\usepackage{booktabs}
\usepackage{siunitx}
\usepackage[dvipsnames]{xcolor}
\usepackage{verbatim}
\usepackage[english]{babel}
\usepackage{chngcntr}
\usepackage{adjustbox}
\usepackage{float}
\usepackage{graphicx}
\usepackage{scalerel,stackengine}
\usepackage{titlesec}
\usepackage{parskip}
\usepackage{mathtools}
\usepackage{multirow,tabularx}
\usepackage{notoccite}
\usepackage{natbib}
\usepackage{mathrsfs}
\usepackage{lineno}
\usepackage{tensor}
\usepackage{accents}
\usepackage{etoolbox}
\usepackage{xstring}
\usepackage{xspace}
\usepackage{hyperref}
\hypersetup{%
     colorlinks = true,%
     linkcolor = Blue,%
     citecolor = Blue,%
     filecolor = Blue,%
     urlcolor = Blue%
     }%
\usepackage[capitalize]{cleveref}

\allowdisplaybreaks

\def\nn{\nonumber}
\newcommand{\w}[1]{\\[0.#1cm]}

\newrobustcmd{\PSALTer}{\textit{PSALTer}\xspace}

\newrobustcmd{\pea}[1]{%
	\emph{#1}\textbf{\ \ \ ---}
}
\titleformat{\paragraph}[runin]{\normalfont\normalsize\bfseries}{\emph\theparagraph}{1em}{\pea}

\newrobustcmd{\typeone}[1]{%
	{I}	
}
\newrobustcmd{\typetwo}[1]{%
	{II}	
}
\newrobustcmd{\typethree}[1]{%
	{III}	
}
\newrobustcmd{\typefour}[1]{%
	{IV}	
}
\newrobustcmd{\typefive}[1]{%
	{V}	
}
\newrobustcmd{\typesix}[1]{%
	{VI}	
}

\newrobustcmd{\first}[1]{%
	{1\textsuperscript{st}}	
}
\newrobustcmd{\second}[1]{%
	{2\textsuperscript{nd}}	
}
\newrobustcmd{\third}[1]{%
	{3\textsuperscript{rd}}	
}
\newrobustcmd{\fourth}[1]{%
	{4\textsuperscript{th}}	
}

\newrobustcmd{\IsOff}[1]{%
	{\ =0\ \land\ }	
}

\newrobustcmd{\MAGg}[1]{%
	\tensor{g}{#1}
}

\newrobustcmd{\MAGd}[1]{%
	\tensor*{\delta}{#1}
}
\newrobustcmd{\MAGl}[1]{%
	\tensor{\xi}{#1}
}

\newrobustcmd{\MAGD}[1]{%
  \tensor{\Delta}{#1}
}
\newrobustcmd{\MAGA}[1]{%
  \tensor{A}{#1}
}
\newrobustcmd{\MAGF}[1]{%
	\tensor{\mathcal{F}}{#1}
}
\newrobustcmd{\MAGFt}[1]{%
	\tensor{\tilde{\mathcal{F}}}{#1}
}
\newrobustcmd{\MAGFa}[1]{%
	\tensor{\mathcal{F}}{^{(14)}#1}
}
\newrobustcmd{\MAGFb}[1]{%
	\tensor{\mathcal{F}}{^{(13)}#1}
}
\newrobustcmd{\MAGT}[1]{%
	\tensor{\mathcal{T}}{#1}
}
\newrobustcmd{\MAGTt}[1]{%
	\tensor{\tilde{\mathcal{T}}}{#1}
}
\newrobustcmd{\MAGQ}[1]{%
	\tensor{\mathcal{Q}}{#1}
}
\newrobustcmd{\MAGQt}[1]{%
	\tensor{\tilde{\mathcal{Q}}}{#1}
}
\newrobustcmd{\MAGTh}[1]{%
	\tensor{\hat{\mathcal{T}}}{#1}
}
\newrobustcmd{\MAGV}[1]{%
	\tensor{\mathcal{V}}{#1}
}

\newrobustcmd{\alp}[1]{%
       {\alpha}	
}

\newrobustcmd{\bet}[1]{%
  \tensor[^{(#1)}]{\mu}{}
}

\stackMath 
\newcommand\OmitIndices[1]{%
\savestack{\tmpbox}{\stretchto{%
\scaleto{%
\scalerel*[\widthof{\ensuremath{#1}}]{\kern-.6pt\curlywedge\kern-.6pt}%
{\rule[-\textheight/2]{1ex}{\textheight}}
}{\textheight}%
}{0.5ex}}%
\stackon[1pt]{#1}{\tmpbox}%
}

\newrobustcmd{\LPV}[1]{%
	\IfEqCase{#1}{%
	{}{L_{\text{PV}}}%
	}%
	[{L_{\text{PV}}\left(#1\right)}]%
}

\newrobustcmd{\F}[2][placeholder]{%
	\IfEqCase{#1}{%
	{placeholder}{\tensor{T}{#2}}%
	{1}{\tensor[^{(1)}]{F}{#2}}%
	{2}{\tensor[^{(2)}]{F}{#2}}%
	{3}{\tensor[^{(3)}]{F}{#2}}%
	}%
	[\packageError{cosmicclass}{Symbol #1 is not an irreducible part!}{}]%
}

\newrobustcmd{\GenericVector}[1]{%
	\smash{{#1}_{\tensor[^{{(2)}}]{\hspace{-1pt}\lambda}{}}^{J^P}}%
}
\newrobustcmd{\GenericTensor}[1]{%
	\smash{{#1}_{\tensor[^{{(1)}}]{\hspace{-1pt}\lambda}{}}^{J^P}}%
}

\newrobustcmd{\g}[1]{%
	\tensor{g}{#1}%
}
\newrobustcmd{\rcCon}[1]{%
	\tensor*{\Gamma}{#1}%
}
\newrobustcmd{\rCon}[1]{%
	\tensor*{\mathring{\Gamma}}{#1}%
}
\newrobustcmd{\Con}[1]{%
	\tensor{\Gamma}{#1}%
}
\newrobustcmd{\B}[1]{%
	\tensor{B}{#1}%
}
\newrobustcmd{\PD}[1]{%
	\tensor{\partial}{#1}%
}
\newrobustcmd{\rD}[1]{%
	\tensor{\mathring{\nabla}}{#1}%
}
\newrobustcmd{\rcD}[1]{%
	\tensor{\nabla}{#1}%
}
\newrobustcmd{\rR}[1]{%
	\tensor{\mathring{R}}{#1}%
}
\newrobustcmd{\rcR}[1]{%
	\tensor{R}{#1}%
}

\renewrobustcmd{\F}[2][placeholder]{%
	\IfEqCase{#1}{%
	{placeholder}{\tensor{T}{#2}}%
	{1}{\tensor[^{(1)}]{F}{#2}}%
	{2}{\tensor[^{(2)}]{F}{#2}}%
	{3}{\tensor[^{(3)}]{F}{#2}}%
	}%
	[\packageError{cosmicclass}{Symbol #1 is not an irreducible part!}{}]%
}

\newrobustcmd{\Si}[2][placeholder]{%
	\IfEqCase{#1}{%
	{placeholder}{\tensor{T}{#2}}%
	{1}{\tensor[^{(1)}]{S}{#2}}%
	{2}{\tensor[^{(2)}]{S}{#2}}%
	{3}{\tensor[^{(3)}]{S}{#2}}%
	}%
	[\packageError{cosmicclass}{Symbol #1 is not an irreducible part!}{}]%
}

\newrobustcmd{\T}[2][placeholder]{%
	\IfEqCase{#1}{%
	{placeholder}{\tensor{T}{#2}}%
	{1}{\tensor[^{(1)}]{T}{#2}}%
	{2}{\tensor[^{(2)}]{T}{#2}}%
	{3}{\tensor[^{(3)}]{T}{#2}}%
	}%
	[\packageError{cosmicclass}{Symbol #1 is not an irreducible part!}{}]%
}

\newrobustcmd{\mass}[2][placeholder]{%
	\IfEqCase{#1}{%
	{placeholder}{\tensor{m}{#2}}%
	{1}{\tensor[^{(1)}]{m}{#2}}%
	{2}{\tensor[^{(2)}]{m}{#2}}%
	{3}{\tensor[^{(3)}]{m}{#2}}%
	}%
	[\packageError{cosmicclass}{Symbol #1 is not an irreducible part!}{}]%
}

\newrobustcmd{\TLambda}[2][placeholder]{%
	\IfEqCase{#1}{%
	{placeholder}{\tensor{\lambda}{#2}}%
	{1}{\tensor[^{(1)}]{\lambda}{#2}}%
	{2}{\tensor[^{(2)}]{\lambda}{#2}}%
	{3}{\tensor[^{(3)}]{\lambda}{#2}}%
	{Q}{\tensor[^{(Q)}]{\lambda}{#2}}%
	}%
	[\packageError{cosmicclass}{Symbol #1 is not an irreducible part!}{}]%
}

\newrobustcmd{\RLambda}[2][placeholder]{%
	\IfEqCase{#1}{%
	{placeholder}{\tensor{\lambda}{#2}}%
	{1}{\tensor[^1]{\lambda}{#2}}%
	{2}{\tensor[^2]{\lambda}{#2}}%
	{3}{\tensor[^3]{\lambda}{#2}}%
	{R}{\tensor[^{(R)}]{\lambda}{#2}}%
	}%
	[\packageError{cosmicclass}{Symbol #1 is not an irreducible part!}{}]%
}

\newrobustcmd{\QLambda}[2][placeholder]{%
	\IfEqCase{#1}{%
		{placeholder}{\tensor{\hat{\lambda}}{#2}}%
	{1}{\tensor[^1]{\hat{\lambda}}{#2}}%
	{2}{\tensor[^2]{\hat{\lambda}}{#2}}%
	{3}{\tensor[^3]{\hat{\lambda}}{#2}}%
	}%
	[\packageError{cosmicclass}{Symbol #1 is not an irreducible part!}{}]%
}

\newrobustcmd{\Mgra}[1]{%
  {\tensor{M}{_{\text{#1}}}}%
}
\newrobustcmd{\Mpro}[1]{%
  {\tensor{\mathproper{M}}{_{\text{#1}}}}%
}
\newrobustcmd{\Malt}[1]{%
  {\tensor{\mathscr{M}}{_{\text{#1}}}}%
}
\newrobustcmd{\Mkom}[1]{%
  {\tensor{\mathfrak{M}}{_{\text{#1}}}}%
}

\newrobustcmd{\Mtotal}{%
  {\tensor{M}{_{\text{T}}}}%
}

\newrobustcmd{\Qtotal}{%
  {\tensor{Q}{_{\text{T}}}}%
}

\newrobustcmd{\Qtotalcal}{%
  {\tensor{\mathcal{  Q}}{_{\text{T}}}}%
}

\newrobustcmd{\action}[1]{%
  {\tensor{S}{_{\text{#1}}}}%
}

\newrobustcmd{\lagrangian}[1]{%
  {\tensor{L}{_{\text{#1}}}}%
}

\newrobustcmd{\lagrangianprop}[1]{%
  {\tensor{\mathproper{L}}{_{\text{#1}}}}%
}

\newrobustcmd{\epl}{%
  {\tensor{\mathsf{e}}{_+}}%
}

\newrobustcmd{\epe}{%
  {\tensor{\mathsf{e}}{_\perp}}%
}

\newrobustcmd{\qz}{%
  {\text{\color{orange}\cmark}}%
}
\newrobustcmd{\jz}{%
  {\text{\color{red}\xmark}}%
}

\newrobustcmd{\projmatrix}[2][placeholder]{%
  {\tensor*{M}{_{#1}^{#2}}}
}
\newrobustcmd{\projorthhum}[2][placeholder]{%
  {\tensor[^#2]{\smash{\check{\mathcal{  P}}}}{#1}}
}
\newrobustcmd{\projorthhumu}[2][placeholder]{%
  {\tensor[^#2]{\smash{\check{\mathcal{  P}}}}{#1}}
}
\newrobustcmd{\projorth}[2][placeholder]{%
  {\tensor[^#2]{\smash{\hat{\mathcal{  P}}}}{#1}}
}
\newrobustcmd{\projlore}[2][placeholder]{%
  {\tensor[^#2]{\hat{\mathcal{  P}}}{#1}}
}

\newrobustcmd{\gensec}[3][placeholder]{%
  {\tensor*[^#1]{\chi}{^{#2}_{\acu{#3}}}}
}

\newrobustcmd{\glfourr}{%
  {\mathrm{GL}(4,\mathbb{R})}%
}

\newrobustcmd{\sltwoc}{%
  {\mathrm{SL}(2,\mathbb{C})}%
}

\newrobustcmd{\poincare}{%
  {\mathbb{R}^{1,3}\rtimes\mathrm{SO}^+(1,3)}%
}

\newrobustcmd{\poincaref}{%
  {\mathrm{P}(1,3)}%
}

\newrobustcmd{\weyl}{%
  {\mathrm{W}(1,3)}%
}

\newrobustcmd{\conformal}{%
  {\mathrm{C}(1,3)}%
}

\newrobustcmd{\diffeomorphism}{%
  {\mathbb{R}^{1,3}}%
}

\newrobustcmd{\soonethree}{%
  {\mathrm{SO}^+(1,3)}%
}

\newrobustcmd{\othree}{%
  {\mathrm{SO}(3)}%
}

\newrobustcmd{\sothree}{%
  {\mathrm{SO}(3)}%
}

\newrobustcmd{\sotwo}{%
  {\mathrm{SO}(2)}%
}

\newrobustcmd{\suthreec}{%
  {\mathrm{SU}(3)_{\text{c}}}%
}

\newrobustcmd{\sutwol}{%
  {\mathrm{SU}(2)_{\text{L}}}%
}

\newrobustcmd{\uoney}{%
  {\mathrm{U}(1)_{\text{Y}}}%
}

\newrobustcmd{\uone}{%
  {\mathrm{U}(1)}%
}

\newrobustcmd{\uoneem}{%
  {\mathrm{U}(1)_{\text{em}}}%
}

\newrobustcmd{\sutwo}{%
  {\mathrm{SU}(2)}%
}

\newrobustcmd{\eplus}{%
  {\tensor{\mathsf{e}}{_{+}}}%
}

\newrobustcmd{\esf}[1]{%
  {\tensor{\mathsf{e}}{_{#1}}}
}%
\newrobustcmd{\esfu}[1]{%
  {\tensor{\mathsf{e}}{^{#1}}}
}%
\newrobustcmd{\gam}[1]{%
  {\tensor{\gamma}{_{#1}}}
}%
\newrobustcmd{\gamu}[1]{%
  {\tensor{\gamma}{^{#1}}}
}%

\newrobustcmd{\planck}{%
  {m_{\text{p}}}%
}

\newrobustcmd{\Planck}{%
	{M_{\text{Pl}}}%
}

\newrobustcmd{\caligR}{%
  {\mathcal{R}}%
}
\newrobustcmd{\caligT}{%
  {\mathcal{T}}%
}

\newrobustcmd{\pgt}{%
  PGT\textsuperscript{q,+}\ %
}

\newrobustcmd{\unl}[1]{%
  {\mathfrak{#1}}%
}
\newrobustcmd{\ovl}[1]{%
\overline{#1}%
}
\newrobustcmd{\acu}[1]{%
\acute{#1}%
}

\newrobustcmd{\indiq}[2][placeholder]{%
\IfEqCase{#1}{%
  {placeholder}{%
    \IfEqCase{#2}{%
      {1}{\ovl{k}}%
      {2}{\ovl{kl}}%
      {3}{\ovl{klm}}%
    }%
  }%
}[#1]%
}%

\newrobustcmd{\indaq}[2][placeholder]{%
\IfEqCase{#1}{%
  {placeholder}{%
    \IfEqCase{#2}{%
      {1}{\overline{k}}%
      {2}{\overline{kl}}%
      {3}{\overline{klm}}%
    }%
  }%
}[#1]%
}%

\newrobustcmd{\indeq}[2][placeholder]{%
\IfEqCase{#1}{%
  {placeholder}{%
    \IfEqCase{#2}{%
      {1}{k}%
      {2}{kl}%
      {3}{klm}%
    }%
  }%
}[#1]%
}%

\newrobustcmd{\indoq}[2][placeholder]{%
\IfEqCase{#1}{%
  {placeholder}{%
    \IfEqCase{#2}{%
      {1}{\alpha}%
      {2}{\alpha\beta}%
      {3}{\alpha\beta\gamma}%
    }%
  }%
}[#1]%
}%

\newrobustcmd{\fcphi}[1]{%
  \tensor[^{\text{FC}}]{\phi}{_{#1}}%
}

\newrobustcmd{\scphi}[1]{%
  \tensor[^{\text{SC}}]{\phi}{_{#1}}%
}

\newrobustcmd{\fcmul}[1]{%
  \tensor[^{\text{FC}}]{\upsilon}{_{#1}}%
}

\newrobustcmd{\arb}{%
  {\tensor{f}{_{\text{lin}}}}%
}

\newrobustcmd{\scmul}[1]{%
  \tensor[^{\text{SC}}]{\upsilon}{_{#1}}%
}

\newrobustcmd{\foli}[1]{%
\tensor{n}{_{#1}}%
}

\newrobustcmd{\foliu}[1]{%
\tensor{n}{^{#1}}%
}

\newrobustcmd{\covderl}[1]{%
\tensor{\mathcal{D}}{^{\flat}_{\indiq[#1]{1}}}%
}

\newrobustcmd{\covder}[1]{%
\tensor{\mathcal{D}}{_{\indiq[#1]{1}}}%
}
\newrobustcmd{\coder}[1]{%
\tensor{D}{_{\indiq[#1]{1}}}%
}

\newrobustcmd{\deltal}[2]{%
  \tensor*{\delta}{_{\phantom{\flat}}^{\flat}_{#1}^{#2}}%
}

\newrobustcmd{\deltaud}[2]{%
  \tensor*{\delta}{^{#1}_{#2}}%
}
\newrobustcmd{\etau}[1]{%
\tensor{\eta}{^{\indiq[#1]{2}}}%
}
\newrobustcmd{\etaul}[1]{%
\tensor{\eta}{^{\flat}^{\indiq[#1]{2}}}%
}
\newrobustcmd{\etad}[1]{%
\tensor{\eta}{_{\indiq[#1]{2}}}%
}
\newrobustcmd{\etadl}[1]{%
\tensor{\eta}{^{\flat}_{\indiq[#1]{2}}}%
}

\newrobustcmd{\epsul}[1]{%
\tensor{\epsilon}{^{\flat}^{\indiq[#1]{3}}^{\perp}}
}
\newrobustcmd{\epsdl}[1]{%
\tensor{\epsilon}{^{\flat}_{\indiq[#1]{3}}_{\perp}}
}
\newrobustcmd{\epsd}[1]{%
\tensor{\epsilon}{_{\indiq[#1]{3}}_{\perp}}
}
\newrobustcmd{\epsu}[1]{%
\tensor{\epsilon}{^{\indiq[#1]{3}}^{\perp}}
}

\newrobustcmd{\hfl}[2]{%
  \tensor{h}{^{\flat}_{#1}^{#2}}
}

\newrobustcmd{\cgalp}{\tensor{\alpha}{_{\text{CG}}}}

\newrobustcmd{\cbet}[1]{%
  \tensor{\bar{\beta}}{_{#1}}
}
\newrobustcmd{\calp}[1]{%
  \tensor{\bar{\alpha}}{_{#1}}
}

\newrobustcmd{\alpg}[1]{%
  \tensor{\check{\alpha}}{_{#1}}
}
\newrobustcmd{\betg}[1]{%
  \tensor{\check{\beta}}{_{#1}}
}
\newrobustcmd{\calpg}[1]{%
  \tensor{\acu{\alpha}}{_{#1}}
}
\newrobustcmd{\cbetg}[1]{%
  \tensor{\acu{\beta}}{_{#1}}
}

\newrobustcmd{\hub}{%
  {\underline{\mathsf{h}}}
}
\newrobustcmd{\hubm}{%
  {\underline{\mathsf{h}}^{-1}}
}
\newrobustcmd{\hob}{%
  {\bar{\mathsf{h}}}
}
\newrobustcmd{\hobm}{%
  {\bar{\mathsf{h}}^{-1}}
}
\newrobustcmd{\hdet}{%
  {\det \mathsf{h}}
}
\newrobustcmd{\hmdet}{%
  {\det \mathsf{h}^{-1}}
}
\newrobustcmd{\Rsf}{%
  {\mathsf{R}}
}

\newrobustcmd{\alpm}[2][placeholder]{%
  {\tensor*{\hat{\alpha}}{_{#1}^{#2}}}
}
\newrobustcmd{\calpm}[2][placeholder]{%
  \tensor*{\bar{\alpha}}{_{#1}^{#2}}
}

\newrobustcmd{\betm}[2][placeholder]{%
  {\tensor*{\hat{\beta}}{_{#1}^{#2}}}
}
\newrobustcmd{\cbetm}[2][placeholder]{%
  \tensor*{\bar{\beta}}{_{#1}^{#2}}
}

\newrobustcmd{\lamr}{%
  {\zeta_{\mathcal{  R}} }
}
\newrobustcmd{\barlamr}{%
  {\bar{\zeta}_{\mathcal{  R}} }
}
\newrobustcmd{\lamt}{%
  {\zeta_{\mathcal{  T}} }
}
\newrobustcmd{\barlamt}{%
  {\bar{\zeta}_{\mathcal{  T}} }
}

\newrobustcmd{\atmp}[1]{%
  \tensor{\hat{a}}{_{#1}}
}
\newrobustcmd{\btmp}[1]{%
  \tensor{b}{_{#1}}
}

\newrobustcmd{\ctmp}[2][placeholder]{%
  {\tensor*{c}{_{#1}^{#2}}}
}
\newrobustcmd{\dtmp}[2][placeholder]{%
  {\tensor*{d}{_{#1}^{#2}}}
}

\newrobustcmd{\etmp}[1]{%
  \tensor{e}{_{#1}}
}

\newrobustcmd{\batmp}[1]{%
  \tensor{\ovl{a}}{_{#1}}
}
\newrobustcmd{\bbtmp}[1]{%
  \tensor{\ovl{b}}{_{#1}}
}
\newrobustcmd{\bctmp}[1]{%
  \tensor{\ovl{c}}{_{#1}}
}
\newrobustcmd{\bdtmp}[1]{%
  \tensor{\ovl{d}}{_{#1}}
}
\newrobustcmd{\betmp}[1]{%
  \tensor{\ovl{e}}{_{#1}}
}

\newrobustcmd{\ptl}[1]{%
  \tensor{\partial}{#1}
}

\newrobustcmd{\etaf}[1]{%
  \tensor{\eta}{#1}
}

\newrobustcmd{\epsf}[1]{%
  \tensor{\epsilon}{#1}
}

\newrobustcmd{\RSO}[2][placeholder]{%
  {\tensor[^{#2}]{\mathcal{  R}}{#1}}
}
\newrobustcmd{\TSO}[2][placeholder]{%
  {\tensor[^{#2}]{\mathcal{  T}}{#1}}
}
\newrobustcmd{\FSO}[2][placeholder]{%
  {\tensor[^{#2}]{\mathcal{  F}}{#1}}
}
\newrobustcmd{\spinSO}[2][placeholder]{%
  {\tensor[^{#2}]{\sigma}{#1}}
}
\newrobustcmd{\RLambdaSO}[2][placeholder]{%
  {\tensor[^{#2}]{\zeta}{#1}}
}
\newrobustcmd{\TLambdaSO}[2][placeholder]{%
  {\tensor[^{#2}]{\zeta}{#1}}
}
\newrobustcmd{\KSO}[2][placeholder]{%
  {\tensor[^{#2}]{\mathcal{  K}}{#1}}
}

\newrobustcmd{\bper}[2][placeholder]{%
\IfEqCase{#2}{%
  {s}{\tensor{\mathfrak{s}}{#1}}%
  {a}{\tensor{\mathfrak{a}}{#1}}%
  {sbar}{\tensor{\bar{\mathfrak{s}}}{#1}}%
}[\packageError{cosmicclass}{Unidentified Critical Case: #1}{}]%
}

\newrobustcmd{\Jl}{%
  {J^{\flat}}%
}%

\newrobustcmd{\Nl}{%
  {N^{\flat}}%
}%

\newrobustcmd{\haml}[2][placeholder]{%
\IfEqCase{#2}{%
{mom0p}{\tensor{\mathcal{H}}{^{\flat}_{\perp}}}%
{mom1m}{\tensor{\mathcal{H}}{^{\flat}_{\indoq[#1]{1}}}}%
{rot1p}{\tensor{\mathcal{H}}{^{\flat}_{\indaq[#1]{2}}}}%
{rot1m}{\tensor{\mathcal{H}}{^{\flat}_{\perp}_{\indaq[#1]{1}}}}%
}[\packageError{cosmicclass}{Unidentified Critical Case: #1}{}]%
}

\newrobustcmd{\arc}[2][placeholder]{%
\IfEqCase{#2}{%
{B1p}{\tensor{\vartheta}{_{\perp\indiq[#1]{2}}}}%
{B2m}{\tensor[^{\text{T}}]{\vartheta}{_{\indiq[#1]{3}}}}%
{A0m}{\tensor[^{\text{P}}]{\vartheta}{}}%
{A1p}{\tensor{\overset{\wedge}{\vartheta}}{_{\perp\indiq[#1]{2}}}}%
{A1m}{\tensor{\overset{\rightharpoonup}{\vartheta}}{_{\indiq[#1]{1}}}}%
{A2p}{\tensor{\overset{\sim}{\vartheta}}{_{\perp\indiq[#1]{2}}}}%
{A2m}{\tensor[^{\text{T}}]{\vartheta}{_{\perp\indiq[#1]{3}}}}%
}[\packageError{cosmicclass}{Unidentified Critical Case: #1}{}]%
}

\newrobustcmd{\pic}[2][placeholder]{%
\IfEqCase{#2}{%
{B0p}{\varphi}%
{B1p}{\tensor{\overset{\wedge}{\varphi}}{_{\indiq[#1]{2}}}}%
{B1m}{\tensor{\varphi}{_{\perp\indiq[#1]{1}}}}%
{B2p}{\tensor{\overset{\sim}{\varphi}}{_{\indiq[#1]{2}}}}%
{A0p}{\tensor{\varphi}{_\perp}}%
{A0m}{\tensor[^{\text{P}}]{\varphi}{}}%
{A1p}{\tensor{\overset{\wedge}{\varphi}}{_{\perp\indiq[#1]{2}}}}%
{A1m}{\tensor{\overset{\rightharpoonup}{\varphi}}{_{\indiq[#1]{1}}}}%
{A2p}{\tensor{\overset{\sim}{\varphi}}{_{\perp\indiq[#1]{2}}}}%
{A2m}{\tensor[^{\text{T}}]{\varphi}{_{\indiq[#1]{3}}}}%
}[\packageError{cosmicclass}{Unidentified Critical Case: #1}{}]%
}

\newrobustcmd{\picu}[2][placeholder]{%
\IfEqCase{#2}{%
{B0p}{\varphi}%
{B1p}{\tensor{\smash{\overset{\wedge}{\varphi}}}{^{\indiq[#1]{2}}}}%
{B1m}{\tensor{\varphi}{^{\perp\indiq[#1]{1}}}}%
{B2p}{\tensor{\smash{\overset{\sim}{\varphi}}}{^{\indiq[#1]{2}}}}%
{A0p}{\tensor{\varphi}{_\perp}}%
{A0m}{\tensor[^{\text{P}}]{\varphi}{}}%
{A1p}{\tensor{\smash{\overset{\wedge}{\varphi}}}{^{\perp\indiq[#1]{2}}}}%
{A1m}{\tensor{\smash{\overset{\rightharpoonup}{\varphi}}}{^{\indiq[#1]{1}}}}%
{A2p}{\tensor{\smash{\overset{\sim}{\varphi}}}{^{\perp\indiq[#1]{2}}}}%
{A2m}{\tensor[^{\text{T}}]{\varphi}{^{\indiq[#1]{3}}}}%
}[\packageError{cosmicclass}{Unidentified Critical Case: #1}{}]%
}

\newrobustcmd{\picl}[2][placeholder]{%
\IfEqCase{#2}{%
{B0p}{\tensor{\varphi}{^{\flat}}}%
{B1p}{\tensor{\smash{\overset{\wedge}{\varphi}}}{^{\flat}_{\indiq[#1]{2}}}}%
{B1m}{\tensor{\varphi}{^{\flat}_{\perp}_{\indiq[#1]{1}}}}%
{B2p}{\tensor{\smash{\overset{\sim}{\varphi}}}{^{\flat}_{\indiq[#1]{2}}}}%
{A0p}{\tensor{\varphi}{_\perp}^{\flat}}%
{A0m}{\tensor[^{\text{P}}]{\varphi}{^{\flat}}}%
{A1p}{\tensor{\smash{\overset{\wedge}{\varphi}}}{^{\flat}_{\perp\indiq[#1]{2}}}}%
{A1m}{\tensor{\smash{\overset{\rightharpoonup}{\varphi}}}{^{\flat}_{\indiq[#1]{1}}}}%
{A2p}{\tensor{\smash{\overset{\sim}{\varphi}}}{^{\flat}_{\perp\indiq[#1]{2}}}}%
{A2m}{\tensor[^{\text{T}}]{\varphi}{^{\flat}_{\indiq[#1]{3}}}}%
}[\packageError{cosmicclass}{Unidentified Critical Case: #1}{}]%
}

\newrobustcmd{\mull}[2][placeholder]{%
\IfEqCase{#2}{%
{B0p}{\tensor{u}{^{\flat}}}%
{B1p}{\tensor{\smash{\overset{\wedge}{u}}}{^{\flat}_{\indiq[#1]{2}}}}%
{B1m}{\tensor{u}{^{\flat}_{\perp}_{\indiq[#1]{1}}}}%
{B2p}{\tensor{\smash{\overset{\sim}{u}}}{^{\flat}_{\indiq[#1]{2}}}}%
{A0p}{\tensor{u}{_\perp}^{\flat}}%
{A0m}{\tensor[^{\text{P}}]{u}{^{\flat}}}%
{A1p}{\tensor{\smash{\overset{\wedge}{u}}}{^{\flat}_{\perp\indiq[#1]{2}}}}%
{A1m}{\tensor{\smash{\overset{\rightharpoonup}{u}}}{^{\flat}_{\indiq[#1]{1}}}}%
{A2p}{\tensor{\smash{\overset{\sim}{u}}}{^{\flat}_{\perp\indiq[#1]{2}}}}%
{A2m}{\tensor[^{\text{T}}]{u}{^{\flat}_{\indiq[#1]{3}}}}%
}[\packageError{cosmicclass}{Unidentified Critical Case: #1}{}]%
}

\newrobustcmd{\mul}[2][placeholder]{%
\IfEqCase{#2}{%
{B0p}{\tensor{u}{}}%
{B1p}{\tensor{\smash{\overset{\wedge}{u}}}{_{\indiq[#1]{2}}}}%
{B1m}{\tensor{u}{_{\perp}_{\indiq[#1]{1}}}}%
{B2p}{\tensor{\smash{\overset{\sim}{u}}}{_{\indiq[#1]{2}}}}%
{A0p}{\tensor{u}{_\perp}}%
{A0m}{\tensor[^{\text{P}}]{u}{}}%
{A1p}{\tensor{\smash{\overset{\wedge}{u}}}{_{\perp\indiq[#1]{2}}}}%
{A1m}{\tensor{\smash{\overset{\rightharpoonup}{u}}}{_{\indiq[#1]{1}}}}%
{A2p}{\tensor{\smash{\overset{\sim}{u}}}{_{\perp\indiq[#1]{2}}}}%
{A2m}{\tensor[^{\text{T}}]{u}{_{\indiq[#1]{3}}}}%
}[\packageError{cosmicclass}{Unidentified Critical Case: #1}{}]%
}

\newrobustcmd{\PiP}[2][placeholder]{%
\IfEqCase{#2}{%
{B0p}{\hat{\pi}}%
{B1p}{\tensor{\overset{\wedge}{\hat{\pi}}}{_{\indiq[#1]{2}}}}%
{B1m}{\tensor{\hat{\pi}}{_{\perp\indiq[#1]{1}}}}%
{B2p}{\tensor{\overset{\sim}{\hat{\pi}}}{_{\indiq[#1]{2}}}}%
{A0p}{\tensor{\hat{\pi}}{_\perp}}%
{A0m}{\tensor[^{\text{P}}]{\hat{\pi}}{}}%
{A1p}{\tensor{\overset{\wedge}{\hat{\pi}}}{_{\perp\indiq[#1]{2}}}}%
{A1m}{\tensor{\overset{\rightharpoonup}{\hat{\pi}}}{_{\indiq[#1]{1}}}}%
{A2p}{\tensor{\overset{\sim}{\hat{\pi}}}{_{\perp\indiq[#1]{2}}}}%
{A2m}{\tensor[^{\text{T}}]{\hat{\pi}}{_{\indiq[#1]{3}}}}%
}[\packageError{cosmicclass}{Unidentified Critical Case: #1}{}]%
}

\newrobustcmd{\PiPu}[2][placeholder]{%
\IfEqCase{#2}{%
{B0p}{\hat{\pi}}%
{B1p}{\tensor{\smash{\overset{\wedge}{\hat{\pi}}}}{^{\indiq[#1]{2}}}}%
{B1m}{\tensor{\smash{\hat{\pi}}}{^{\perp\indiq[#1]{1}}}}%
{B2p}{\tensor{\smash{\overset{\sim}{\hat{\pi}}}}{^{\indiq[#1]{2}}}}%
{A0p}{\tensor{\smash{\hat{\pi}}}{^\perp}}%
{A0m}{\tensor[^{\text{P}}]{\smash{\hat{\pi}}}{}}%
{A1p}{\tensor{\smash{\overset{\wedge}{\hat{\pi}}}}{^{\perp\indiq[#1]{2}}}}%
{A1m}{\tensor{\smash{\overset{\rightharpoonup}{\hat{\pi}}}}{^{\indiq[#1]{1}}}}%
{A2p}{\tensor{\smash{\overset{\sim}{\hat{\pi}}}}{^{\perp\indiq[#1]{2}}}}%
{A2m}{\tensor[^{\text{T}}]{\smash{\hat{\pi}}}{^{\indiq[#1]{3}}}}%
}[\packageError{cosmicclass}{Unidentified Critical Case: #1}{}]%
}

\newrobustcmd{\sicl}[2][placeholder]{%
\IfEqCase{#2}{%
{B0p}{\tensor{\chi}{^{\flat}}}%
{B1p}{\tensor{\smash{\overset{\wedge}{\chi}}}{^{\flat}_{\indiq[#1]{2}}}}%
{B1m}{\tensor{\chi}{^{\flat}_{\perp}_{\indiq[#1]{1}}}}%
{B2p}{\tensor{\smash{\overset{\sim}{\chi}}}{^{\flat}_{\indiq[#1]{2}}}}%
{A0p}{\tensor{\chi}{^{\flat}_\perp}}%
{A0m}{\tensor[^{\text{P}}]{\chi}{^{\flat}}}%
{A1p}{\tensor{\smash{\overset{\wedge}{\chi}}}{^{\flat}_{\perp\indiq[#1]{2}}}}%
{A1m}{\tensor{\smash{\overset{\rightharpoonup}{\chi}}}{^{\flat}_{\indiq[#1]{1}}}}%
{A2p}{\tensor{\smash{\overset{\sim}{\chi}}}{^{\flat}_{\perp\indiq[#1]{2}}}}%
{A2m}{\tensor[^{\text{T}}]{\chi}{^{\flat}_{\indiq[#1]{3}}}}%
}[\packageError{cosmicclass}{Unidentified Critical Case: #1}{}]%
}

\newrobustcmd{\ticl}[2][placeholder]{%
\IfEqCase{#2}{%
{B0p}{\tensor{\zeta}{^{\flat}}}%
{B1p}{\tensor{\smash{\overset{\wedge}{\zeta}}}{^{\flat}_{\indiq[#1]{2}}}}%
{B1m}{\tensor{\zeta}{^{\flat}_{\perp}_{\indiq[#1]{1}}}}%
{B2p}{\tensor{\smash{\overset{\sim}{\zeta}}}{^{\flat}_{\indiq[#1]{2}}}}%
{A0p}{\tensor{\zeta}{^{\flat}_\perp}}%
{A0m}{\tensor[^{\text{P}}]{\zeta}{^{\flat}}}%
{A1p}{\tensor{\smash{\overset{\wedge}{\zeta}}}{^{\flat}_{\perp\indiq[#1]{2}}}}%
{A1m}{\tensor{\smash{\overset{\rightharpoonup}{\zeta}}}{^{\flat}_{\indiq[#1]{1}}}}%
{A2p}{\tensor{\smash{\overset{\sim}{\zeta}}}{^{\flat}_{\perp\indiq[#1]{2}}}}%
{A2m}{\tensor[^{\text{T}}]{\zeta}{^{\flat}_{\indiq[#1]{3}}}}%
}[\packageError{cosmicclass}{Unidentified Critical Case: #1}{}]%
}

\newrobustcmd{\PiPl}[2][placeholder]{%
\IfEqCase{#2}{%
{B0p}{\tensor{\hat{\pi}}{^{\flat}}}%
{B1p}{\tensor{\smash{\overset{\wedge}{\hat{\pi}}}}{^{\flat}_{\indiq[#1]{2}}}}%
{B1m}{\tensor{\hat{\pi}}{^{\flat}_{\perp}_{\indiq[#1]{1}}}}%
{B2p}{\tensor{\smash{\overset{\sim}{\hat{\pi}}}}{^{\flat}_{\indiq[#1]{2}}}}%
{A0p}{\tensor{\hat{\pi}}{_\perp}^{\flat}}%
{A0m}{\tensor[^{\text{P}}]{\hat{\pi}}{^{\flat}}}%
{A1p}{\tensor{\smash{\overset{\wedge}{\hat{\pi}}}}{^{\flat}_{\perp\indiq[#1]{2}}}}%
{A1m}{\tensor{\smash{\overset{\rightharpoonup}{\hat{\pi}}}}{^{\flat}_{\indiq[#1]{1}}}}%
{A2p}{\tensor{\smash{\overset{\sim}{\hat{\pi}}}}{^{\flat}_{\perp\indiq[#1]{2}}}}%
{A2m}{\tensor[^{\text{T}}]{\hat{\pi}}{^{\flat}_{\indiq[#1]{3}}}}%
}[\packageError{cosmicclass}{Unidentified Critical Case: #1}{}]%
}

\newrobustcmd{\sic}[2][placeholder]{%
\IfEqCase{#2}{%
{B0p}{\chi}%
{B1p}{\tensor{\overset{\wedge}{\chi}}{_{\indiq[#1]{2}}}}%
{B1m}{\tensor{\chi}{_{\perp\indiq[#1]{1}}}}%
{B2p}{\tensor{\overset{\sim}{\chi}}{_{\indiq[#1]{2}}}}%
{A0p}{\tensor{\chi}{_\perp}}%
{A0m}{\tensor[^{\text{P}}]{\chi}{}}%
{A1p}{\tensor{\overset{\wedge}{\chi}}{_{\perp\indiq[#1]{2}}}}%
{A1m}{\tensor{\overset{\rightharpoonup}{\chi}}{_{\indiq[#1]{1}}}}%
{A2p}{\tensor{\overset{\sim}{\chi}}{_{\perp\indiq[#1]{2}}}}%
{A2m}{\tensor[^{\text{T}}]{\chi}{_{\indiq[#1]{3}}}}%
}[\packageError{cosmicclass}{Unidentified Critical Case: #1}{}]%
}

\newrobustcmd{\lorsicpar}[2][placeholder]{%
\IfEqCase{#2}{%
{B0m}{\tensor*[^{\text{P}}]{\smash{\underline{\chi}}}{^{\parallel}}}%
{B1p}{\tensor*{\smash{\overset{\wedge}{\chi}}}{^{\parallel}_{\indiq[#1]{2}}}}%
{B1m}{\tensor*{\chi}{^{\parallel}_{\perp\indiq[#1]{1}}}}%
{B2m}{\tensor*[^{\text{T}}]{\smash{\underline{\chi}}}{^{\parallel}_{\indiq[#1]{3}}}}%
{A0p}{\tensor*{\chi}{^{\parallel}_\perp}}%
{A0m}{\tensor*[^{\text{P}}]{\chi}{^{\parallel}}}%
{A1p}{\tensor*{\smash{\overset{\wedge}{\chi}}}{^{\parallel}_{\perp\indiq[#1]{2}}}}%
{A1m}{\tensor*{\smash{\overset{\rightharpoonup}{\chi}}}{^{\parallel}_{\indiq[#1]{1}}}}%
{A2p}{\tensor*{\smash{\overset{\sim}{\chi}}}{^{\parallel}_{\perp\indiq[#1]{2}}}}%
{A2m}{\tensor*[^{\text{T}}]{\chi}{^{\parallel}_{\indiq[#1]{3}}}}%
}[\packageError{cosmicclass}{Unidentified Critical Case: #1}{}]%
}

\newrobustcmd{\lorsicpir}[2][placeholder]{%
\IfEqCase{#2}{%
{B0p}{\tensor*{\chi}{^{\vDash}}}%
{B1p}{\tensor*{\smash{\overset{\wedge}{\chi}}}{^{\vDash}_{\indiq[#1]{2}}}}%
{B1m}{\tensor*{\chi}{^{\vDash}_{\perp\indiq[#1]{1}}}}%
{B2p}{\tensor*{\smash{\overset{\sim}{\chi}}}{^{\vDash}_{\indiq[#1]{2}}}}%
{A0p}{\tensor*{\chi}{^{\vDash}_\perp}}%
{A0m}{\tensor*[^{\text{P}}]{\chi}{^{\vDash}}}%
{A1p}{\tensor*{\smash{\overset{\wedge}{\chi}}}{^{\vDash}_{\perp\indiq[#1]{2}}}}%
{A1m}{\tensor*{\smash{\overset{\rightharpoonup}{\chi}}}{^{\vDash}_{\indiq[#1]{1}}}}%
{A2p}{\tensor*{\smash{\overset{\sim}{\chi}}}{^{\vDash}_{\perp\indiq[#1]{2}}}}%
{A2m}{\tensor*[^{\text{T}}]{\chi}{^{\vDash}_{\indiq[#1]{3}}}}%
}[\packageError{cosmicclass}{Unidentified Critical Case: #1}{}]%
}

\newrobustcmd{\lorsicper}[2][placeholder]{%
\IfEqCase{#2}{%
{B0p}{\tensor*{\chi}{^{\perp}}}%
{B1p}{\tensor*{\smash{\overset{\wedge}{\chi}}}{^{\perp}_{\indiq[#1]{2}}}}%
{B1m}{\tensor*{\chi}{^{\perp}_{\perp\indiq[#1]{1}}}}%
{B2p}{\tensor*{\smash{\overset{\sim}{\chi}}}{^{\perp}_{\indiq[#1]{2}}}}%
{A0p}{\tensor*{\chi}{^{\perp}_\perp}}%
{A0m}{\tensor*[^{\text{P}}]{\chi}{^{\perp}}}%
{A1p}{\tensor*{\smash{\overset{\wedge}{\chi}}}{^{\perp}_{\perp\indiq[#1]{2}}}}%
{A1m}{\tensor*{\smash{\overset{\rightharpoonup}{\chi}}}{^{\perp}_{\indiq[#1]{1}}}}%
{A2p}{\tensor*{\smash{\overset{\sim}{\chi}}}{^{\perp}_{\perp\indiq[#1]{2}}}}%
{A2m}{\tensor*[^{\text{T}}]{\chi}{^{\perp}_{\indiq[#1]{3}}}}%
}[\packageError{cosmicclass}{Unidentified Critical Case: #1}{}]%
}

\newrobustcmd{\Tl}[2][placeholder]{%
\IfEqCase{#2}{%
{B0p}{\tensor{\chi}{^{\flat}}}%
{B1p}{\tensor{\smash{\overset{\wedge}{\chi}}}{^{\flat}_{\indiq[#1]{2}}}}%
{B1m}{\tensor{\chi}{^{\flat}_{\perp}_{\indiq[#1]{1}}}}%
{B2p}{\tensor{\smash{\overset{\sim}{\chi}}}{^{\flat}_{\indiq[#1]{2}}}}%
{A0p}{\tensor{\chi}{^{\flat}_\perp}}%
{A0m}{\tensor[^{\text{P}}]{\mathcal{T}}{^{\flat}}}%
{A1p}{\tensor{\smash{\overset{\wedge}{\chi}}}{^{\flat}_{\perp\indiq[#1]{2}}}}%
{A1m}{\tensor{\smash{\overset{\rightharpoonup}{\mathcal{T}}}}{^{\flat}_{\indiq[#1]{1}}}}%
{A2p}{\tensor{\smash{\overset{\sim}{\chi}}}{^{\flat}_{\perp\indiq[#1]{2}}}}%
{A2m}{\tensor[^{\text{T}}]{\mathcal{T}}{^{\flat}_{\indiq[#1]{3}}}}%
}[\tensor{\mathcal{T}}{^{\flat}_{\indiq[#1]{3}}}]%
}

\newrobustcmd{\cT}[2][placeholder]{%
\IfEqCase{#2}{%
{B1p}{\tensor{\mathcal{T}}{_{\perp\indiq[#1]{2}}}}%
{B1m}{\tensor{\overset{\rightharpoonup}{\mathcal{T}}}{_{\indiq[#1]{1}}}}%
{A0m}{\tensor[^{\text{P}}]{\mathcal{T}}{}}%
{A2m}{\tensor[^{\text{T}}]{\mathcal{T}}{_{\indiq[#1]{3}}}}%
}[\packageError{cosmicclass}{Unidentified Critical Case: #1}{}]%
}

\newrobustcmd{\cTLambda}[2][placeholder]{%
\IfEqCase{#2}{%
{B1p}{\tensor{\zeta}{_{\perp\indiq[#1]{2}}}}%
{B1m}{\tensor{\overset{\rightharpoonup}{\zeta}}{_{\indiq[#1]{1}}}}%
{A0m}{\tensor[^{\text{P}}]{\zeta}{}}%
{A2m}{\tensor[^{\text{T}}]{\zeta}{_{\indiq[#1]{3}}}}%
}[\packageError{cosmicclass}{Unidentified Critical Case: #1}{}]%
}

\newrobustcmd{\cTpic}[2][placeholder]{%
\IfEqCase{#2}{%
{B1p}{\tensor{\phi}{_{\perp\indiq[#1]{2}}}}%
{B1m}{\tensor{\overset{\rightharpoonup}{\phi}}{_{\indiq[#1]{1}}}}%
{A0m}{\tensor[^{\text{P}}]{\phi}{}}%
{A2m}{\tensor[^{\text{T}}]{\phi}{_{\indiq[#1]{3}}}}%
}[\packageError{cosmicclass}{Unidentified Critical Case: #1}{}]%
}

\newrobustcmd{\cTpicl}[2][placeholder]{%
\IfEqCase{#2}{%
{B1p}{\tensor{\phi}{^{\flat}_{\perp\indiq[#1]{2}}}}%
{B1m}{\tensor{\overset{\rightharpoonup}{\phi}}{^{\flat}_{\indiq[#1]{1}}}}%
{A0m}{\tensor[^{\text{P}}]{\phi}{^{\flat}}}%
{A2m}{\tensor[^{\text{T}}]{\phi}{^{\flat}_{\indiq[#1]{3}}}}%
}[\packageError{cosmicclass}{Unidentified Critical Case: #1}{}]%
}

\newrobustcmd{\cTPiP}[2][placeholder]{%
\IfEqCase{#2}{%
{B1p}{\tensor{\varpi}{_{\perp\indiq[#1]{2}}}}%
{B1m}{\tensor{\overset{\rightharpoonup}{\varpi}}{_{\indiq[#1]{1}}}}%
{A0m}{\tensor[^{\text{P}}]{\varpi}{}}%
{A2m}{\tensor[^{\text{T}}]{\varpi}{_{\indiq[#1]{3}}}}%
}[\packageError{cosmicclass}{Unidentified Critical Case: #1}{}]%
}

\newrobustcmd{\cTl}[2][placeholder]{%
\IfEqCase{#2}{%
{B1p}{\tensor{\mathcal{T}}{^{\flat}_{\perp\indiq[#1]{2}}}}%
{B1m}{\tensor{\smash{\overset{\rightharpoonup}{\mathcal{T}}}}{^{\flat}_{\indiq[#1]{1}}}}%
{A0m}{\tensor[^{\text{P}}]{\mathcal{T}}{^{\flat}}}%
{A2m}{\tensor[^{\text{T}}]{\mathcal{T}}{^{\flat}_{\indiq[#1]{3}}}}%
}[\packageError{cosmicclass}{Unidentified Critical Case: #1}{}]%
}

\newrobustcmd{\cTu}[2][placeholder]{%
\IfEqCase{#2}{%
{B1p}{\tensor{\mathcal{T}}{^{\perp\indiq[#1]{2}}}}%
{B1m}{\tensor{\smash{\overset{\rightharpoonup}{\mathcal{T}}}}{^{\indiq[#1]{1}}}}%
{A0m}{\tensor[^{\text{P}}]{\mathcal{T}}{}}%
{A2m}{\tensor[^{\text{T}}]{\mathcal{T}}{^{\indiq[#1]{3}}}}%
}[\packageError{cosmicclass}{Unidentified Critical Case: #1}{}]%
}

\newrobustcmd{\ncTLambda}[2][placeholder]{%
\IfEqCase{#2}{%
{B0p}{\tensor{\zeta}{^{\indiq[#1]{1}}_{\indiq[#1]{1}\perp}}}%
{B1p}{\tensor{\zeta}{_{[\indiq[#1]{2}]\perp}}}%
{B1m}{\tensor{\zeta}{_{\perp\indiq[#1]{1}\perp}}}%
{B2p}{\tensor{\zeta}{_{\langle\indiq[#1]{2}\rangle\perp}}}%
}[\packageError{cosmicclass}{Unidentified Critical Case: #1}{}]%
}

\newrobustcmd{\ncTmul}[2][placeholder]{%
\IfEqCase{#2}{%
{B0p}{\tensor{\upsilon}{^{\indiq[#1]{1}}_{\indiq[#1]{1}\perp}}}%
{B1p}{\tensor{\upsilon}{_{[\indiq[#1]{2}]\perp}}}%
{B1m}{\tensor{\upsilon}{_{\perp\indiq[#1]{1}\perp}}}%
{B2p}{\tensor{\upsilon}{_{\langle\indiq[#1]{2}\rangle\perp}}}%
}[\packageError{cosmicclass}{Unidentified Critical Case: #1}{}]%
}

\newrobustcmd{\ncTpic}[2][placeholder]{%
\IfEqCase{#2}{%
{B0p}{\tensor{\phi}{^{\indiq[#1]{1}}_{\indiq[#1]{1}\perp}}}%
{B1p}{\tensor{\phi}{_{[\indiq[#1]{2}]\perp}}}%
{B1m}{\tensor{\phi}{_{\perp\indiq[#1]{1}\perp}}}%
{B2p}{\tensor{\phi}{_{\langle\indiq[#1]{2}\rangle\perp}}}%
}[\packageError{cosmicclass}{Unidentified Critical Case: #1}{}]%
}

\newrobustcmd{\ncTpicl}[2][placeholder]{%
\IfEqCase{#2}{%
  {B0p}{\tensor{\phi}{^{\flat}^{\indiq[#1]{1}}_{\indiq[#1]{1}\perp}}}%
{B1p}{\tensor{\phi}{^{\flat}_{[\indiq[#1]{2}]\perp}}}%
{B1m}{\tensor{\phi}{^{\flat}_{\perp\indiq[#1]{1}\perp}}}%
{B2p}{\tensor{\phi}{^{\flat}_{\langle\indiq[#1]{2}\rangle\perp}}}%
}[\packageError{cosmicclass}{Unidentified Critical Case: #1}{}]%
}

\newrobustcmd{\ncTPiP}[2][placeholder]{%
\IfEqCase{#2}{%
{B0p}{\tensor{\varpi}{^{\indiq[#1]{1}}_{\indiq[#1]{1}\perp}}}%
{B1p}{\tensor{\varpi}{_{[\indiq[#1]{2}]\perp}}}%
{B1m}{\tensor{\varpi}{_{\perp\indiq[#1]{1}\perp}}}%
{B2p}{\tensor{\varpi}{_{\langle\indiq[#1]{2}\rangle\perp}}}%
}[\packageError{cosmicclass}{Unidentified Critical Case: #1}{}]%
}

\newrobustcmd{\ncT}[2][placeholder]{%
\IfEqCase{#2}{%
{B0p}{\tensor{\mathcal{T}}{^{\indiq[#1]{1}}_{\indiq[#1]{1}\perp}}}%
{B1p}{\tensor{\mathcal{T}}{_{[\indiq[#1]{2}]\perp}}}%
{B1m}{\tensor{\mathcal{T}}{_{\perp\indiq[#1]{1}\perp}}}%
{B2p}{\tensor{\mathcal{T}}{_{\langle\indiq[#1]{2}\rangle\perp}}}%
}[\packageError{cosmicclass}{Unidentified Critical Case: #1}{}]%
}

\newrobustcmd{\cR}[2][placeholder]{%
\IfEqCase{#2}{%
{A0p}{\tensor{\underline{\mathcal{R}}}{}}%
{A0m}{\tensor[^{\text{P}}]{\mathcal{R}}{_{\perp\circ}}}%
{A1p}{\tensor{\underline{\mathcal{R}}}{_{[\indiq[#1]{2}]}}}%
{A1m}{\tensor{\mathcal{R}}{_{\perp\indiq[#1]{1}}}}%
{A2p}{\tensor{\underline{\mathcal{R}}}{_{\langle\indiq[#1]{2}\rangle}}}%
{A2m}{\tensor[^{\text{T}}]{\mathcal{R}}{_{\perp\indiq[#1]{3}}}}%
}[\packageError{cosmicclass}{Unidentified Critical Case: #1}{}]%
}

\newrobustcmd{\cRLambda}[2][placeholder]{%
\IfEqCase{#2}{%
{A0p}{\tensor{\underline{\zeta}}{}}%
{A0m}{\tensor[^{\text{P}}]{\zeta}{_{\perp\circ}}}%
{A1p}{\tensor{\underline{\zeta}}{_{[\indiq[#1]{2}]}}}%
{A1m}{\tensor{\zeta}{_{\perp\indiq[#1]{1}}}}%
{A2p}{\tensor{\underline{\zeta}}{_{\langle\indiq[#1]{2}\rangle}}}%
{A2m}{\tensor[^{\text{T}}]{\zeta}{_{\perp\indiq[#1]{3}}}}%
}[\packageError{cosmicclass}{Unidentified Critical Case: #1}{}]%
}

\newrobustcmd{\cRl}[2][placeholder]{%
\IfEqCase{#2}{%
  {A0p}{\tensor{\underline{\mathcal{R}}}{^{\flat}}}%
{A0m}{\tensor[^{\text{P}}]{\mathcal{R}}{^{\flat}_{\perp\circ}}}%
{A1p}{\tensor{\underline{\mathcal{R}}}{^{\flat}_{[\indiq[#1]{2}]}}}%
{A1m}{\tensor{\mathcal{R}}{^{\flat}_{\perp\indiq[#1]{1}}}}%
{A2p}{\tensor{\underline{\mathcal{R}}}{^{\flat}_{\langle\indiq[#1]{2}\rangle}}}%
{A2m}{\tensor[^{\text{T}}]{\mathcal{R}}{^{\flat}_{\perp\indiq[#1]{3}}}}%
}[\packageError{cosmicclass}{Unidentified Critical Case: #1}{}]%
}

\newrobustcmd{\cRu}[2][placeholder]{%
\IfEqCase{#2}{%
{A0p}{\tensor{\underline{\mathcal{R}}}{}}%
{A0m}{\tensor[^{\text{P}}]{\mathcal{R}}{_{\perp\circ}}}%
{A1p}{\tensor{\underline{\mathcal{R}}}{^{[\indiq[#1]{2}]}}}%
{A1m}{\tensor{\mathcal{R}}{^{\perp\indiq[#1]{1}}}}%
{A2p}{\tensor{\underline{\mathcal{R}}}{^{\langle\indiq[#1]{2}\rangle}}}%
{A2m}{\tensor[^{\text{T}}]{\mathcal{R}}{^{\perp\indiq[#1]{3}}}}%
}[\packageError{cosmicclass}{Unidentified Critical Case: #1}{}]%
}

\newrobustcmd{\ncR}[2][placeholder]{%
\IfEqCase{#2}{%
  {A0p}{\tensor{\mathcal{R}}{_{\perp\perp}}}%
{A0m}{\tensor[^{\text{P}}]{\mathcal{R}}{_{\circ\perp}}}%
{A1p}{\tensor{\mathcal{R}}{_{\perp[\indiq[#1]{2}]\perp}}}%
{A1m}{\tensor{\mathcal{R}}{_{\indiq[#1]{1}\perp}}}%
{A2p}{\tensor{\mathcal{R}}{_{\perp\langle\indiq[#1]{2}\rangle\perp}}}%
{A2m}{\tensor[^{\text{T}}]{\mathcal{R}}{_{\indiq[#1]{3}\perp}}}%
}[\packageError{cosmicclass}{Unidentified Critical Case: #1}{}]%
}

\newrobustcmd{\ncRLambda}[2][placeholder]{%
\IfEqCase{#2}{%
  {A0p}{\tensor{\zeta}{_{\perp\perp}}}%
{A0m}{\tensor[^{\text{P}}]{\zeta}{_{\circ\perp}}}%
{A1p}{\tensor{\zeta}{_{\perp[\indiq[#1]{2}]\perp}}}%
{A1m}{\tensor{\zeta}{_{\indiq[#1]{1}\perp}}}%
{A2p}{\tensor{\zeta}{_{\perp\langle\indiq[#1]{2}\rangle\perp}}}%
{A2m}{\tensor[^{\text{T}}]{\zeta}{_{\indiq[#1]{3}\perp}}}%
}[\packageError{cosmicclass}{Unidentified Critical Case: #1}{}]%
}

\newrobustcmd{\Proj}[2][placeholder]{%
\IfEqCase{#2}{%
  {A2m}{\tensor[^{\text{T}}]{\check{\mathcal{P}}}{#1}}%
}[\packageError{cosmicclass}{Unidentified Critical Case: #1}{}]%
}

\newrobustcmd{\Projl}[2][placeholder]{%
\IfEqCase{#2}{%
  {A2m}{\tensor[^{\text{T}}]{\check{\mathcal{P}}}{^{\flat}#1}}%
}[\packageError{cosmicclass}{Unidentified Critical Case: #1}{}]%
}

\newrobustcmd{\fA}{%
  {\tensor{\mathcal{  A}}{_{\acu{u}}}}%
}
\newrobustcmd{\fB}{%
  {\tensor{\mathcal{  B}}{_{\acu{v}}}}%
}
\newrobustcmd{\fC}{%
  {\tensor{\mathcal{  C}}{^{\acu{v}}}}%
}
\newrobustcmd{\fphi}{%
  {\tensor{\phi}{^{\acu{w}}}}%
}
\newrobustcmd{\fpi}{%
  {\tensor{\pi}{_{\acu{w}}}}%
}

\newrobustcmd{\covard}[2]{%
  {\frac{\bar{\delta}#1}{\bar{\delta}#2}}
}
\newrobustcmd{\copard}[2]{%
  {\frac{\bar{\partial}#1}{\bar{\partial}#2}}
}
\newrobustcmd{\pard}[2]{%
  {\frac{\partial #1}{\partial #2}}
}

\newrobustcmd{\PPM}[1]{%
  {\left[\tensor*{\mathsf{M}}{_{\ }^{\left(\text{#1}\right)}}\right]}%
}

\begin{document}

\title{Particle spectra of general Ricci-type Palatini or metric-affine theories}

\author{W. Barker}
\email{wb263@cam.ac.uk}
\affiliation{Astrophysics Group, Cavendish Laboratory, JJ Thomson Avenue, Cambridge CB3 0HE, UK}
\affiliation{Kavli Institute for Cosmology, Madingley Road, Cambridge CB3 0HA, UK}

\author{C. Marzo}
\email{carlo.marzo@kbfi.ee}
\affiliation{Laboratory for High Energy and Computational Physics, NICPB, R\"{a}vala 10, Tallinn 10143, Estonia}

\begin{abstract}
	In the context of weak-field metric-affine (i.e. Palatini) gravity near Minkowski spacetime, we compute the particle spectra in the simultaneous presence of all independent contractions quadratic in Ricci-type tensors. Apart from the full metric-affine geometry, we study kinematic limits with vanishing torsion (i.e. a symmetric connection) and vanishing non-metricity (i.e. a metric connection, which is physically indistinguishable from Poincar\'e gauge theory at the level of the particle spectrum). We present a detailed report on how spin-parity projection operators can be used to derive systematically and unambiguously the character of the propagating states. The unitarity constraints derived from the requirements of tachyon- and ghost-freedom are obtained. We show that, even in the presence of all Ricci-type operators, only a narrow selection of viable theories emerges by a tuning.
\end{abstract}

\maketitle

\section{Introduction}
\begin{figure}
\includegraphics[width=1\linewidth]{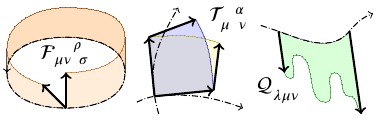}
	\caption{The effects of metric-affine curvature $\MAGF{_{\mu\nu}^\rho_\sigma}$ in~\cref{FDef} -- vector rotation after parallel transport in a closed loop -- torsion $\MAGT{_\mu^\alpha_\nu}$ in~\cref{tqdef} -- non-closure of parallelograms formed from (infinitesimally) parallel-transported vectors -- and non-metricity $\MAGQ{_{\lambda\mu\nu}}$ in~\cref{tqdef} -- change in vector length under parallel transport.}
\label{Fig1}
\end{figure}

The success of the curvature-based geometrical formulation of gravity has stimulated a search for the dynamical interpretation of geometrical properties analogous to curvature. This program has been directed to solve, or mitigate, some of the main shortcomings of the current status quo, mainly the lack of a perturbatively renormalizable quantum theory of gravity, and the phenomenological need for a dark sector. While hopes to address the renormalization issue have, so far, had to rely on non-traditional routes~\cite{Salvio:2014soa,Alvarez-Gaume:2015rwa,Stelle:1976gc,Anselmi:2017ygm,Anselmi:2017yux,Ohta:2015zwa, Percacci:2009fh, Bender:2007wu, Mannheim:2011ds, Donoghue:2021cza, Donoghue:2021eto,Karananas:2023zgg,Karananas:2021gco}, the possibility of interesting phenomenology from particles of geometrical origin is frequently used in cosmology and dark-sector model-building~\cite{Iosifidis:2023but, Shapiro:2001rz, Mondal:2023cxx, Rigouzzo:2023sbb, Fomin:2023tsn, Gonzalez-Espinoza:2020jss, Arcos:2004tzt, Karananas:2021zkl, Karananas:2018nrj, Gialamas:2022xtt, Mikura:2020qhc, Shimada:2018lnm, Azri:2018qux, Iosifidis:2022xvp,Dioguardi:2021fmr,Dioguardi:2022oqu,Dioguardi:2023jwa}. Metric-affine gravity (MAG)~\cite{Percacci:2020ddy,Hehl:1994ue, BeltranJimenez:2019acz,Baldazzi:2021kaf,Neville:1978bk,Neville:1979rb,Nezhad:2023dys,Annala:2022gtl,Rasanen:2022ijc,Ito:2021ssc,Annala:2021zdt,Enckell:2020lvn,Rasanen:2018ihz,Enckell:2018hmo,Rasanen:2017ivk,Melichev:2023lpn,Baldazzi:2021kaf,Mikura:2024mji,Mikura:2023ruz,Percacci:2020bzf,Mikura:2021ldx,Mikura:2021clt,He:2022xef} represents the principal realization of this program. It broadens the matter content by considering the affine connection as an independent three-index object, giving rise to torsion and non-metricity (see~\cref{Fig1}) as dynamical fields. This results in a notable growth in computational complexity, both in the number of allowed operators and in the profiling of the multiple particle states carried in by the affine connection. Such a broad parameter space is expected, as is often the case, to narrow under the pressure of field-theoretical self-consistency constraints, such as unitarity and elimination of tachyons~\cite{VanNieuwenhuizen:1973fi,Sezgin:1979zf,Sezgin:1981xs,Neville:1979rb,Neville:1978bk,Percacci:2020ddy,Marzo:2021iok,Mikura:2023ruz,Mikura:2024mji,Karananas:2014pxa,Lin:2018awc,Lin:2020phk,Annala:2022gtl,Barker:2023fem}. The imposition of these requirements is a highly non-trivial task and has often necessitated severe simplifications to arrive at positive scenarios. Also, the analysis often has to rely on a very indirect route, without directly accessing the pole structure of the propagator.

In this paper, we use the arena of Ricci-type MAG to illustrate, in a detailed step-by-step fashion, how the formalism of spin-parity projectors can unambiguously and straightforwardly reveal the nature of the (tree-level) particle spectrum. By \emph{Ricci-type} we refer to all the operators that may be added to the Einstein--Hilbert Lagrangian which are quadratic in the rank-two traces of metric-affine curvature (in MAG there are nine such operators, whilst in standard GR there is only one). Building on early work (see e.g.~\cite{Vitagliano:2010pq,Allemandi:2004wn,Li:2007xw,Olmo:2009xy,Barragan:2010qb,Bauer:2011sft}), this particular space of operators was first properly charted by Annala and Räsänen~\cite{Annala:2022gtl}. The reparameterisation-based methods used in that work are particularly innovative, but they are only able to access a `punctuated' bulk of the full parameter space due to certain \emph{degeneracy conditions} which must be avoided. In the present work we build on these foundations by applying the spin-parity projection formalism: our method does not come with any restriction on the parameters of the model. Indeed, if the set of projector operators is known, the approach adopted can be applied to every tensor-valued Lagrangian admitting a Minkowskian background expansion. We have authored a new Wolfram Language implementation of this procedure for all such theories, including the MAG: \emph{Particle Spectrum for any Tensor Lagrangian} (\PSALTer{}). The \PSALTer{} software will be properly presented in a dedicated paper~\cite{Barker:2024}\footnote{See~\cite{Barker:2023bmr} for a recent application of \PSALTer{} to theories proposed in~\cite{Partanen:2023dkt,Partanen:2023sjn}.}, but in this paper we use it to confirm all our results (see~\cref{SecMetApp,SecTorApp,SecGenApp}). The body this paper is set out as follows. In~\cref{SecProj} we develop the underlying theory by setting out the MAG conventions in~\cref{MAGConv} and briefly presenting our spectral algorithm in~\cref{SatPropSec} and the computer implementation in~\cref{PSALTerSec}. In~\cref{Results} we present all of the results, respectively for the vanishing non-metricity kinematic limit in~\cref{SecMet} and the vanishing torsion kinematic limit in~\cref{SecTor}, and for the full MAG in~\cref{SecGen}. Conclusions follow in~\cref{Conclusions}. We will use the `West Coast' signature ${(+,-,-,-)}$, other conventions will be introduced as needed.

\section{Theoretical development}\label{SecProj}
The requirements of locality and Lorentz invariance select tensor fields as the building-blocks for most of the theoretical speculations about high-energy physics. The price to pay for using them is in the risk of uncontrolled unitarity violations. It is a profound realization that the consistent adjustments to recover unitarity restrict us to Maxwell's theory, for rank-one fields, as well as Einstein's theory of gravity, when applied to models of symmetric rank-two fields~\cite{Gupta:1954zz,Deser:1963zzc,Deser:1969wk,VanNieuwenhuizen:1973fi,Fang:1978rc}. The continuation of this program within the intricate scenarios of higher-rank fields, as well as multi-field quadratic Lagrangians, is technically prohibitive. Many indirect shortcuts have supported claims of healthy particle propagation, but the proliferation of indices often prevents a direct approach to the pole structure of the propagator. The spin-parity projection approach to spectral analysis has been most thoroughly expostulated by Lin, Hobson and Lasenby in~\cite{Lin:2018awc}, with follow-up papers in~\cite{Lin:2019ugq,Lin:2020phk}. Further information about the method can also be found in (e.g.)~\cite{Karananas:2014pxa,Percacci:2020ddy,VanNieuwenhuizen:1973fi,Sezgin:1979zf,Sezgin:1981xs,Kuhfuss:1986rb,Neville:1979rb} --- but it is~\cite{Lin:2018awc} which provides the most concise introduction for our purposes.

\subsection{Spin-parity kinematics of metric-affine gravity}\label{MAGConv}
To have full control over the particle spectrum we introduce operators which project the Lorentz index structure onto the labels $J$ and $P$ of spin and parity, enumerating the irreducible representation under the $\mathrm{SU}(2)$ little group. In general, further reduction to the $\mathrm{U}(1)$ little group of massless particles and helicity states can be done. This results in no practical improvement, making the decomposition into $\mathrm{SU}(2)$ representations generic enough. In the case of MAG, the needed set of projection operators has been recently completed~\cite{Percacci:2020ddy,Mendonca:2019gco}, developing the past studies of~\cite{Rivers1964,VanNieuwenhuizen:1973fi,Neville:1978bk,Neville:1979rb,Sezgin:1979zf,Sezgin:1981xs,Karananas:2014pxa}. We present here a summary of the main ideas regarding the use of projection operators for the computation of the poles and residues of the propagator. We refer to~\cite{Lin:2018awc,Sezgin:1979zf,Percacci:2020ddy,Marzo:2021esg} for further details.

The independent MAG connection $\MAGA{_\mu^\nu_\rho}$ carries in 64 new degrees of freedom (d.o.f), which are not present in the derived Levi--Civita connection $\Con{_\mu^\nu_\rho}\equiv\Con{_{(\mu|}^\nu_{|\rho)}}\equiv\tensor{g}{^{\nu\lambda}}\big(\PD{_{(\mu}}\g{_{\rho)\lambda}}-\frac{1}{2}\PD{_{\lambda}}\g{_{\mu\rho}}\big)$. These new d.o.f are often partitioned into the torsion and non-metricity tensors
\begin{equation}
	\MAGT{_\mu^\alpha_\nu}\equiv 2\MAGA{_{[\mu|}^\alpha_{|\nu]}},\quad
	\MAGQ{_{\lambda\mu\nu}}\equiv-\tensor{\partial}{_\lambda}\MAGg{_{\mu\nu}}+2\MAGA{_\lambda^\alpha_{(\mu|}}\MAGg{_{\alpha|\nu)}}.\label{tqdef}
\end{equation}
The two tensors in~\cref{tqdef} are geometric counterparts to the metric-affine (i.e. non-Riemannian) curvature
\begin{equation}
	\MAGF{_{\mu\nu}^\rho_\sigma}\equiv 2\left(\tensor{\partial}{_{[\mu}}\MAGA{_{\nu]}^\rho_\sigma}+\MAGA{_{[\mu|}^\rho_\alpha}\MAGA{_{|\nu]}^\alpha_\sigma}\right).\label{FDef}
\end{equation}
The influences of these three geometric properties on vectors which are being parallel-transported through the spacetime are shown in~\cref{Fig1}. We will be particularly interested in the three \emph{Ricci-type} contractions of the metric-affine curvature 
\begin{align}
		\MAGF{_{\mu\nu}}&\equiv\MAGF{_{\mu\nu\alpha}^\alpha},\quad
		\MAGFa{_{\mu\nu}}\equiv\MAGF{_{\alpha\mu\nu}^\alpha},\quad
		\MAGFb{_{\mu\nu}}\equiv\MAGF{_{\alpha\mu}^\alpha_{\nu}},\label{Contractions}
\end{align}
by which we just mean the contractions with two free indices. The quantity $\MAGF{_{\mu\nu}}$ is the \emph{homothetic} curvature~\cite{BeltranJimenez:2016wxw,Iosifidis:2018jwu}, while $\MAGFb{_{\mu\nu}}$ and $\MAGFa{_{\mu\nu}}$ are variously \emph{pseudo}-Ricci tensors~\cite{Percacci:2020ddy}, or $\MAGFa{_{\mu\nu}}$ is the \emph{co}-Ricci~\cite{BeltranJimenez:2016wxw}. The Riemannian curvature of course yields only one Ricci-type contraction: the homothetic curvature vanishes identically in the absence of non-metricity, and the (co-)Ricci tensors coincide in the absence of torsion. We define $\MAGF{}\equiv\MAGF{_{\mu\nu}^{\mu\nu}}$ as the Ricci scalar: there is still only one such scalar in metric-affine geometry.

In the \emph{first-order} or \emph{Palatini} parameterisation of MAG, we take the 10 d.o.f in $\g{_{\mu\nu}}$ and the $64$ new d.o.f in $\MAGA{_\mu^\nu_\rho}$ to be fundamental fields. An advantage of the first-order parameterisation is that the MAG field strength tensor in~\cref{FDef} is free from second derivatives. In the \emph{second-order} or \emph{post-Riemannian} parameterisation, we keep $\g{_{\mu\nu}}$ but treat the tensor-valued difference $\MAGD{_\mu^\nu_\rho}\equiv\MAGA{_\mu^\nu_\rho}-\Con{_\mu^\nu_\rho}$ (sometimes termed the \emph{distortion}) as fundamental, effectively partitioning the 64 d.o.f among the $24$ d.o.f in $\MAGT{_\mu^\alpha_\nu}\equiv\MAGT{_{[\mu|}^\alpha_{|\nu]}}$ plus the $40$ d.o.f in $\MAGQ{_{\lambda\mu\nu}}\equiv\MAGQ{_{\lambda(\mu\nu)}}$ according to
\begin{equation}\label{TorNonMe}
	\MAGT{_\mu^\alpha_\nu}\equiv 2\MAGD{_{[\mu|}^\alpha_{|\nu]}},\quad
	\MAGQ{_{\lambda\mu\nu}}\equiv 2\MAGD{_{\lambda(\mu\nu)}}.
\end{equation}
In the second-order parameterisation~\cref{FDef} is expanded into the Riemannian curvature (which naturally has second-derivatives in $\g{_{\mu\nu}}$) and many other terms which are (Levi--Civita) covariant derivatives and second powers of $\MAGT{_\mu^\alpha_\nu}$ and $\MAGQ{_{\lambda\mu\nu}}$. The second-order parameterisation has an advantage in revealing the true nature of all MAG-type theories: every MAG theory is a (non-)minimal coupling of standard metric-based gravity to an asymmetric rank-three matter field $\MAGD{_\mu^\rho_\nu}$. We are free to work with either set of variables due to re-parameterisation invariance of the physics.

Working in the first-order formulation, the weak-field regime near to Minkowski spacetime can be captured by an inherently perturbative $\MAGA{_\mu^\nu_\rho}$ and a metric perturbation $\g{_{\mu\nu}}\equiv\tensor{\eta}{_{\mu\nu}}+\tensor{h}{_{\mu\nu}}$. These perturbations carry multiple massive particle states
\begin{subequations}
\begin{align} 
	A_{\mu \nu \rho} &\supset  \Big\{ 3_1^- , 2_1^+ , 2_2^+ , 2_3^+ , 2_1^- , 2_2^- , 1^+_1 , 1^+_2 , 1^+_3  , 1^-_1 , 1^-_2 ,  \nn \\
	&\hspace{25pt} 1^-_3 , 1^-_4 , 1^-_5 , 1^-_6 , 0^+_1 , 0^+_2 , 0^+_3 , 0^+_4 , 0^-_1 \Big\}, \label{ParticleDec1} \w2 
	h_{\mu \nu} &\supset  \Big\{ 2_4^+ , 1^-_7 , 0^+_5 , 0^+_6 \Big\},\label{ParticleDec2}
\end{align}
\end{subequations}
where we used the compact notation $J^P_j$ in referring to the j-th representation of spin $J$ and parity $P$. Enumeration convention is adopted from~\cite{Percacci:2020ddy}. From~\cref{ParticleDec1,ParticleDec2} the 64 and 10 d.o.f can be recovered respectively by summing the multiplicities $2J+1$ over all states. The \PSALTer{} notation for these states differs from the subscript notation in~\cref{ParticleDec1,ParticleDec2}, and full definitions of these states are provided in~\cref{MetricAffineGravity}.

\subsection{Saturated propagator and particle spectra}\label{SatPropSec}

Following~\cite{Marzo:2021esg}, we use a synthetic notation to describe the quadratic (i.e. perturbative) action in momentum space
\begin{align} \label{ActionMom}
	&S[\Phi] =  \frac{1}{2} \int \mathrm{d}^4 k \, \bigg( \Phi(-k) \,  \mathcal \, K(k) \, \Phi(k) + \nn \w2
& \hspace{2cm} + \mathcal{J}(-k)\Phi(k) + \mathcal{J}(k)\Phi(-k) \bigg)\, , 
\end{align}
where  $K(k)$ is the Fourier-transformed kinetic term (wave operator) and we have introduced a linear coupling between the fields, collectively labelled as $\Phi(k)$, and a source $\mathcal J(k)$. Connecting this to the specific formulation in~\cref{MAGConv}, we identify $\Phi$ as the collection of perturbative fields $\tensor{h}{_{\mu\nu}}$ and $\MAGA{_\mu^\rho_\nu}$. Within the quadratic approximation, the indices on these fields are raised and lowered using $\tensor{\eta}{^{\mu\nu}}$ and $\tensor{\eta}{_{\mu\nu}}$, which are non-dynamical, and the Greek indices refer to Cartesian coordinates on the Minkowski background\footnote{Indeed, it should be explicitly stated that along with these raising and lowering rules $\MAGA{_\mu^\rho_\nu}$ is a dynamical \emph{tensor} field in the quadratic approximation. Geometrically, $\MAGA{_\mu^\rho_\nu}$ is a connection, but the physics does not know or care about the geometric foundations of the theory: only the representations of the particle states are important. Of course, $\MAGD{_\mu^\rho_\nu}$ is already geometrically a tensor in the second-order formulation of MAG. The re-parameterisation is shown explicitly in~\cref{Reparam}, and the key point is that the linearised Levi--Civita connection is also tensorial at lowest order in perturbations because the partial derivative $\PD{_\mu}$ is covariant at that order.}. Conjugate to $\Phi$, the fields $\mathcal{J}$ in MAG are the symmetric matter stress-energy tensor $\tensor{T}{^{\mu\nu}}$ and the rank-three current $\tensor{W}{^\mu_\rho^\nu}$. This latter current is sometimes called the \emph{hypermomentum}~\cite{Hehl:1977fj,Iosifidis:2023kyf,Iosifidis:2023but,Iosifidis:2022xvp,Iosifidis:2021nra,Iosifidis:2020upr,Iosifidis:2020gth}. In the second-order formulation, a separate current must be defined as conjugate to $\MAGD{_\mu^\rho_\nu}$ --- we do not bother to ascribe it another symbol.

The propagating states will appear in the form of isolated poles for the propagator $\mathcal D(k)$ defined through the equation $K(k) \cdot \mathcal D(k) \equiv \hat 1$. The use of projection operators $P^{i,i}_{\left\lbrace J,P\right\rbrace}$ drastically simplifies the solution of this inversion problem.
By exploiting the defining properties
\begin{subequations}
\begin{align} 
&\sum_{J,P,i} P^{i,i}_{\left\lbrace J,P\right\rbrace}{}_{\mu_1 \mu_2 \cdots \mu_n}^{\quad\quad \nu_1 \nu_2 \cdots \nu_n} \equiv \displaystyle{\hat 1}{}_{\mu_1 \mu_2 \cdots \mu_n}^{\quad\quad \nu_1 \nu_2 \cdots \nu_n} , \label{AlgebraOP1} \\ 
&\nn \\
& P^{i,k}_{\left\lbrace J,P\right\rbrace}{}_{\mu_1 \mu_2 \cdots \mu_n}^{\quad\quad \rho_1 \rho_2 \cdots \rho_n} \,\, P^{j,w}_{\left\lbrace J',P'\right\rbrace}{}_{\rho_1 \rho_2 \cdots \rho_n}^{\quad\quad \nu_1 \nu_2 \cdots \nu_n} \nn \\
& \hspace{50pt}\equiv\delta_{k, j}\, \delta_{J, J'}\, \delta_{P, P'}\, P^{i,w}_{\left\lbrace J,P\right\rbrace}{}_{\mu_1 \mu_2 \cdots \mu_n}^{\quad\quad \nu_1 \nu_2 \cdots \nu_n} , \label{AlgebraOP2} \\
&\nn \\
& P^{i,j}_{\left\lbrace J,P\right\rbrace}{}_{\mu_1 \mu_2 \cdots \mu_n}^{\quad\quad \nu_1 \nu_2 \cdots \nu_n} \equiv 
\left(P^{j,i}_{\left\lbrace J,P\right\rbrace}{}^{\nu_1 \nu_2 \cdots \nu_n}_{\quad \quad \mu_1 \mu_2 \cdots \mu_n}\right)^*  ,\label{AlgebraOP3}
\end{align}
\end{subequations}
it is possible to decompose the kinetic term as 
\begin{align}
	&\int \mathrm{d}^4 k \,  \Phi(-k) \, K(k) \, \Phi(k) \, \nn \\
	& = \int \mathrm{d}^4 k \, \Phi(-k) \, \sum_{J,P,i,j} \bigg( a^{\left\lbrace J,P \right\rbrace}_{i,j}  P^{i,j}_{\left\lbrace J,P \right\rbrace} \bigg) \, \Phi(k)  \,\,, \label{ADefinition}
\end{align}
where the tortuous fabric of the indices is reformulated in terms of the simpler spin-parity matrices $a^{\left\lbrace J,P \right\rbrace}_{i,j}$, obtained by tracing over the Lorentz indices
\begin{align}a_{i,j}^{\left\lbrace J,P \right\rbrace} \equiv \frac{1}{2 J - 1} \,\, \mathrm{Tr}\, P^{i,j}_{\left\lbrace J,P \right\rbrace} \, \mathcal K (k) \,.
\end{align}
The orthonormality of the projection operators in~\cref{AlgebraOP2} reduces the computation of the propagator $\mathcal D(k)$ to an inversion problem for the matrices $a^{\left\lbrace J,P \right\rbrace}_{i,j}$   
\begin{align}\label{BDefinition}
&\mathcal D(k) \equiv \sum_{J,P,i,j}  b^{\left\lbrace J,P \right\rbrace}_{i,j}  P^{i,j}_{\left\lbrace J,P \right\rbrace} \,, \quad  b^{\left\lbrace J,P \right\rbrace}_{i,j} \equiv \left(a^{\left\lbrace J,P \right\rbrace}_{i,j}\right)^{-1}  \, .
\end{align}
When the degeneracy of the spin-parity matrices hampers this inversion, the model displays gauge symmetry. It is a bonus of this formalism that full control over the gauge symmetries is provided by the inspection of such simple objects as the null-vectors $X_i^{r = 1,2,\cdots, n}$ of $a^{\left\lbrace J,P \right\rbrace}_{i,j}$, so that manipulations of linear algebra replace tensor operations. In turn, this promotes a code-friendly implementation of the spectral problem. The matrix structure not only allows the identification of symmetries which are already present in the model, but it also allows parameter-tunings to be identified which lead to the emergence of new symmetries. Finally, the gauge-invariant \emph{saturated} propagator is obtained by restricting the inversion to the non-degenerate subspace in the spin-parity matrices, and contracting the inverse matrix with constrained sources $\tilde J(k)$
\begin{subequations}
\begin{gather} \label{SatProp}
 \mathcal D_S(k) \equiv  \tilde{J}^*\left(k\right) \bigg(\sum_{J,P,i,j}  \tilde{b}^{\left\lbrace J,P \right\rbrace}_{i,j}  P^{i,j}_{\left\lbrace J,P \right\rbrace}\bigg)  \tilde{J}\left(k\right), \\
 X^{*}{}^r_j \,\, P^{i,j}_{\left\lbrace J,P \right\rbrace} \,\, \tilde{J} \left(k\right) \equiv 0 \,, \quad\,\,r = 1,2,\cdots, n\,\, .\label{SatProp2}
\end{gather} 
\end{subequations}

\subsection{Computer implementation}\label{PSALTerSec}
The saturated propagator in~\cref{SatProp} represents the arrival point of our computation. Previous efforts in MAG, even within the projection operators approach, generally relied on an indirect determination for the signs of the residues of the poles in $\mathcal D_S(k)$. Such methods avoid the difficult computation of the source constraint equations~\cref{SatProp2} in the massless limit. However, as detailed in~\cite{Marzo:2021esg,Barker:2024,Lin:2018awc}, the constraint equations can be fully resolved by choosing a suitable reference frame (with $k^2 = 0$ as a limit of $k^2 >0 $). This technique removes the main theoretical challenges encountered in the computation of the spectrum leaving, as the only limitation, the technical bounds in manipulating large expressions. The simplifications afforded by the use of projection operators, as well as the unambiguous computational procedure, have encouraged the development of opportune tools to promptly and automatically tackle the particle spectrum. Some of the most recent results on this subject~\cite{Barker:2023bmr,Barker:2023fem} and the core of this work's conclusions, are obtained with the use of~\PSALTer{}~\cite{Barker:2024}, a Wolfram Language implementation of these ideas. The \PSALTer{} software can automatically return the spectrum of any tensor-valued field theory up to rank-three.

The \PSALTer{} analyses of key theories to be considered in~\cref{Results} are presented in~\cref{AnnalaRasanenColumn4,ZeroTorsionGeneralFirstOrder,ZeroTorsionGeneralSecondOrder,ZeroTorsionNo2pNo1pSimpleFirstOrder,ZeroTorsionNo2pNo1pSimpleSecondOrder,ZeroTorsionNo2pNo1pComplicatedFirstOrder,ZeroTorsionNo2pNo1pComplicatedSecondOrder,UnrestrictedGeneralFirstOrder,UnrestrictedGeneralSecondOrder}. These are vector graphics which contain the following information;
\begin{itemize}
	\item The linearised (quadratic) action in~\cref{ActionMom}, in the position-space representation. This expression is the only input to the \PSALTer{} software, and it is not necessary to perform any kind of decomposition of fields into irreducible parts.
	\item Automatically computed: the elements $a^{\left\lbrace J,P \right\rbrace}_{i,j}$ in~\cref{ADefinition} which encode the wave operator of the theory, provided as one matrix for each spin sector. Because the theories considered here are not parity-violating, it is reassuring to see that the matrices are always block-diagonal across parity-even (red) and parity-odd (blue) sub-sectors: the mixed-parity (purple) blocks are empty.
	\item Automatically computed: the inverse $b^{\left\lbrace J,P \right\rbrace}_{i,j}$ matrices in~\cref{BDefinition} which encode the saturated propagator of the theory. We believe our implementation to be the first which uses Moore--Penrose inversion~\cite{1955PCPS...51..406P,1956PCPS...52...17P} (i.e. a uniquely defined gauge fixing) to obtain these coefficients.
	\item Automatically computed: the source constraints $X_i^r$ in~\cref{SatProp2}, which are guaranteed to encode all the gauge symmetries present in the theory.
	\item Automatically computed: the spectrum of all massive and massless particles present in the theory. This includes information about the particle spin $J$, parity $P$, pole residue and mass. In the case of massless particles, there is no physical notion of spin which survives, but the number of independent polarisations is given.
	\item Automatically computed: the overall unitarity conditions which must be imposed on the Lagrangian coupling coefficients. These conditions are derived from the above pole residues and masses, so as to support the no-ghost and no-tachyon criteria. There is of course no guarantee that such conditions exist, so the calculation is time-limited to ten seconds. In case of `timeout', the masses and residues provide all the relevant information for further tuning the theory anyway.
\end{itemize}
Because these various outputs may be extremely cumbersome and have uncertain dimensions after typesetting, \PSALTer{} uses a rectangle-packing algorithm to find the most economical layout for each theory: consequently some of the formulae in~\cref{AnnalaRasanenColumn4,ZeroTorsionGeneralFirstOrder,ZeroTorsionGeneralSecondOrder,ZeroTorsionNo2pNo1pSimpleFirstOrder,ZeroTorsionNo2pNo1pSimpleSecondOrder,ZeroTorsionNo2pNo1pComplicatedFirstOrder,ZeroTorsionNo2pNo1pComplicatedSecondOrder,UnrestrictedGeneralFirstOrder,UnrestrictedGeneralSecondOrder} are rotated on the page. The various $\mathrm{SO}(3)$ irrep definitions are provided separately, in~\cref{PoincareGaugeTheory,ZeroTorsionPalatini,MetricAffineGravity}. Each of these figures defines a `kinematic module' for the software: a declaration of the fundamental tensor fields and their conjugate sources which are present in a class of theories. Within each module, the spectral analysis of infinitely many distinct models can be performed, depending on the admixture of operators in the quadratic action.

Two steps of the analysis are computationally expensive: the Moore--Penrose inversion and the evaluation of massless residues. When the theory contains more than two or three independent Lagrangian couplings (parameters) and tensor fields of rank three or more, these calculations start to pose a highly non-trivial computer algebra problem. Consequently, many subroutines in \PSALTer{} are automatically parallelised, to take advantage of the available infrastructure. For expedience, analysis of each theory in this work was performed using a dedicated compute node consisting of 112 Intel\textsuperscript{\textregistered} \emph{Sapphire Rapids} CPUs, or 64 AMD\textsuperscript{\textregistered} \emph{Ryzen Threadripper} CPUs, depending on availability. The former setup is close to the current state of the art in high-performance computing (HPC). The resulting throughput is very fast, and in fact each theory would only have taken approximately 20 minutes to process on a modern PC with four CPU cores.

\section{Results}\label{Results}
Using the building-blocks in~\cref{Contractions} the most general Ricci-type MAG in the first-order formulation is
\begin{align} \label{ActionMAG}
	S[g,&A]=-\frac{1}{2}\int\mathrm{d}^4x\sqrt{-g}\bigg[-a_0\MAGF{}
\nn\\
&
				+\MAGFb{^{\mu\nu}}
				\Big(
					c_7\MAGFb{_{\mu\nu}}
					+c_8\MAGFb{_{\nu\mu}}
				\Big)
\nn\\
&
				+\MAGFa{^{\mu\nu}}
				\Big(
					c_9\MAGFa{_{\mu\nu}}
					+c_{10}\MAGFa{_{\nu\mu}}
				\Big)
\nn\\
&
				+\MAGFa{^{\mu\nu}}
				\Big(
					c_{11}\MAGFb{_{\mu\nu}}
					+c_{12}\MAGFb{_{\nu\mu}}
				\Big)
				\nn\\&
				+\MAGF{^{\mu\nu}}
					\Big(
					c_{13}\MAGF{_{\mu\nu}}
					+c_{14}\MAGFb{_{\mu\nu}}
					+c_{15}\MAGFa{_{\mu\nu}}
					\Big)\bigg],
\end{align}
where we borrow the numbering of Lagrangian couplings directly from~\cite{Percacci:2020ddy}\footnote{In that work, a more general MAG action is considered, in which a total of 28 invariant operators are present in the Lagrangian density.}. As emphasised in~\cref{MAGConv}, re-parameterisation invariance leads to the equivalence $S[g,A]\cong S[g,\Delta]$ with the second-order formalism. The computational algorithm sketched in~\cref{SatPropSec} can obviously handle both representations of the dynamical fields and our tests have adopted both approaches as a further self-consistency check. While the final outcome does not change, the particular form of the intermediate steps does. In this regard, we find the second-order basis more convenient for enumerating the spin-parity states in kinematically restricted version of the MAG, due to the index symmetries in~\cref{TorNonMe}. 

\subsection{Zero non-metricity} \label{SecMet}
The imposition of zero non-metricity is easily realized in a formalism with explicit distortion. From~\cref{TorNonMe} it is clear that a two-index-antisymmetric rank-three field $\MAGD{_{\lambda \mu \nu}}  \equiv - \MAGD{_{\lambda \nu \mu}}$ nullifies $Q_{\lambda \mu \nu}\equiv 0$. 
This achieves a reduction of the spin-parity sectors of~\cref{ParticleDec1} %
\begin{align} \label{ParticleDecNonMet}
	&\MAGD{_{\mu \nu \rho}} \supset  \Big\{2_3^+ , 2_2^- , 1^+_2 , 1^+_3,  1^-_3 ,  1^-_6 , 0^+_3, 0^-_1 \Big\},
\end{align}
reflected by a further redundancy in the number of independent Ricci-type contractions 
\begin{align}
	\MAGF{_{\mu \nu }}\equiv 0, \quad \MAGFa{_{\mu\nu}} \equiv - \MAGFb{_{\mu\nu}}\, . 
\end{align}
The action~\cref{ActionMAG} is therefore simplified into
\begin{align}  \label{ActionMetricMAG}
	S[g,&A]=-\frac{1}{2}\int\mathrm{d}^4x\sqrt{-g}\bigg[-a_0\MAGF{}
\nn\\
&
				+\MAGFb{^{\mu\nu}}
				\Big(
					g_7\MAGFb{_{\mu\nu}}
					+g_8\MAGFb{_{\nu\mu}}
				\Big)\bigg].
\end{align}
Note that we follow~\cite{Percacci:2020ddy} in re-labelling the dimensionless coefficients from $c_i$ to $g_i$ when passing from~\cref{ActionMAG} to~\cref{ActionMetricMAG} by kinematic restriction. The spectrum of this simple three-parameter model can be promptly, and unambiguously profiled within our formalism. Just within this section (but not within~\cref{SecTor,SecGen}) the accompanying \PSALTer{} analysis in~\cref{SecMetApp} will be made in the Poincar\'e gauge theory (PGT) formulation of zero-non-metricity MAG. The kinematic structure of PGT is more extensive than its MAG counterpart due to extra \emph{antisymmetric} parts of the tetrad fields (which are nullified by an extra Lorentz gauge symmetry). This kinematic structure is presented in~\cref{PoincareGaugeTheory}, but is otherwise analogous to that in~\cref{ParticleDec2,ParticleDecNonMet}. For now, we proceed with our discussion as if we were working in the MAG formulation. To access the singular structure of the propagator we first identify the spin-parity sectors of the kinetic term. The PGT matrices are shown in~\cref{SecMetApp}, in~\cref{AnnalaRasanenColumn4}. We can immediately recognize in the absence of the $1^-_7$ and $0^+_6$ sectors the hallmark of diffeomorphism invariance. All the information concerning the quadratic terms is encoded in such matrices and, from the arguments of~\cref{SecProj}, a direct link exists between the zeroes of their determinants and the singularities of $\tilde D_S(k)$. The shape of the residue and the position of the singularity will determine the nature of the propagating particles. Once the degeneracies of $a^{\left\lbrace 1,- \right\rbrace}_{i,j}$ and $a^{\left\lbrace 0,+ \right\rbrace}_{i,j}$ are removed, we immediately find that no massive poles are present. We can therefore extract the known result~\cite{Annala:2022gtl}:
\\\\ \emph{No massive states propagate in linearised zero-non-metricity Ricci-type theories}. \\\\
The massless poles are present in the $2^+$ and $0^+$ sectors. Again, these are known traits of graviton propagation. To confirm that the graviton is present we explore the form of these massless poles in the final, gauge invariant, propagator. The constraints to be imposed are read off the null vectors of $a^{\left\lbrace 1,- \right\rbrace}_{i,j}$ and $a^{\left\lbrace 0,+ \right\rbrace}_{i,j}$. By choosing the light-like frame $k^{\mu} = \big(\mathcal{E}, 0, 0, \mathcal{E} \big) $ the sources are constrained as
\begin{align} \label{GravConst}
&  T^{0 0} = T^{0 3} , \quad T^{1 3} = T^{0 1} , \quad T^{2 3} = T^{0 2} , \quad T^{3 3} = T^{0 3}.
\end{align}
By substituting them into the propagator, we find the following structure 
\begin{align} \label{SatPropMet}
	\lim_{k^2 \mapsto 0}\mathcal D_S(k)  = \frac{1}{k^2} \left(\frac{2}{-a_0}\right)\sum_{i = 1,2}|S_i|^2 \, ,
\end{align}
where the $S_i$ are linearly independent combinations of the sources. Thus,~\cref{SatPropMet} reveals the two (helicity) states with a residue $\propto - 1/ a_0$. We conclude by refining the previous statement  \\\\ \emph{Linearised zero-non-metricity Ricci-type theories propagate only a healthy massless graviton with $a_0 < 0 $}. 

\subsection{Zero torsion}\label{SecTor}
The zero-torsion case offers a stronger challenge to our methods, displaying a larger parameter space and non-trivial interplay among different operators. We will show that the spin-parity formalism, followed by direct access to the propagator, gives full access to the tree-level spectrum without imposing simplifying restrictions. First, in terms of the distortion tensor, the zero-torsion condition in~\cref{TorNonMe} is achieved by working with a two-index-symmetric rank-three field $\MAGD{_{\lambda \mu \nu}} \equiv \MAGD{_{\nu \mu \lambda}}$. The available particle content in the symmetric distortion is then
\begin{align} \label{ParticleDecT}
&\MAGD{_{\mu \nu \rho}} \supset \Big\{ 3_1^- , 2_1^+ , 2_2^+ , 2_1^- , 1^+_1 , 1^-_1,  1^-_2 ,
\nn\\
	&\hspace{40pt}
	1^-_4,  1^-_5 , 0^+_1, 0^+_2, 0^+_4 \Big\},
\end{align}
which causes  $\MAGF{_{\mu\nu}}\equiv-2\MAGFb{_{[\mu\nu]}}$, leaving seven independent combinations
\begin{align} \label{ActionTLessMAG}
	S[g,&A]=-\frac{1}{2}\int\mathrm{d}^4x\sqrt{-g}\bigg[-a_0\MAGF{}
\nn\\
&
				+\MAGFb{^{\mu\nu}}
				\Big(
					h_7\MAGFb{_{\mu\nu}}
					+h_8\MAGFb{_{\nu\mu}}
				\Big)
\nn\\
&
				+\MAGFa{^{\mu\nu}}
				\Big(
					h_9\MAGFa{_{\mu\nu}}
					+h_{10}\MAGFa{_{\nu\mu}}
				\Big)
\nn\\
&
				+\MAGFa{^{\mu\nu}}
				\Big(
					h_{11}\MAGFb{_{\mu\nu}}
					+h_{12}\MAGFb{_{\nu\mu}}
				\Big)\bigg].
\end{align}
Note once again that we follow~\cite{Percacci:2020ddy} in re-labelling the dimensionless coefficients from $c_i$ to $h_i$ when passing from~\cref{ActionMAG} to~\cref{ActionTLessMAG} by kinematic restriction. The kinematic structures in~\cref{ParticleDec2,ParticleDecT} are reflected in the \PSALTer{} definitions of the states, which are fully defined in~\cref{ZeroTorsionPalatini}. The computation of the spin-parity matrices produces the remaining output in~\cref{SecTorApp}, with the general case~\cref{ActionTLessMAG} displayed in~\cref{ZeroTorsionGeneralFirstOrder}. After simple inspections of the determinants, we find that it is impossible to impose further gauge symmetries, unless $a_0$, the only dimensionful parameter, is set to zero. This would nullify the graviton propagation and, therefore, we discard this option. Taking into account the constraints of diffeomorphism invariance as illustrated in~\cref{SecMet}, we can compute the positions of the zeroes for all the determinants. Already for the $2^+$ sector we find that a state of mass 
\begin{equation}\label{Mass2p}
    m^2_{2^+} = - \frac{a_0}{h_{10} + h_9},
\end{equation}
is allowed to propagate whenever $h_{10} + h_9 \neq 0$. The computation of the limit $k^2 \mapsto m^2_{2^+}$ gives rise to the following pole residue, where we suppress the positive-definite quadratic form in the $2^+$ sources
\begin{align}
	\lim_{k^2 \mapsto m_{2^+}^2}\mathcal D_S(k)\propto 2\bigg[&4h_{10}^2+h_{11}^2+h_{12}^2+2h_{11}\big(h_{12}-2h_9\big)
	\nn\\
	&
	-4h_{12}h_9+4h_9^2-2h_9a_0
	\nn\\
	&
	-2h_{10}\big(2h_{11}+2h_{12}-4h_9+a_0\big)\bigg]
	\nn\\
	&
	\times\bigg[\big(h_{10}+h_9\big)^2a_0\bigg]^{-1}.
 \label{Res2p}
\end{align}
The positivity of the spin-two mass in~\cref{Mass2p} and the residue in~\cref{Res2p} select possible real values of the parameters involved. Among these, the requirement is seen for a positive $a_0$. This shows that a massive spin-two is incompatible with the healthy propagation of the graviton: we must discard it. This, as can be seen from the determinants, can be accomplished by demanding $h_{10}=-h_9$ or the stronger $h_{10}=h_9 =0$. 
The consequences of these different choices can be appreciated by observing that the theory also propagates a massive spin-one particle whose mass and residue is given by
\begin{align} \label{sp1piu}
&    m^2_{1^+} = \frac{a_0}{h_{9} - h_{10}} \, , \quad  \lim_{k^2 \mapsto m_{1^+}^2}\mathcal D_S(k)\propto \frac{4}{h_{10} - h_{9}} \, .
\end{align}
We can therefore explore the possibility of keeping such a state by adopting $h_{10}  = - h_9 \neq 0$ and studying the consequences for the rest of the spectrum. It is easy to show that this requires $h_{10}  = - h_9 > 0$ and $a_0 < 0$, so that such propagation can indeed be afforded without spoiling the gravitational priorities of the model.

A more alarming scenario is presented by the degenerate spin $1^-$ sector, where the restricted determinant shows a quartic equation in the momentum. Imposing, for instance, $h_{10} = - h_9$ we would find
\begin{align} \label{Det1mTL}
     & |a_{i,j}^{\{1,-\}} | = \frac{a_0^2}{32} \Big( 5 \left((h_{12}-h_{11})^2 + 8 h_9 (h_8 - h_7) \right)k^4 +  \nn \\ 
     & - 4 a_0 \left( 3 h_{11} - 3 h_{12} - h_7 + h_8 + 12 h_9 \right) k^2 + 12 a_0^2 \Big) \, . 
\end{align}
The determinant in~\cref{Det1mTL} can be read off the denominators of the $b^{\left\lbrace 1,- \right\rbrace}_{i,j}$ matrix elements\footnote{Unfortunately, the $b^{\left\lbrace 1,- \right\rbrace}_{i,j}$ matrices are very large expressions, so \PSALTer{} frequently suppresses them when attempting to typeset the results for publication. Although~\cref{Det1mTL} cannot therefore be confirmed from~\cref{SecTorApp}, the full results are available in the Wolfram \emph{Mathematica} notebook file from which \PSALTer{} is run: this document, along with the source script, is made available in the supplement~\cite{SupplementalMaterials}.}. Bona fide unitarity demands the removal of the quartic term. The absence of a ghostly massive spin-two particle and the spin-one dipole motivates the defining constraint over the parameter space. Many (apparently) different solutions can be found by asking to solve such constraints in terms of different subsets of the couplings. It is a great advantage that the algorithmic disposition of the spin-parity approach allows a simple scan over the broad space of solutions. We proceed by considering the two separate branches obtained from $h_{10}  = - h_9 > 0$ and, for each of the two possibilities, gather the different solutions yielded by nullifying the quadratic coefficient of $k^4$ in~\cref{Det1mTL}. The theories associated with this scan are presented in~\cite{SupplementalMaterials}. Collecting all the masses and residues for, besides the graviton, the two massive spin-one states of opposite parity we find, in all cases that
\\\\ \emph{Linearised zero-torsion Ricci-type theories do not admit simultaneous propagation of massive spin-one states of opposite parity}. 
\\\\
We can naturally continue by asking for the dismissal of the full $1^-$ propagation by setting to zero the coefficient of $k^2$ in~\cref{Det1mTL}. Again, we do this by gathering all the relevant equations and solving them for all the possible subsets of the free parameters, and again we refer to~\cite{SupplementalMaterials}. We find that healthy solutions are available, although not for all the given cases. In the healthy scenarios, we rediscover the mass/residue ratio of~\cref{sp1piu} rephrased in terms of the available couplings. 

Finally, we can investigate the case $h_9 = h_{10} = 0$ which removes the propagation of the massive spin-one state of positive parity~\cref{sp1piu} and, simultaneously, of the massive spin-two. Under such circumstances, the cancellation of the dipole propagation simplifies to demanding $(h_{11}-h_{12})^2 = 0$ in~\cref{Det1mTL}, namely $h_{11} = h_{12} \neq 0$ (see~\cref{ZeroTorsionNo2pNo1pComplicatedFirstOrder,ZeroTorsionNo2pNo1pComplicatedSecondOrder}) and $h_{11} = h_{12} = 0$ (see~\cref{ZeroTorsionNo2pNo1pSimpleFirstOrder,ZeroTorsionNo2pNo1pSimpleSecondOrder}). 
For the second, simpler scenario, we find in the second-order case~\cref{ZeroTorsionNo2pNo1pSimpleSecondOrder}
\begin{align} \label{sp1meno}
&    m^2_{1^-} = \frac{ -3 a_0}{h_{7} - h_{8}} \, , \quad  \lim_{k^2 \mapsto m_{1^-}^2}\mathcal D_S(k)\propto \frac{34}{h_{7} - h_{8}} \, .
\end{align}
Such a simple setup, which clearly presents a viable propagation, is only slightly modified when considering the branch $h_{11} = h_{12} \neq 0$, with only the residue's form being affected. Again, ghost- and tachyon-freedom can be accounted for. We conclude, accordingly, stating that, besides the graviton:
\\\\ \emph{Either a healthy massive vector of negative parity or positive parity propagates in linearised zero-torsion Ricci-type theories}. 

\subsection{Generic case}\label{SecGen}
\subsubsection{General properties}
The transition to the case of an unconstrained affine connection presents an obvious growth in computational complexity induced by the multiple components of ~\cref{ParticleDec1,ParticleDec2} and, consequently, by the independence of all three Ricci-type tensors. The challenges of the associated spectral problem are quite visible in the cumbersome spin-parity matrices of~\cref{SecGenApp}. The general spectrum associated with~\cref{ActionMAG} is shown in~\cref{UnrestrictedGeneralFirstOrder,UnrestrictedGeneralSecondOrder}, respectively for the first- and second-order formulations of the theory. The inclusion of all the components considerably changes the nature of the unconstrained spectrum. First, we notice how the massive state of spin-two is no longer present, the determinant having a simple proportionality to $k^2$. Similarly,~\cref{Det1mTL} is now of the form\footnote{Again, the $b^{\left\lbrace 1,- \right\rbrace}_{i,j}$ matrices are suppressed in~\cref{SecGenApp}, for full results see the supplement~\cite{SupplementalMaterials}.}
\begin{subequations}
\begin{align} 
 |a_{i,j}^{\{1,-\}} | &=  \tfrac{1}{4} a_0^4 k^2 \Bigl(f_1 a_0  - 5 f_2 k^2\Bigr) ,\label{onemsec}\\
	f_0&\equiv c_{10} -  c_{11} + c_{12} - 16 c_{13} + 4 c_{14} + 4 c_{15}  \nn \\
	&\ \ \ -  c_7+ c_8 -  c_9),  \\ 
	f_1 &\equiv  (c_{14}+c_{15})^2 +  4 c_{13} (c_{10} -  c_{11} + c_{12} \nn\\
	&\ \ \ -  c_7 + c_8 -  c_9),
\end{align}
\end{subequations}
and the dangerous dipole of~\cref{Det1mTL} leaves space for a massless vector. That this is indeed the case, and that the pole is not a spurious feature of the determinant, is demonstrated by the direct computation of the saturated propagator in the massless limit. For this computation we have to account for a further, associated peculiarity encountered in this scenario. The rank of the spin $0^+$ sector is now reduced by two, signalling the emergence of an Abelian symmetry. The presence of this symmetry was predicted by Iosifidis and Koivisto~\cite{Iosifidis:2018zwo} --- it appears whenever squares of the full metric-affine curvature are added to the Einstein--Hilbert term, and is a remnant of the full projective symmetry of that term. It is instructive to explicitly show what this entails in terms of source constraints in the light-like frame $k = \big(\mathcal{E}, 0, 0, \mathcal{E}\big)$. We find, together with~\cref{GravConst}, the following reduction
\begin{gather} 
W^{000} - W^{011} - W^{022} - W^{033}   
\nn\\
	\hspace{50pt}	=  W^{300} - W^{311} - W^{322} - W^{333} \, .\label{AbConst}
\end{gather}
Accounting for~\cref{AbConst} we recognize four independent states in the massless limit of the saturated propagator. Two of these are precisely the helicity states of the graviton, proportional to $-1/a_0$ and recognizable in~\cref{SatPropMet}. The residue of the other two states, while signalling unambiguously the propagation connected to the massless spin-one state, has an extremely convoluted form due to the concurrence of many different parameters in its definition. Nevertheless, the requirement for its positivity has the manageable structure
\begin{align}
	&16 c_{13} - 4 (c_{14} +  c_{15}) + c_{11} - c_{12}  \nn\\ 
	&\hspace{30pt}+ c_7 - c_8 + c_9 - c_{10} > 0.  
\end{align}
When searching for massive propagation, similarly to the torsionless case, both spin-one sectors source one state each, with masses 
\begin{subequations}
\begin{align}
	m^2_{1^-} &= \Big[ {a_0} (c_{10}-c_{11}+c_{12}-16 c_{13}+4 c_{14}+4 c_{15}
	\nn\\
	&\ \ \ 
	-c_{7}+c_{8}-c_{9})\Big]\times\Big[5 \big(4 c_{13} (c_{10}-c_{11}+c_{12}
	\nn\\
	&\ \ \ 
	-c_{7}+c_{8}-c_{9})+(c_{14}+c_{15})^2\big)\Big]^{-1}, \\
	m^2_{1^+} &=  -\frac{{a_0}}{c_{10}-c_{11}+c_{12}-c_{7}+c_{8}-c_{9}} \, .\label{1pMass}
\end{align}
\end{subequations}
The overall survey of the propagating states points, therefore, to three additional particles populating the spectrum besides the graviton. To solve the spectral problem we analyze the conditions for their simultaneous propagation.  

\subsubsection{Allowing the massless vector}
The presence of a spin-one massless state can be included in our analysis. The related phenomenological concerns can then be seen as suggesting incompleteness, rather than an inconsistency: a mechanism to provide a mass gap is expected. Under such a hypothesis, we can investigate the coexistence of such a state with the others. The computation of the residues of the massive states is carried through the constrained propagator. The explicit effect of the different gauge symmetries on the sources is extracted in the rest-mass frame $k = \big(m, 0, 0, 0\big)$ and gives 
\begin{align} \label{GravConstMass}
&  T^{0 0} = T^{0 1} = T^{0 2} = T^{0 3}  = 0 ,  
\end{align}
for diffeomorphism invariance, and 
\begin{align} \label{AbConstMass}
&  W^{000} - W^{011} - W^{022} - W^{033}  = 0 \, ,
\end{align}
for the extra Abelian symmetry. 
When testing the sign of each residue, as well as the masses, with the requirement $a_0 < 0$, we immediately find an obstruction within the $1^-$ sector. Having committed to retaining the massless propagation, we must simplify the model by removing its massive counterpart. We proceed, therefore, by considering all the eleven solutions of $f_1 = 0$ in~\cref{onemsec} and recomputing the residues for the remaining propagating states.   Once more, no viable solutions are found (see~\cite{SupplementalMaterials}). Finally, we kill the massive $1^+$ propagation in~\cref{1pMass} by adding the further condition 
\begin{align}\label{No1pConstraint}
    c_{10} - c_{11} + c_{12} - c_{7} + c_8 - c_9 = 0.
\end{align}
To coherently include both constraints we consider pairs of parameters which are solutions of the corresponding equation system. Twelve solutions are found (see~\cite{SupplementalMaterials}), all of them with positive residues for the surviving massless sector. 
We can therefore draw the following: 
\\\\ \emph{A healthy massless vector of negative parity propagates in linearised generic Ricci-type theories}. 

Once again, we can make contact with the literature. We notice that~\cref{No1pConstraint} does not eliminate $c_{13}$, which controls the square of the homothetic curvature. It is known that when this operator is added to the Einstein--Hilbert term in full MAG geometry, the resulting theory cannot be distinguished from the vacuum Einstein--Maxwell theory~\cite{Hehl:1999sb,PhysRevD.56.7769} --- the extra massless vector in this case is identified with our $1^-$ state.
\subsubsection{Removing the massless vector}
The only way to dispose of the massless vector is to introduce a further degeneracy in the $1^-$ sector. This can be enforced by solving for $f_1 = f_2 = 0$ in~\cref{onemsec}. Once more, the spin-parity approach grants us the possibility to explore the results in a systematic way. The analysis is made more complicated by the peculiar challenges met in this scenario, where each solution of the $f_1 = f_2 = 0$ system affects the form of the gauge symmetry, thus necessitating, each time, a recomputation of all the main features of the theory. Despite the demanding computational task (see~\cite{SupplementalMaterials}), the outcome turned out to be the same for all the (twenty) different solutions defined in terms of pairs of independent parameters. We can, consequently, present the results for this scenario by focusing on a particular solution:
\begin{subequations}
\begin{align} \label{asolution}
c_9 &= c_{10}-c_{11} + c_{12} +  16 c_{13}  - c_{7} +  c_{8}, \\
	c_{15} &= 8 c_{13} - c_{14} \, ,\label{asolution2}
\end{align}
\end{subequations}
producing the following propagator poles
\begin{align} \label{masscorr}
& m^2_{1^-} = \frac{{a_0} }{20 c_{13}} , \quad  m^2_{1^+} =  \frac{{a_0}}{16 c_{13}} \, . 
\end{align}
The correlation among the masses of the two spin-one sectors is not an accident of the chosen solution but illustrates a common feature: the strict proportionality $m^2_{1^-}/m^2_{1^+} = 4/5 $. Such correlation signals the impossibility of removing the propagation of one state without interfering with the other. To assess the nature of the states we need to compute the saturated propagator in the presence of the augmented gauge constraints defined, for instance, by~\cref{asolution,asolution2}. For the massive limit of $k = \big(m, 0, 0, 0\big)$ for some $m$ we must consider, on top of~\cref{GravConstMass} and~\cref{AbConstMass} the extra degeneracy of the $1^-$ sector. Being generated by a three-dimensional vector, these source constraints take the form
\begin{align} \label{FinalConstMass}
&  W^{00i} - W^{0i0} =  \frac{a_0}{m^2}\frac{W^{i00} - W^{i11} - W^{i22} -W^{i33}}{c_{11} - c_{12} - 4 c_{14} + 2 c_{7} - 2 c_{8}}  \,.
\end{align}
The odd appearance of couplings within the definitions of the symmetry is a curious feature of this analysis. This, however, does not present any theoretical downsides if we consider that such parameters will be normalized to pure numbers in the quadratic, final Lagrangian. The massive limit confirms the presence of the propagation of three states as expected for massive spin-one particles. For the two different sectors, and the representative selection of dependent parameters shown in~\cref{asolution,asolution2}, 
we find the corresponding residue 
\begin{subequations}
\begin{align}
	\lim_{k^2 \mapsto m_{1^-}^2}&\mathcal D_S(k) \propto \frac{1}{50 c_{13}} \Big[ 11 
	+ \Big(50 (6 c_{13}- c_{14}) (c_{12} - c_{11} 
	\nn\\
	&\ \ \ 
	+ 2 (6 c_{13} + c_{14} - c_7 + c_8))\Big)\times\Big((c_{12}-c_{11} 
	\nn\\
	&\ \ \ 
	+ 4 c_{14} -2 c_7 + 2 c_8)^2\Big)^{-1}\Big],
	\\
	\lim_{k^2 \mapsto m_{1^+}^2}&\mathcal D_S(k) \propto -\frac{1}{4096 \, c_{13}^3} \Big[c_{11}^2-2 c_{11} (c_{12}+2 (8 c_{13}-c_{7} 
	\nn\\
	&\ \ \ 
      +c_{8}))+c_{12}^2+4 c_{12} (8 c_{13}-c_{7}+c_{8})+768 c_{13}^2
	\nn\\
	&\ \ \ 
      + 64 c_{13} (c_{8}-c_{7})+4 (c_{7}-c_{8})^2 \Big]\, .
\end{align}
\end{subequations}
While the requirement of positivity of $\lim_{k^2 \mapsto m_{1^-}^2}\mathcal D_S(k)$ does not challenge the unitarity of the graviton sector, nor the tachyon-freedom conditions over the vector masses, this is not the case for $\lim_{k^2 \mapsto m_{1^+}^2}\mathcal D_S(k) > 0 $, which calls for $c_{13} < 0$. Again, different choices in solving for $f_1 = f_2 = 0$ do affect the form of the residues, but no simultaneous solutions are found (see~\cite{SupplementalMaterials}). The correlation existing for the propagation of both massive vectors in~\cref{masscorr} prohibits, therefore, both massive states from appearing in a healthy spectrum. 

\section{Conclusions}\label{Conclusions}
It is difficult to overestimate the importance of accommodating the absence of ghosts and tachyons in quantum field theory. Control over unitarity is key to understanding the shape of possible new theories and future extensions of the current models. Spin-parity projectors provide a computational framework for fully controlling the propagation of quadratic Lagrangian, which lends itself well to computer implementation. Once the needed operators are collected, the spectral problem is basically solved~\cite{Lin:2018awc,Lin:2020phk}. The output is unambiguous, given the direct access to the propagator, and does not rely upon intricate field redefinitions or the introduction of spurious fields to achieve reductions to known cases. In this work, we have adopted the spin-parity formalism to illustrate its reach and the capacity to tackle a broader set of operators than previously possible. We have made a thorough survey of the Ricci-type MAG operator space, but our analysis is not intended to be exhaustive. The point we are making is that if further special cases turn out to be of interest in the future, then it will be economical to test them using our approach. Recently, some spectral analyses of the PGT and Weyl gauge theory (WGT) have been made~\cite{Lin:2018awc,Lin:2019ugq,Lin:2020phk}, which really \emph{are} exhaustive. The trick to making exhaustive surveys is to recursively search over the root system of the wave operator determinant. This would make an appealing (and apparently straightforward) extension to our current \PSALTer{} program, but we defer it to future work.

There are two key limitations to our approach. Firstly, the authors of~\cite{Annala:2022gtl} are able to extend their analyses to particle spectra on Friedmann backgrounds: we \emph{cannot} do this. There is some hope for the extension of the spectral algorithm to de Sitter backgrounds in the near future~\cite{Sengor:2022kji}, but further applications to non-maximally-symmetric spacetimes are currently speculative~\cite{Tekin:2016vli}. Secondly, the theories in~\cref{ActionMAG,ActionMetricMAG,ActionTLessMAG} may propagate more species in their full nonlinear dynamics than are revealed in the spectral analysis. This is already known to happen in the case of the theory in~\cref{ActionTLessMAG}, for which the $1^+$ and $1^-$ torsion modes are strongly coupled near Minkowski spacetime~\cite{Barker:2023fem}. When this happens, it means that the model is inherently non-perturbative around Minkowski spacetime, so the quadratic approximation in~\cref{ActionMom} is just a fictional model which has nothing to do with the actual physics. It is hard to see how this cannot be a pathology (with or without ghost-tachyon-freedom of the strongly coupled modes), and the only sure way to diagnose it is via a nonlinear Hamiltonian analysis~\cite{Barker:2021oez}. It is possible that the methods of~\cite{Annala:2022gtl} are also sensitive to strong coupling, if for example propagating d.o.f are lost as the Friedmann background is deformed into the Minkowski background. However, it is not clear that such an approach would always detect the problem when it exists. Attempts at computer algebra Hamiltonian analysis were made in~\cite{Barker:2022kdk}, but the implementation was not theory-agnostic (restricted to PGT). The \PSALTer{} software showcased here is theory-agnostic by design. It can be downloaded from~\href{https://github.com/wevbarker/PSALTer}{github.com/wevbarker/PSALTer}\footnote{In the longer term we hope that \PSALTer{} will become an official contribution to the \emph{xAct} project~\cite{Martin-Garcia:2007bqa,Martin-Garcia:2008yei,Martin-Garcia:2008ysv,Brizuela:2008ra,Nutma:2013zea}.}.

\begin{acknowledgements}
We are particularly indebted to Jaakko Annala and Syksy Räsänen for multiple in-person discussions, and to Damianos Iosifidis, Roberto Percacci and Sebastian Zell for useful correspondence.

W.~B. is grateful for the kind hospitality of Leiden University and the Lorentz Institute, and the support of Girton College, Cambridge. The work of C.~M. was supported by the Estonian Research Council grants PRG1677, RVTT3, RVTT7, and the CoE program TK202 ``\emph{Fundamental Universe}''.

This work used the Newton server, access to which was provisioned by Will Handley, and funded through an ERC grant.

This work used the DiRAC Data Intensive service (CSD3 \href{www.csd3.cam.ac.uk}{www.csd3.cam.ac.uk}) at the University of Cambridge, managed by the University of Cambridge University Information Services on behalf of the STFC DiRAC HPC Facility (\href{www.dirac.ac.uk}{www.dirac.ac.uk}). The DiRAC component of CSD3 at Cambridge was funded by BEIS, UKRI and STFC capital funding and STFC operations grants. DiRAC is part of the UKRI Digital Research Infrastructure.
\end{acknowledgements}

\bibliography{main,INSPIRE}

\begin{thebibliography}{109}%
\makeatletter
\providecommand \@ifxundefined [1]{%
 \@ifx{#1\undefined}
}%
\providecommand \@ifnum [1]{%
 \ifnum #1\expandafter \@firstoftwo
 \else \expandafter \@secondoftwo
 \fi
}%
\providecommand \@ifx [1]{%
 \ifx #1\expandafter \@firstoftwo
 \else \expandafter \@secondoftwo
 \fi
}%
\providecommand \natexlab [1]{#1}%
\providecommand \enquote  [1]{``#1''}%
\providecommand \bibnamefont  [1]{#1}%
\providecommand \bibfnamefont [1]{#1}%
\providecommand \citenamefont [1]{#1}%
\providecommand \href@noop [0]{\@secondoftwo}%
\providecommand \href [0]{\begingroup \@sanitize@url \@href}%
\providecommand \@href[1]{\@@startlink{#1}\@@href}%
\providecommand \@@href[1]{\endgroup#1\@@endlink}%
\providecommand \@sanitize@url [0]{\catcode `\\12\catcode `\$12\catcode
  `\&12\catcode `\#12\catcode `\^12\catcode `\_12\catcode `\%12\relax}%
\providecommand \@@startlink[1]{}%
\providecommand \@@endlink[0]{}%
\providecommand \url  [0]{\begingroup\@sanitize@url \@url }%
\providecommand \@url [1]{\endgroup\@href {#1}{\urlprefix }}%
\providecommand \urlprefix  [0]{URL }%
\providecommand \Eprint [0]{\href }%
\providecommand \doibase [0]{https://doi.org/}%
\providecommand \selectlanguage [0]{\@gobble}%
\providecommand \bibinfo  [0]{\@secondoftwo}%
\providecommand \bibfield  [0]{\@secondoftwo}%
\providecommand \translation [1]{[#1]}%
\providecommand \BibitemOpen [0]{}%
\providecommand \bibitemStop [0]{}%
\providecommand \bibitemNoStop [0]{.\EOS\space}%
\providecommand \EOS [0]{\spacefactor3000\relax}%
\providecommand \BibitemShut  [1]{\csname bibitem#1\endcsname}%
\let\auto@bib@innerbib\@empty
\bibitem [{\citenamefont {Salvio}\ and\ \citenamefont
  {Strumia}(2014)}]{Salvio:2014soa}%
  \BibitemOpen
  \bibfield  {author} {\bibinfo {author} {\bibfnamefont {A.}~\bibnamefont
  {Salvio}}\ and\ \bibinfo {author} {\bibfnamefont {A.}~\bibnamefont
  {Strumia}},\ }\bibfield  {title} {\bibinfo {title} {{Agravity}},\ }\href
  {https://doi.org/10.1007/JHEP06(2014)080} {\bibfield  {journal} {\bibinfo
  {journal} {JHEP}\ }\textbf {\bibinfo {volume} {06}},\ \bibinfo {pages}
  {080}},\ \Eprint {https://arxiv.org/abs/1403.4226} {arXiv:1403.4226 [hep-ph]}
  \BibitemShut {NoStop}%
\bibitem [{\citenamefont {Alvarez-Gaume}\ \emph {et~al.}(2016)\citenamefont
  {Alvarez-Gaume}, \citenamefont {Kehagias}, \citenamefont {Kounnas},
  \citenamefont {L\"ust},\ and\ \citenamefont
  {Riotto}}]{Alvarez-Gaume:2015rwa}%
  \BibitemOpen
  \bibfield  {author} {\bibinfo {author} {\bibfnamefont {L.}~\bibnamefont
  {Alvarez-Gaume}}, \bibinfo {author} {\bibfnamefont {A.}~\bibnamefont
  {Kehagias}}, \bibinfo {author} {\bibfnamefont {C.}~\bibnamefont {Kounnas}},
  \bibinfo {author} {\bibfnamefont {D.}~\bibnamefont {L\"ust}},\ and\ \bibinfo
  {author} {\bibfnamefont {A.}~\bibnamefont {Riotto}},\ }\bibfield  {title}
  {\bibinfo {title} {{Aspects of Quadratic Gravity}},\ }\href
  {https://doi.org/10.1002/prop.201500100} {\bibfield  {journal} {\bibinfo
  {journal} {Fortsch. Phys.}\ }\textbf {\bibinfo {volume} {64}},\ \bibinfo
  {pages} {176} (\bibinfo {year} {2016})},\ \Eprint
  {https://arxiv.org/abs/1505.07657} {arXiv:1505.07657 [hep-th]} \BibitemShut
  {NoStop}%
\bibitem [{\citenamefont {Stelle}(1977)}]{Stelle:1976gc}%
  \BibitemOpen
  \bibfield  {author} {\bibinfo {author} {\bibfnamefont {K.~S.}\ \bibnamefont
  {Stelle}},\ }\bibfield  {title} {\bibinfo {title} {{Renormalization of Higher
  Derivative Quantum Gravity}},\ }\href
  {https://doi.org/10.1103/PhysRevD.16.953} {\bibfield  {journal} {\bibinfo
  {journal} {Phys. Rev. D}\ }\textbf {\bibinfo {volume} {16}},\ \bibinfo
  {pages} {953} (\bibinfo {year} {1977})}\BibitemShut {NoStop}%
\bibitem [{\citenamefont {Anselmi}(2017)}]{Anselmi:2017ygm}%
  \BibitemOpen
  \bibfield  {author} {\bibinfo {author} {\bibfnamefont {D.}~\bibnamefont
  {Anselmi}},\ }\bibfield  {title} {\bibinfo {title} {{On the quantum field
  theory of the gravitational interactions}},\ }\href
  {https://doi.org/10.1007/JHEP06(2017)086} {\bibfield  {journal} {\bibinfo
  {journal} {JHEP}\ }\textbf {\bibinfo {volume} {06}},\ \bibinfo {pages}
  {086}},\ \Eprint {https://arxiv.org/abs/1704.07728} {arXiv:1704.07728
  [hep-th]} \BibitemShut {NoStop}%
\bibitem [{\citenamefont {Anselmi}\ and\ \citenamefont
  {Piva}(2017)}]{Anselmi:2017yux}%
  \BibitemOpen
  \bibfield  {author} {\bibinfo {author} {\bibfnamefont {D.}~\bibnamefont
  {Anselmi}}\ and\ \bibinfo {author} {\bibfnamefont {M.}~\bibnamefont {Piva}},\
  }\bibfield  {title} {\bibinfo {title} {{A new formulation of Lee-Wick quantum
  field theory}},\ }\href {https://doi.org/10.1007/JHEP06(2017)066} {\bibfield
  {journal} {\bibinfo  {journal} {JHEP}\ }\textbf {\bibinfo {volume} {06}},\
  \bibinfo {pages} {066}},\ \Eprint {https://arxiv.org/abs/1703.04584}
  {arXiv:1703.04584 [hep-th]} \BibitemShut {NoStop}%
\bibitem [{\citenamefont {Ohta}\ and\ \citenamefont
  {Percacci}(2016)}]{Ohta:2015zwa}%
  \BibitemOpen
  \bibfield  {author} {\bibinfo {author} {\bibfnamefont {N.}~\bibnamefont
  {Ohta}}\ and\ \bibinfo {author} {\bibfnamefont {R.}~\bibnamefont
  {Percacci}},\ }\bibfield  {title} {\bibinfo {title} {{Ultraviolet Fixed
  Points in Conformal Gravity and General Quadratic Theories}},\ }\href
  {https://doi.org/10.1088/0264-9381/33/3/035001} {\bibfield  {journal}
  {\bibinfo  {journal} {Class. Quant. Grav.}\ }\textbf {\bibinfo {volume}
  {33}},\ \bibinfo {pages} {035001} (\bibinfo {year} {2016})},\ \Eprint
  {https://arxiv.org/abs/1506.05526} {arXiv:1506.05526 [hep-th]} \BibitemShut
  {NoStop}%
\bibitem [{\citenamefont {Percacci}\ and\ \citenamefont
  {Zanusso}(2010)}]{Percacci:2009fh}%
  \BibitemOpen
  \bibfield  {author} {\bibinfo {author} {\bibfnamefont {R.}~\bibnamefont
  {Percacci}}\ and\ \bibinfo {author} {\bibfnamefont {O.}~\bibnamefont
  {Zanusso}},\ }\bibfield  {title} {\bibinfo {title} {{One loop beta functions
  and fixed points in Higher Derivative Sigma Models}},\ }\href
  {https://doi.org/10.1103/PhysRevD.81.065012} {\bibfield  {journal} {\bibinfo
  {journal} {Phys. Rev. D}\ }\textbf {\bibinfo {volume} {81}},\ \bibinfo
  {pages} {065012} (\bibinfo {year} {2010})},\ \Eprint
  {https://arxiv.org/abs/0910.0851} {arXiv:0910.0851 [hep-th]} \BibitemShut
  {NoStop}%
\bibitem [{\citenamefont {Bender}\ and\ \citenamefont
  {Mannheim}(2008)}]{Bender:2007wu}%
  \BibitemOpen
  \bibfield  {author} {\bibinfo {author} {\bibfnamefont {C.~M.}\ \bibnamefont
  {Bender}}\ and\ \bibinfo {author} {\bibfnamefont {P.~D.}\ \bibnamefont
  {Mannheim}},\ }\bibfield  {title} {\bibinfo {title} {{No-ghost theorem for
  the fourth-order derivative Pais-Uhlenbeck oscillator model}},\ }\href
  {https://doi.org/10.1103/PhysRevLett.100.110402} {\bibfield  {journal}
  {\bibinfo  {journal} {Phys. Rev. Lett.}\ }\textbf {\bibinfo {volume} {100}},\
  \bibinfo {pages} {110402} (\bibinfo {year} {2008})},\ \Eprint
  {https://arxiv.org/abs/0706.0207} {arXiv:0706.0207 [hep-th]} \BibitemShut
  {NoStop}%
\bibitem [{\citenamefont {Mannheim}(2012)}]{Mannheim:2011ds}%
  \BibitemOpen
  \bibfield  {author} {\bibinfo {author} {\bibfnamefont {P.~D.}\ \bibnamefont
  {Mannheim}},\ }\bibfield  {title} {\bibinfo {title} {{Making the Case for
  Conformal Gravity}},\ }\href {https://doi.org/10.1007/s10701-011-9608-6}
  {\bibfield  {journal} {\bibinfo  {journal} {Found. Phys.}\ }\textbf {\bibinfo
  {volume} {42}},\ \bibinfo {pages} {388} (\bibinfo {year} {2012})},\ \Eprint
  {https://arxiv.org/abs/1101.2186} {arXiv:1101.2186 [hep-th]} \BibitemShut
  {NoStop}%
\bibitem [{\citenamefont {Donoghue}\ and\ \citenamefont
  {Menezes}(2022)}]{Donoghue:2021cza}%
  \BibitemOpen
  \bibfield  {author} {\bibinfo {author} {\bibfnamefont {J.~F.}\ \bibnamefont
  {Donoghue}}\ and\ \bibinfo {author} {\bibfnamefont {G.}~\bibnamefont
  {Menezes}},\ }\bibfield  {title} {\bibinfo {title} {{On quadratic gravity}},\
  }\href {https://doi.org/10.1393/ncc/i2022-22026-7} {\bibfield  {journal}
  {\bibinfo  {journal} {Nuovo Cim. C}\ }\textbf {\bibinfo {volume} {45}},\
  \bibinfo {pages} {26} (\bibinfo {year} {2022})},\ \Eprint
  {https://arxiv.org/abs/2112.01974} {arXiv:2112.01974 [hep-th]} \BibitemShut
  {NoStop}%
\bibitem [{\citenamefont {Donoghue}\ and\ \citenamefont
  {Menezes}(2021)}]{Donoghue:2021eto}%
  \BibitemOpen
  \bibfield  {author} {\bibinfo {author} {\bibfnamefont {J.~F.}\ \bibnamefont
  {Donoghue}}\ and\ \bibinfo {author} {\bibfnamefont {G.}~\bibnamefont
  {Menezes}},\ }\bibfield  {title} {\bibinfo {title} {{Ostrogradsky instability
  can be overcome by quantum physics}},\ }\href
  {https://doi.org/10.1103/PhysRevD.104.045010} {\bibfield  {journal} {\bibinfo
   {journal} {Phys. Rev. D}\ }\textbf {\bibinfo {volume} {104}},\ \bibinfo
  {pages} {045010} (\bibinfo {year} {2021})},\ \Eprint
  {https://arxiv.org/abs/2105.00898} {arXiv:2105.00898 [hep-th]} \BibitemShut
  {NoStop}%
\bibitem [{\citenamefont {Karananas}\ \emph {et~al.}(2023)\citenamefont
  {Karananas}, \citenamefont {Shaposhnikov},\ and\ \citenamefont
  {Zell}}]{Karananas:2023zgg}%
  \BibitemOpen
  \bibfield  {author} {\bibinfo {author} {\bibfnamefont {G.~K.}\ \bibnamefont
  {Karananas}}, \bibinfo {author} {\bibfnamefont {M.}~\bibnamefont
  {Shaposhnikov}},\ and\ \bibinfo {author} {\bibfnamefont {S.}~\bibnamefont
  {Zell}},\ }\bibfield  {title} {\bibinfo {title} {{Scale invariant
  Einstein-Cartan gravity and flat space conformal symmetry}},\ }\href
  {https://doi.org/10.1007/JHEP11(2023)171} {\bibfield  {journal} {\bibinfo
  {journal} {JHEP}\ }\textbf {\bibinfo {volume} {11}},\ \bibinfo {pages}
  {171}},\ \Eprint {https://arxiv.org/abs/2307.11151} {arXiv:2307.11151
  [hep-th]} \BibitemShut {NoStop}%
\bibitem [{\citenamefont {Karananas}\ \emph
  {et~al.}(2021{\natexlab{a}})\citenamefont {Karananas}, \citenamefont
  {Shaposhnikov}, \citenamefont {Shkerin},\ and\ \citenamefont
  {Zell}}]{Karananas:2021gco}%
  \BibitemOpen
  \bibfield  {author} {\bibinfo {author} {\bibfnamefont {G.~K.}\ \bibnamefont
  {Karananas}}, \bibinfo {author} {\bibfnamefont {M.}~\bibnamefont
  {Shaposhnikov}}, \bibinfo {author} {\bibfnamefont {A.}~\bibnamefont
  {Shkerin}},\ and\ \bibinfo {author} {\bibfnamefont {S.}~\bibnamefont
  {Zell}},\ }\bibfield  {title} {\bibinfo {title} {{Scale and Weyl invariance
  in Einstein-Cartan gravity}},\ }\href
  {https://doi.org/10.1103/PhysRevD.104.124014} {\bibfield  {journal} {\bibinfo
   {journal} {Phys. Rev. D}\ }\textbf {\bibinfo {volume} {104}},\ \bibinfo
  {pages} {124014} (\bibinfo {year} {2021}{\natexlab{a}})},\ \Eprint
  {https://arxiv.org/abs/2108.05897} {arXiv:2108.05897 [hep-th]} \BibitemShut
  {NoStop}%
\bibitem [{\citenamefont {Iosifidis}(2023)}]{Iosifidis:2023but}%
  \BibitemOpen
  \bibfield  {author} {\bibinfo {author} {\bibfnamefont {D.}~\bibnamefont
  {Iosifidis}},\ }\bibfield  {title} {\bibinfo {title} {{Metric-Affine
  Cosmologies: kinematics of Perfect (Ideal) Cosmological Hyperfluids and first
  integrals}},\ }\href {https://doi.org/10.1088/1475-7516/2023/09/045}
  {\bibfield  {journal} {\bibinfo  {journal} {JCAP}\ }\textbf {\bibinfo
  {volume} {09}},\ \bibinfo {pages} {045}},\ \Eprint
  {https://arxiv.org/abs/2301.09868} {arXiv:2301.09868 [gr-qc]} \BibitemShut
  {NoStop}%
\bibitem [{\citenamefont {Shapiro}(2002)}]{Shapiro:2001rz}%
  \BibitemOpen
  \bibfield  {author} {\bibinfo {author} {\bibfnamefont {I.~L.}\ \bibnamefont
  {Shapiro}},\ }\bibfield  {title} {\bibinfo {title} {{Physical aspects of the
  space-time torsion}},\ }\href {https://doi.org/10.1016/S0370-1573(01)00030-8}
  {\bibfield  {journal} {\bibinfo  {journal} {Phys. Rept.}\ }\textbf {\bibinfo
  {volume} {357}},\ \bibinfo {pages} {113} (\bibinfo {year} {2002})},\ \Eprint
  {https://arxiv.org/abs/hep-th/0103093} {arXiv:hep-th/0103093} \BibitemShut
  {NoStop}%
\bibitem [{\citenamefont {Mondal}\ and\ \citenamefont
  {Chakraborty}(2023)}]{Mondal:2023cxx}%
  \BibitemOpen
  \bibfield  {author} {\bibinfo {author} {\bibfnamefont {V.}~\bibnamefont
  {Mondal}}\ and\ \bibinfo {author} {\bibfnamefont {S.}~\bibnamefont
  {Chakraborty}},\ }\bibfield  {title} {\bibinfo {title} {{Lorentzian quantum
  cosmology with torsion}},\ }\href@noop {} {\  (\bibinfo {year} {2023})},\
  \Eprint {https://arxiv.org/abs/2305.01690} {arXiv:2305.01690 [gr-qc]}
  \BibitemShut {NoStop}%
\bibitem [{\citenamefont {Rigouzzo}\ and\ \citenamefont
  {Zell}(2023)}]{Rigouzzo:2023sbb}%
  \BibitemOpen
  \bibfield  {author} {\bibinfo {author} {\bibfnamefont {C.}~\bibnamefont
  {Rigouzzo}}\ and\ \bibinfo {author} {\bibfnamefont {S.}~\bibnamefont
  {Zell}},\ }\bibfield  {title} {\bibinfo {title} {{Coupling metric-affine
  gravity to the standard model and dark matter fermions}},\ }\href
  {https://doi.org/10.1103/PhysRevD.108.124067} {\bibfield  {journal} {\bibinfo
   {journal} {Phys. Rev. D}\ }\textbf {\bibinfo {volume} {108}},\ \bibinfo
  {pages} {124067} (\bibinfo {year} {2023})},\ \Eprint
  {https://arxiv.org/abs/2306.13134} {arXiv:2306.13134 [gr-qc]} \BibitemShut
  {NoStop}%
\bibitem [{\citenamefont {Fomin}\ \emph {et~al.}(2023)\citenamefont {Fomin},
  \citenamefont {Chervon},\ and\ \citenamefont {Bolshakova}}]{Fomin:2023tsn}%
  \BibitemOpen
  \bibfield  {author} {\bibinfo {author} {\bibfnamefont {I.~V.}\ \bibnamefont
  {Fomin}}, \bibinfo {author} {\bibfnamefont {S.~V.}\ \bibnamefont {Chervon}},\
  and\ \bibinfo {author} {\bibfnamefont {K.~A.}\ \bibnamefont {Bolshakova}},\
  }\bibfield  {title} {\bibinfo {title} {{Modified inflationary models based on
  scalar-torsion gravity}},\ }\href@noop {} {\  (\bibinfo {year} {2023})},\
  \Eprint {https://arxiv.org/abs/2312.01142} {arXiv:2312.01142 [gr-qc]}
  \BibitemShut {NoStop}%
\bibitem [{\citenamefont {Gonzalez-Espinoza}\ and\ \citenamefont
  {Otalora}(2021)}]{Gonzalez-Espinoza:2020jss}%
  \BibitemOpen
  \bibfield  {author} {\bibinfo {author} {\bibfnamefont {M.}~\bibnamefont
  {Gonzalez-Espinoza}}\ and\ \bibinfo {author} {\bibfnamefont {G.}~\bibnamefont
  {Otalora}},\ }\bibfield  {title} {\bibinfo {title} {{Cosmological dynamics of
  dark energy in scalar-torsion $f(T,\phi )$ gravity}},\ }\href
  {https://doi.org/10.1140/epjc/s10052-021-09270-x} {\bibfield  {journal}
  {\bibinfo  {journal} {Eur. Phys. J. C}\ }\textbf {\bibinfo {volume} {81}},\
  \bibinfo {pages} {480} (\bibinfo {year} {2021})},\ \Eprint
  {https://arxiv.org/abs/2011.08377} {arXiv:2011.08377 [gr-qc]} \BibitemShut
  {NoStop}%
\bibitem [{\citenamefont {Arcos}\ and\ \citenamefont
  {Pereira}(2004)}]{Arcos:2004tzt}%
  \BibitemOpen
  \bibfield  {author} {\bibinfo {author} {\bibfnamefont {H.~I.}\ \bibnamefont
  {Arcos}}\ and\ \bibinfo {author} {\bibfnamefont {J.~G.}\ \bibnamefont
  {Pereira}},\ }\bibfield  {title} {\bibinfo {title} {{Torsion gravity: A
  Reappraisal}},\ }\href {https://doi.org/10.1142/S0218271804006462} {\bibfield
   {journal} {\bibinfo  {journal} {Int. J. Mod. Phys. D}\ }\textbf {\bibinfo
  {volume} {13}},\ \bibinfo {pages} {2193} (\bibinfo {year} {2004})},\ \Eprint
  {https://arxiv.org/abs/gr-qc/0501017} {arXiv:gr-qc/0501017} \BibitemShut
  {NoStop}%
\bibitem [{\citenamefont {Karananas}\ \emph
  {et~al.}(2021{\natexlab{b}})\citenamefont {Karananas}, \citenamefont
  {Shaposhnikov}, \citenamefont {Shkerin},\ and\ \citenamefont
  {Zell}}]{Karananas:2021zkl}%
  \BibitemOpen
  \bibfield  {author} {\bibinfo {author} {\bibfnamefont {G.~K.}\ \bibnamefont
  {Karananas}}, \bibinfo {author} {\bibfnamefont {M.}~\bibnamefont
  {Shaposhnikov}}, \bibinfo {author} {\bibfnamefont {A.}~\bibnamefont
  {Shkerin}},\ and\ \bibinfo {author} {\bibfnamefont {S.}~\bibnamefont
  {Zell}},\ }\bibfield  {title} {\bibinfo {title} {{Matter matters in
  Einstein-Cartan gravity}},\ }\href
  {https://doi.org/10.1103/PhysRevD.104.064036} {\bibfield  {journal} {\bibinfo
   {journal} {Phys. Rev. D}\ }\textbf {\bibinfo {volume} {104}},\ \bibinfo
  {pages} {064036} (\bibinfo {year} {2021}{\natexlab{b}})},\ \Eprint
  {https://arxiv.org/abs/2106.13811} {arXiv:2106.13811 [hep-th]} \BibitemShut
  {NoStop}%
\bibitem [{\citenamefont {Karananas}(2018)}]{Karananas:2018nrj}%
  \BibitemOpen
  \bibfield  {author} {\bibinfo {author} {\bibfnamefont {G.~K.}\ \bibnamefont
  {Karananas}},\ }\bibfield  {title} {\bibinfo {title} {{On the strong-CP
  problem and its axion solution in torsionful theories}},\ }\href
  {https://doi.org/10.1140/epjc/s10052-018-5972-0} {\bibfield  {journal}
  {\bibinfo  {journal} {Eur. Phys. J. C}\ }\textbf {\bibinfo {volume} {78}},\
  \bibinfo {pages} {480} (\bibinfo {year} {2018})},\ \Eprint
  {https://arxiv.org/abs/1805.08781} {arXiv:1805.08781 [hep-th]} \BibitemShut
  {NoStop}%
\bibitem [{\citenamefont {Gialamas}\ and\ \citenamefont
  {Tamvakis}(2023)}]{Gialamas:2022xtt}%
  \BibitemOpen
  \bibfield  {author} {\bibinfo {author} {\bibfnamefont {I.~D.}\ \bibnamefont
  {Gialamas}}\ and\ \bibinfo {author} {\bibfnamefont {K.}~\bibnamefont
  {Tamvakis}},\ }\bibfield  {title} {\bibinfo {title} {{Inflation in
  metric-affine quadratic gravity}},\ }\href
  {https://doi.org/10.1088/1475-7516/2023/03/042} {\bibfield  {journal}
  {\bibinfo  {journal} {JCAP}\ }\textbf {\bibinfo {volume} {03}},\ \bibinfo
  {pages} {042}},\ \Eprint {https://arxiv.org/abs/2212.09896} {arXiv:2212.09896
  [gr-qc]} \BibitemShut {NoStop}%
\bibitem [{\citenamefont {Mikura}\ \emph {et~al.}(2020)\citenamefont {Mikura},
  \citenamefont {Tada},\ and\ \citenamefont {Yokoyama}}]{Mikura:2020qhc}%
  \BibitemOpen
  \bibfield  {author} {\bibinfo {author} {\bibfnamefont {Y.}~\bibnamefont
  {Mikura}}, \bibinfo {author} {\bibfnamefont {Y.}~\bibnamefont {Tada}},\ and\
  \bibinfo {author} {\bibfnamefont {S.}~\bibnamefont {Yokoyama}},\ }\bibfield
  {title} {\bibinfo {title} {{Conformal inflation in the metric-affine
  geometry}},\ }\href {https://doi.org/10.1209/0295-5075/132/39001} {\bibfield
  {journal} {\bibinfo  {journal} {EPL}\ }\textbf {\bibinfo {volume} {132}},\
  \bibinfo {pages} {39001} (\bibinfo {year} {2020})},\ \Eprint
  {https://arxiv.org/abs/2008.00628} {arXiv:2008.00628 [hep-th]} \BibitemShut
  {NoStop}%
\bibitem [{\citenamefont {Shimada}\ \emph {et~al.}(2019)\citenamefont
  {Shimada}, \citenamefont {Aoki},\ and\ \citenamefont
  {Maeda}}]{Shimada:2018lnm}%
  \BibitemOpen
  \bibfield  {author} {\bibinfo {author} {\bibfnamefont {K.}~\bibnamefont
  {Shimada}}, \bibinfo {author} {\bibfnamefont {K.}~\bibnamefont {Aoki}},\ and\
  \bibinfo {author} {\bibfnamefont {K.-i.}\ \bibnamefont {Maeda}},\ }\bibfield
  {title} {\bibinfo {title} {{Metric-affine Gravity and Inflation}},\ }\href
  {https://doi.org/10.1103/PhysRevD.99.104020} {\bibfield  {journal} {\bibinfo
  {journal} {Phys. Rev. D}\ }\textbf {\bibinfo {volume} {99}},\ \bibinfo
  {pages} {104020} (\bibinfo {year} {2019})},\ \Eprint
  {https://arxiv.org/abs/1812.03420} {arXiv:1812.03420 [gr-qc]} \BibitemShut
  {NoStop}%
\bibitem [{\citenamefont {Azri}\ and\ \citenamefont
  {Demir}(2018)}]{Azri:2018qux}%
  \BibitemOpen
  \bibfield  {author} {\bibinfo {author} {\bibfnamefont {H.}~\bibnamefont
  {Azri}}\ and\ \bibinfo {author} {\bibfnamefont {D.}~\bibnamefont {Demir}},\
  }\bibfield  {title} {\bibinfo {title} {{Induced Affine Inflation}},\ }\href
  {https://doi.org/10.1103/PhysRevD.97.044025} {\bibfield  {journal} {\bibinfo
  {journal} {Phys. Rev. D}\ }\textbf {\bibinfo {volume} {97}},\ \bibinfo
  {pages} {044025} (\bibinfo {year} {2018})},\ \Eprint
  {https://arxiv.org/abs/1802.00590} {arXiv:1802.00590 [gr-qc]} \BibitemShut
  {NoStop}%
\bibitem [{\citenamefont {Iosifidis}\ \emph {et~al.}(2024)\citenamefont
  {Iosifidis}, \citenamefont {Myrzakulov},\ and\ \citenamefont
  {Ravera}}]{Iosifidis:2022xvp}%
  \BibitemOpen
  \bibfield  {author} {\bibinfo {author} {\bibfnamefont {D.}~\bibnamefont
  {Iosifidis}}, \bibinfo {author} {\bibfnamefont {R.}~\bibnamefont
  {Myrzakulov}},\ and\ \bibinfo {author} {\bibfnamefont {L.}~\bibnamefont
  {Ravera}},\ }\bibfield  {title} {\bibinfo {title} {{Cosmology of
  Metric-Affine R+\ensuremath{\beta}R2 Gravity with Pure Shear
  Hypermomentum}},\ }\href {https://doi.org/10.1002/prop.202300003} {\bibfield
  {journal} {\bibinfo  {journal} {Fortsch. Phys.}\ }\textbf {\bibinfo {volume}
  {72}},\ \bibinfo {pages} {2300003} (\bibinfo {year} {2024})},\ \Eprint
  {https://arxiv.org/abs/2301.00669} {arXiv:2301.00669 [gr-qc]} \BibitemShut
  {NoStop}%
\bibitem [{\citenamefont {Dioguardi}\ \emph
  {et~al.}(2022{\natexlab{a}})\citenamefont {Dioguardi}, \citenamefont
  {Racioppi},\ and\ \citenamefont {Tomberg}}]{Dioguardi:2021fmr}%
  \BibitemOpen
  \bibfield  {author} {\bibinfo {author} {\bibfnamefont {C.}~\bibnamefont
  {Dioguardi}}, \bibinfo {author} {\bibfnamefont {A.}~\bibnamefont
  {Racioppi}},\ and\ \bibinfo {author} {\bibfnamefont {E.}~\bibnamefont
  {Tomberg}},\ }\bibfield  {title} {\bibinfo {title} {{Slow-roll inflation in
  Palatini F(R) gravity}},\ }\href {https://doi.org/10.1007/JHEP06(2022)106}
  {\bibfield  {journal} {\bibinfo  {journal} {JHEP}\ }\textbf {\bibinfo
  {volume} {06}},\ \bibinfo {pages} {106}},\ \Eprint
  {https://arxiv.org/abs/2112.12149} {arXiv:2112.12149 [gr-qc]} \BibitemShut
  {NoStop}%
\bibitem [{\citenamefont {Dioguardi}\ \emph
  {et~al.}(2022{\natexlab{b}})\citenamefont {Dioguardi}, \citenamefont
  {Racioppi},\ and\ \citenamefont {Tomberg}}]{Dioguardi:2022oqu}%
  \BibitemOpen
  \bibfield  {author} {\bibinfo {author} {\bibfnamefont {C.}~\bibnamefont
  {Dioguardi}}, \bibinfo {author} {\bibfnamefont {A.}~\bibnamefont
  {Racioppi}},\ and\ \bibinfo {author} {\bibfnamefont {E.}~\bibnamefont
  {Tomberg}},\ }\bibfield  {title} {\bibinfo {title} {{Beyond (and back to)
  Palatini quadratic gravity and inflation}},\ }\href@noop {} {\  (\bibinfo
  {year} {2022}{\natexlab{b}})},\ \Eprint {https://arxiv.org/abs/2212.11869}
  {arXiv:2212.11869 [gr-qc]} \BibitemShut {NoStop}%
\bibitem [{\citenamefont {Dioguardi}\ and\ \citenamefont
  {Racioppi}(2023)}]{Dioguardi:2023jwa}%
  \BibitemOpen
  \bibfield  {author} {\bibinfo {author} {\bibfnamefont {C.}~\bibnamefont
  {Dioguardi}}\ and\ \bibinfo {author} {\bibfnamefont {A.}~\bibnamefont
  {Racioppi}},\ }\bibfield  {title} {\bibinfo {title} {{Palatini $F(R,X)$: a
  new framework for inflationary attractors}},\ }\href@noop {} {\  (\bibinfo
  {year} {2023})},\ \Eprint {https://arxiv.org/abs/2307.02963}
  {arXiv:2307.02963 [gr-qc]} \BibitemShut {NoStop}%
\bibitem [{\citenamefont {Percacci}\ and\ \citenamefont
  {Sezgin}(2020)}]{Percacci:2020ddy}%
  \BibitemOpen
  \bibfield  {author} {\bibinfo {author} {\bibfnamefont {R.}~\bibnamefont
  {Percacci}}\ and\ \bibinfo {author} {\bibfnamefont {E.}~\bibnamefont
  {Sezgin}},\ }\bibfield  {title} {\bibinfo {title} {{New class of ghost- and
  tachyon-free metric affine gravities}},\ }\href
  {https://doi.org/10.1103/PhysRevD.101.084040} {\bibfield  {journal} {\bibinfo
   {journal} {Phys. Rev. D}\ }\textbf {\bibinfo {volume} {101}},\ \bibinfo
  {pages} {084040} (\bibinfo {year} {2020})},\ \Eprint
  {https://arxiv.org/abs/1912.01023} {arXiv:1912.01023 [hep-th]} \BibitemShut
  {NoStop}%
\bibitem [{\citenamefont {Hehl}\ \emph {et~al.}(1995)\citenamefont {Hehl},
  \citenamefont {McCrea}, \citenamefont {Mielke},\ and\ \citenamefont
  {Ne'eman}}]{Hehl:1994ue}%
  \BibitemOpen
  \bibfield  {author} {\bibinfo {author} {\bibfnamefont {F.~W.}\ \bibnamefont
  {Hehl}}, \bibinfo {author} {\bibfnamefont {J.~D.}\ \bibnamefont {McCrea}},
  \bibinfo {author} {\bibfnamefont {E.~W.}\ \bibnamefont {Mielke}},\ and\
  \bibinfo {author} {\bibfnamefont {Y.}~\bibnamefont {Ne'eman}},\ }\bibfield
  {title} {\bibinfo {title} {{Metric affine gauge theory of gravity: Field
  equations, Noether identities, world spinors, and breaking of dilation
  invariance}},\ }\href {https://doi.org/10.1016/0370-1573(94)00111-F}
  {\bibfield  {journal} {\bibinfo  {journal} {Phys. Rept.}\ }\textbf {\bibinfo
  {volume} {258}},\ \bibinfo {pages} {1} (\bibinfo {year} {1995})},\ \Eprint
  {https://arxiv.org/abs/gr-qc/9402012} {arXiv:gr-qc/9402012} \BibitemShut
  {NoStop}%
\bibitem [{\citenamefont {Beltr\'an~Jim\'enez}\ and\ \citenamefont
  {Delhom}(2019)}]{BeltranJimenez:2019acz}%
  \BibitemOpen
  \bibfield  {author} {\bibinfo {author} {\bibfnamefont {J.}~\bibnamefont
  {Beltr\'an~Jim\'enez}}\ and\ \bibinfo {author} {\bibfnamefont
  {A.}~\bibnamefont {Delhom}},\ }\bibfield  {title} {\bibinfo {title} {{Ghosts
  in metric-affine higher order curvature gravity}},\ }\href
  {https://doi.org/10.1140/epjc/s10052-019-7149-x} {\bibfield  {journal}
  {\bibinfo  {journal} {Eur. Phys. J. C}\ }\textbf {\bibinfo {volume} {79}},\
  \bibinfo {pages} {656} (\bibinfo {year} {2019})},\ \Eprint
  {https://arxiv.org/abs/1901.08988} {arXiv:1901.08988 [gr-qc]} \BibitemShut
  {NoStop}%
\bibitem [{\citenamefont {Baldazzi}\ \emph {et~al.}(2022)\citenamefont
  {Baldazzi}, \citenamefont {Melichev},\ and\ \citenamefont
  {Percacci}}]{Baldazzi:2021kaf}%
  \BibitemOpen
  \bibfield  {author} {\bibinfo {author} {\bibfnamefont {A.}~\bibnamefont
  {Baldazzi}}, \bibinfo {author} {\bibfnamefont {O.}~\bibnamefont {Melichev}},\
  and\ \bibinfo {author} {\bibfnamefont {R.}~\bibnamefont {Percacci}},\
  }\bibfield  {title} {\bibinfo {title} {{Metric-Affine Gravity as an effective
  field theory}},\ }\href {https://doi.org/10.1016/j.aop.2022.168757}
  {\bibfield  {journal} {\bibinfo  {journal} {Annals Phys.}\ }\textbf {\bibinfo
  {volume} {438}},\ \bibinfo {pages} {168757} (\bibinfo {year} {2022})},\
  \Eprint {https://arxiv.org/abs/2112.10193} {arXiv:2112.10193 [gr-qc]}
  \BibitemShut {NoStop}%
\bibitem [{\citenamefont {Neville}(1978)}]{Neville:1978bk}%
  \BibitemOpen
  \bibfield  {author} {\bibinfo {author} {\bibfnamefont {D.~E.}\ \bibnamefont
  {Neville}},\ }\bibfield  {title} {\bibinfo {title} {{A Gravity Lagrangian
  With Ghost Free Curvature**2 Terms}},\ }\href
  {https://doi.org/10.1103/PhysRevD.18.3535} {\bibfield  {journal} {\bibinfo
  {journal} {Phys. Rev. D}\ }\textbf {\bibinfo {volume} {18}},\ \bibinfo
  {pages} {3535} (\bibinfo {year} {1978})}\BibitemShut {NoStop}%
\bibitem [{\citenamefont {Neville}(1980)}]{Neville:1979rb}%
  \BibitemOpen
  \bibfield  {author} {\bibinfo {author} {\bibfnamefont {D.~E.}\ \bibnamefont
  {Neville}},\ }\bibfield  {title} {\bibinfo {title} {{Gravity Theories With
  Propagating Torsion}},\ }\href {https://doi.org/10.1103/PhysRevD.21.867}
  {\bibfield  {journal} {\bibinfo  {journal} {Phys. Rev. D}\ }\textbf {\bibinfo
  {volume} {21}},\ \bibinfo {pages} {867} (\bibinfo {year} {1980})}\BibitemShut
  {NoStop}%
\bibitem [{\citenamefont {Nezhad}\ and\ \citenamefont
  {Rasanen}(2024)}]{Nezhad:2023dys}%
  \BibitemOpen
  \bibfield  {author} {\bibinfo {author} {\bibfnamefont {H.~B.}\ \bibnamefont
  {Nezhad}}\ and\ \bibinfo {author} {\bibfnamefont {S.}~\bibnamefont
  {Rasanen}},\ }\bibfield  {title} {\bibinfo {title} {{Scalar fields with
  derivative coupling to curvature in the Palatini and the metric
  formulation}},\ }\href {https://doi.org/10.1088/1475-7516/2024/02/009}
  {\bibfield  {journal} {\bibinfo  {journal} {JCAP}\ }\textbf {\bibinfo
  {volume} {02}},\ \bibinfo {pages} {009}},\ \Eprint
  {https://arxiv.org/abs/2307.04618} {arXiv:2307.04618 [gr-qc]} \BibitemShut
  {NoStop}%
\bibitem [{\citenamefont {Annala}\ and\ \citenamefont
  {Rasanen}(2023)}]{Annala:2022gtl}%
  \BibitemOpen
  \bibfield  {author} {\bibinfo {author} {\bibfnamefont {J.}~\bibnamefont
  {Annala}}\ and\ \bibinfo {author} {\bibfnamefont {S.}~\bibnamefont
  {Rasanen}},\ }\bibfield  {title} {\bibinfo {title} {{Stability of
  non-degenerate Ricci-type Palatini theories}},\ }\href
  {https://doi.org/10.1088/1475-7516/2023/04/014} {\bibfield  {journal}
  {\bibinfo  {journal} {JCAP}\ }\textbf {\bibinfo {volume} {04}},\ \bibinfo
  {pages} {014}},\ \bibinfo {note} {[Erratum: JCAP 08, E02 (2023)]},\ \Eprint
  {https://arxiv.org/abs/2212.09820} {arXiv:2212.09820 [gr-qc]} \BibitemShut
  {NoStop}%
\bibitem [{\citenamefont {Rasanen}\ and\ \citenamefont
  {Verbin}(2022)}]{Rasanen:2022ijc}%
  \BibitemOpen
  \bibfield  {author} {\bibinfo {author} {\bibfnamefont {S.}~\bibnamefont
  {Rasanen}}\ and\ \bibinfo {author} {\bibfnamefont {Y.}~\bibnamefont
  {Verbin}},\ }\bibfield  {title} {\bibinfo {title} {{Palatini formulation for
  gauge theory: implications for slow-roll inflation}}\ }\href
  {https://doi.org/10.21105/astro.2211.15584} {10.21105/astro.2211.15584}
  (\bibinfo {year} {2022}),\ \Eprint {https://arxiv.org/abs/2211.15584}
  {arXiv:2211.15584 [astro-ph.CO]} \BibitemShut {NoStop}%
\bibitem [{\citenamefont {Ito}\ \emph {et~al.}(2022)\citenamefont {Ito},
  \citenamefont {Khater},\ and\ \citenamefont {Rasanen}}]{Ito:2021ssc}%
  \BibitemOpen
  \bibfield  {author} {\bibinfo {author} {\bibfnamefont {A.}~\bibnamefont
  {Ito}}, \bibinfo {author} {\bibfnamefont {W.}~\bibnamefont {Khater}},\ and\
  \bibinfo {author} {\bibfnamefont {S.}~\bibnamefont {Rasanen}},\ }\bibfield
  {title} {\bibinfo {title} {{Tree-level unitarity in Higgs inflation in the
  metric and the Palatini formulation}},\ }\href
  {https://doi.org/10.1007/JHEP06(2022)164} {\bibfield  {journal} {\bibinfo
  {journal} {JHEP}\ }\textbf {\bibinfo {volume} {06}},\ \bibinfo {pages}
  {164}},\ \Eprint {https://arxiv.org/abs/2111.05621} {arXiv:2111.05621
  [astro-ph.CO]} \BibitemShut {NoStop}%
\bibitem [{\citenamefont {Annala}\ and\ \citenamefont
  {Rasanen}(2021)}]{Annala:2021zdt}%
  \BibitemOpen
  \bibfield  {author} {\bibinfo {author} {\bibfnamefont {J.}~\bibnamefont
  {Annala}}\ and\ \bibinfo {author} {\bibfnamefont {S.}~\bibnamefont
  {Rasanen}},\ }\bibfield  {title} {\bibinfo {title} {{Inflation with R
  (\ensuremath{\alpha}\ensuremath{\beta}) terms in the Palatini formulation}},\
  }\href {https://doi.org/10.1088/1475-7516/2021/09/032} {\bibfield  {journal}
  {\bibinfo  {journal} {JCAP}\ }\textbf {\bibinfo {volume} {09}},\ \bibinfo
  {pages} {032}},\ \Eprint {https://arxiv.org/abs/2106.12422} {arXiv:2106.12422
  [astro-ph.CO]} \BibitemShut {NoStop}%
\bibitem [{\citenamefont {Enckell}\ \emph {et~al.}(2021)\citenamefont
  {Enckell}, \citenamefont {Nurmi}, \citenamefont {R\"as\"anen},\ and\
  \citenamefont {Tomberg}}]{Enckell:2020lvn}%
  \BibitemOpen
  \bibfield  {author} {\bibinfo {author} {\bibfnamefont {V.-M.}\ \bibnamefont
  {Enckell}}, \bibinfo {author} {\bibfnamefont {S.}~\bibnamefont {Nurmi}},
  \bibinfo {author} {\bibfnamefont {S.}~\bibnamefont {R\"as\"anen}},\ and\
  \bibinfo {author} {\bibfnamefont {E.}~\bibnamefont {Tomberg}},\ }\bibfield
  {title} {\bibinfo {title} {{Critical point Higgs inflation in the Palatini
  formulation}},\ }\href {https://doi.org/10.1007/JHEP04(2021)059} {\bibfield
  {journal} {\bibinfo  {journal} {JHEP}\ }\textbf {\bibinfo {volume} {04}},\
  \bibinfo {pages} {059}},\ \Eprint {https://arxiv.org/abs/2012.03660}
  {arXiv:2012.03660 [astro-ph.CO]} \BibitemShut {NoStop}%
\bibitem [{\citenamefont {Rasanen}(2019)}]{Rasanen:2018ihz}%
  \BibitemOpen
  \bibfield  {author} {\bibinfo {author} {\bibfnamefont {S.}~\bibnamefont
  {Rasanen}},\ }\bibfield  {title} {\bibinfo {title} {{Higgs inflation in the
  Palatini formulation with kinetic terms for the metric}},\ }\href
  {https://doi.org/10.21105/astro.1811.09514} {\bibfield  {journal} {\bibinfo
  {journal} {Open J. Astrophys.}\ }\textbf {\bibinfo {volume} {2}},\ \bibinfo
  {pages} {1} (\bibinfo {year} {2019})},\ \Eprint
  {https://arxiv.org/abs/1811.09514} {arXiv:1811.09514 [gr-qc]} \BibitemShut
  {NoStop}%
\bibitem [{\citenamefont {Enckell}\ \emph {et~al.}(2019)\citenamefont
  {Enckell}, \citenamefont {Enqvist}, \citenamefont {Rasanen},\ and\
  \citenamefont {Wahlman}}]{Enckell:2018hmo}%
  \BibitemOpen
  \bibfield  {author} {\bibinfo {author} {\bibfnamefont {V.-M.}\ \bibnamefont
  {Enckell}}, \bibinfo {author} {\bibfnamefont {K.}~\bibnamefont {Enqvist}},
  \bibinfo {author} {\bibfnamefont {S.}~\bibnamefont {Rasanen}},\ and\ \bibinfo
  {author} {\bibfnamefont {L.-P.}\ \bibnamefont {Wahlman}},\ }\bibfield
  {title} {\bibinfo {title} {{Inflation with $R^2$ term in the Palatini
  formalism}},\ }\href {https://doi.org/10.1088/1475-7516/2019/02/022}
  {\bibfield  {journal} {\bibinfo  {journal} {JCAP}\ }\textbf {\bibinfo
  {volume} {02}},\ \bibinfo {pages} {022}},\ \Eprint
  {https://arxiv.org/abs/1810.05536} {arXiv:1810.05536 [gr-qc]} \BibitemShut
  {NoStop}%
\bibitem [{\citenamefont {Rasanen}\ and\ \citenamefont
  {Wahlman}(2017)}]{Rasanen:2017ivk}%
  \BibitemOpen
  \bibfield  {author} {\bibinfo {author} {\bibfnamefont {S.}~\bibnamefont
  {Rasanen}}\ and\ \bibinfo {author} {\bibfnamefont {P.}~\bibnamefont
  {Wahlman}},\ }\bibfield  {title} {\bibinfo {title} {{Higgs inflation with
  loop corrections in the Palatini formulation}},\ }\href
  {https://doi.org/10.1088/1475-7516/2017/11/047} {\bibfield  {journal}
  {\bibinfo  {journal} {JCAP}\ }\textbf {\bibinfo {volume} {11}},\ \bibinfo
  {pages} {047}},\ \Eprint {https://arxiv.org/abs/1709.07853} {arXiv:1709.07853
  [astro-ph.CO]} \BibitemShut {NoStop}%
\bibitem [{\citenamefont {Melichev}(2023)}]{Melichev:2023lpn}%
  \BibitemOpen
  \bibfield  {author} {\bibinfo {author} {\bibfnamefont {O.~I.}\ \bibnamefont
  {Melichev}},\ }\emph {\bibinfo {title} {{Quantum Aspects of Metric-Affine
  Gravity}}},\ \href@noop {} {Ph.D. thesis},\ \bibinfo  {school} {SISSA}
  (\bibinfo {year} {2023})\BibitemShut {NoStop}%
\bibitem [{\citenamefont {Mikura}\ and\ \citenamefont
  {Percacci}(2024)}]{Mikura:2024mji}%
  \BibitemOpen
  \bibfield  {author} {\bibinfo {author} {\bibfnamefont {Y.}~\bibnamefont
  {Mikura}}\ and\ \bibinfo {author} {\bibfnamefont {R.}~\bibnamefont
  {Percacci}},\ }\bibfield  {title} {\bibinfo {title} {{Some simple theories of
  gravity with propagating nonmetricity}},\ }\href@noop {} {\  (\bibinfo {year}
  {2024})},\ \Eprint {https://arxiv.org/abs/2401.10097} {arXiv:2401.10097
  [gr-qc]} \BibitemShut {NoStop}%
\bibitem [{\citenamefont {Mikura}\ \emph {et~al.}(2023)\citenamefont {Mikura},
  \citenamefont {Naso},\ and\ \citenamefont {Percacci}}]{Mikura:2023ruz}%
  \BibitemOpen
  \bibfield  {author} {\bibinfo {author} {\bibfnamefont {Y.}~\bibnamefont
  {Mikura}}, \bibinfo {author} {\bibfnamefont {V.}~\bibnamefont {Naso}},\ and\
  \bibinfo {author} {\bibfnamefont {R.}~\bibnamefont {Percacci}},\ }\bibfield
  {title} {\bibinfo {title} {{Some simple theories of gravity with propagating
  torsion}},\ }\href@noop {} {\  (\bibinfo {year} {2023})},\ \Eprint
  {https://arxiv.org/abs/2312.10249} {arXiv:2312.10249 [gr-qc]} \BibitemShut
  {NoStop}%
\bibitem [{\citenamefont {Percacci}(2020)}]{Percacci:2020bzf}%
  \BibitemOpen
  \bibfield  {author} {\bibinfo {author} {\bibfnamefont {R.}~\bibnamefont
  {Percacci}},\ }\bibfield  {title} {\bibinfo {title} {{Towards Metric-Affine
  Quantum Gravity}},\ }\href {https://doi.org/10.1142/S0219887820400034}
  {\bibfield  {journal} {\bibinfo  {journal} {Int. J. Geom. Meth. Mod. Phys.}\
  }\textbf {\bibinfo {volume} {17}},\ \bibinfo {pages} {2040003} (\bibinfo
  {year} {2020})},\ \Eprint {https://arxiv.org/abs/2003.09486}
  {arXiv:2003.09486 [gr-qc]} \BibitemShut {NoStop}%
\bibitem [{\citenamefont {Mikura}\ \emph {et~al.}(2021)\citenamefont {Mikura},
  \citenamefont {Tada},\ and\ \citenamefont {Yokoyama}}]{Mikura:2021ldx}%
  \BibitemOpen
  \bibfield  {author} {\bibinfo {author} {\bibfnamefont {Y.}~\bibnamefont
  {Mikura}}, \bibinfo {author} {\bibfnamefont {Y.}~\bibnamefont {Tada}},\ and\
  \bibinfo {author} {\bibfnamefont {S.}~\bibnamefont {Yokoyama}},\ }\bibfield
  {title} {\bibinfo {title} {{Minimal $k$-inflation in light of the conformal
  metric-affine geometry}},\ }\href
  {https://doi.org/10.1103/PhysRevD.103.L101303} {\bibfield  {journal}
  {\bibinfo  {journal} {Phys. Rev. D}\ }\textbf {\bibinfo {volume} {103}},\
  \bibinfo {pages} {L101303} (\bibinfo {year} {2021})},\ \Eprint
  {https://arxiv.org/abs/2103.13045} {arXiv:2103.13045 [hep-th]} \BibitemShut
  {NoStop}%
\bibitem [{\citenamefont {Mikura}\ and\ \citenamefont
  {Tada}(2022)}]{Mikura:2021clt}%
  \BibitemOpen
  \bibfield  {author} {\bibinfo {author} {\bibfnamefont {Y.}~\bibnamefont
  {Mikura}}\ and\ \bibinfo {author} {\bibfnamefont {Y.}~\bibnamefont {Tada}},\
  }\bibfield  {title} {\bibinfo {title} {{On UV-completion of Palatini-Higgs
  inflation}},\ }\href {https://doi.org/10.1088/1475-7516/2022/05/035}
  {\bibfield  {journal} {\bibinfo  {journal} {JCAP}\ }\textbf {\bibinfo
  {volume} {05}}\bibfield  {number} {\bibinfo  {number} { (05)},\ \bibinfo
  {pages} {035}},\ }\Eprint {https://arxiv.org/abs/2110.03925}
  {arXiv:2110.03925 [hep-ph]} \BibitemShut {NoStop}%
\bibitem [{\citenamefont {He}\ \emph {et~al.}(2023)\citenamefont {He},
  \citenamefont {Mikura},\ and\ \citenamefont {Tada}}]{He:2022xef}%
  \BibitemOpen
  \bibfield  {author} {\bibinfo {author} {\bibfnamefont {M.}~\bibnamefont
  {He}}, \bibinfo {author} {\bibfnamefont {Y.}~\bibnamefont {Mikura}},\ and\
  \bibinfo {author} {\bibfnamefont {Y.}~\bibnamefont {Tada}},\ }\bibfield
  {title} {\bibinfo {title} {{Hybrid metric-Palatini Higgs inflation}},\ }\href
  {https://doi.org/10.1088/1475-7516/2023/05/047} {\bibfield  {journal}
  {\bibinfo  {journal} {JCAP}\ }\textbf {\bibinfo {volume} {05}},\ \bibinfo
  {pages} {047}},\ \Eprint {https://arxiv.org/abs/2209.11051} {arXiv:2209.11051
  [hep-th]} \BibitemShut {NoStop}%
\bibitem [{\citenamefont {Van~Nieuwenhuizen}(1973)}]{VanNieuwenhuizen:1973fi}%
  \BibitemOpen
  \bibfield  {author} {\bibinfo {author} {\bibfnamefont {P.}~\bibnamefont
  {Van~Nieuwenhuizen}},\ }\bibfield  {title} {\bibinfo {title} {{On ghost-free
  tensor lagrangians and linearized gravitation}},\ }\href
  {https://doi.org/10.1016/0550-3213(73)90194-6} {\bibfield  {journal}
  {\bibinfo  {journal} {Nucl. Phys. B}\ }\textbf {\bibinfo {volume} {60}},\
  \bibinfo {pages} {478} (\bibinfo {year} {1973})}\BibitemShut {NoStop}%
\bibitem [{\citenamefont {Sezgin}\ and\ \citenamefont {van
  Nieuwenhuizen}(1980)}]{Sezgin:1979zf}%
  \BibitemOpen
  \bibfield  {author} {\bibinfo {author} {\bibfnamefont {E.}~\bibnamefont
  {Sezgin}}\ and\ \bibinfo {author} {\bibfnamefont {P.}~\bibnamefont {van
  Nieuwenhuizen}},\ }\bibfield  {title} {\bibinfo {title} {{New Ghost Free
  Gravity Lagrangians with Propagating Torsion}},\ }\href
  {https://doi.org/10.1103/PhysRevD.21.3269} {\bibfield  {journal} {\bibinfo
  {journal} {Phys. Rev. D}\ }\textbf {\bibinfo {volume} {21}},\ \bibinfo
  {pages} {3269} (\bibinfo {year} {1980})}\BibitemShut {NoStop}%
\bibitem [{\citenamefont {Sezgin}(1981)}]{Sezgin:1981xs}%
  \BibitemOpen
  \bibfield  {author} {\bibinfo {author} {\bibfnamefont {E.}~\bibnamefont
  {Sezgin}},\ }\bibfield  {title} {\bibinfo {title} {{Class of Ghost Free
  Gravity Lagrangians With Massive or Massless Propagating Torsion}},\ }\href
  {https://doi.org/10.1103/PhysRevD.24.1677} {\bibfield  {journal} {\bibinfo
  {journal} {Phys. Rev. D}\ }\textbf {\bibinfo {volume} {24}},\ \bibinfo
  {pages} {1677} (\bibinfo {year} {1981})}\BibitemShut {NoStop}%
\bibitem [{\citenamefont {Marzo}(2022{\natexlab{a}})}]{Marzo:2021iok}%
  \BibitemOpen
  \bibfield  {author} {\bibinfo {author} {\bibfnamefont {C.}~\bibnamefont
  {Marzo}},\ }\bibfield  {title} {\bibinfo {title} {{Radiatively stable ghost
  and tachyon freedom in metric affine gravity}},\ }\href
  {https://doi.org/10.1103/PhysRevD.106.024045} {\bibfield  {journal} {\bibinfo
   {journal} {Phys. Rev. D}\ }\textbf {\bibinfo {volume} {106}},\ \bibinfo
  {pages} {024045} (\bibinfo {year} {2022}{\natexlab{a}})},\ \Eprint
  {https://arxiv.org/abs/2110.14788} {arXiv:2110.14788 [hep-th]} \BibitemShut
  {NoStop}%
\bibitem [{\citenamefont {Karananas}(2015)}]{Karananas:2014pxa}%
  \BibitemOpen
  \bibfield  {author} {\bibinfo {author} {\bibfnamefont {G.~K.}\ \bibnamefont
  {Karananas}},\ }\bibfield  {title} {\bibinfo {title} {{The particle spectrum
  of parity-violating Poincar\'e gravitational theory}},\ }\href
  {https://doi.org/10.1088/0264-9381/32/5/055012} {\bibfield  {journal}
  {\bibinfo  {journal} {Class. Quant. Grav.}\ }\textbf {\bibinfo {volume}
  {32}},\ \bibinfo {pages} {055012} (\bibinfo {year} {2015})},\ \Eprint
  {https://arxiv.org/abs/1411.5613} {arXiv:1411.5613 [gr-qc]} \BibitemShut
  {NoStop}%
\bibitem [{\citenamefont {Lin}\ \emph {et~al.}(2019)\citenamefont {Lin},
  \citenamefont {Hobson},\ and\ \citenamefont {Lasenby}}]{Lin:2018awc}%
  \BibitemOpen
  \bibfield  {author} {\bibinfo {author} {\bibfnamefont {Y.-C.}\ \bibnamefont
  {Lin}}, \bibinfo {author} {\bibfnamefont {M.~P.}\ \bibnamefont {Hobson}},\
  and\ \bibinfo {author} {\bibfnamefont {A.~N.}\ \bibnamefont {Lasenby}},\
  }\bibfield  {title} {\bibinfo {title} {{Ghost and tachyon free Poincar\'e
  gauge theories: A systematic approach}},\ }\href
  {https://doi.org/10.1103/PhysRevD.99.064001} {\bibfield  {journal} {\bibinfo
  {journal} {Phys. Rev. D}\ }\textbf {\bibinfo {volume} {99}},\ \bibinfo
  {pages} {064001} (\bibinfo {year} {2019})},\ \Eprint
  {https://arxiv.org/abs/1812.02675} {arXiv:1812.02675 [gr-qc]} \BibitemShut
  {NoStop}%
\bibitem [{\citenamefont {Lin}\ \emph {et~al.}(2021)\citenamefont {Lin},
  \citenamefont {Hobson},\ and\ \citenamefont {Lasenby}}]{Lin:2020phk}%
  \BibitemOpen
  \bibfield  {author} {\bibinfo {author} {\bibfnamefont {Y.-C.}\ \bibnamefont
  {Lin}}, \bibinfo {author} {\bibfnamefont {M.~P.}\ \bibnamefont {Hobson}},\
  and\ \bibinfo {author} {\bibfnamefont {A.~N.}\ \bibnamefont {Lasenby}},\
  }\bibfield  {title} {\bibinfo {title} {{Ghost- and tachyon-free Weyl gauge
  theories: A systematic approach}},\ }\href
  {https://doi.org/10.1103/PhysRevD.104.024034} {\bibfield  {journal} {\bibinfo
   {journal} {Phys. Rev. D}\ }\textbf {\bibinfo {volume} {104}},\ \bibinfo
  {pages} {024034} (\bibinfo {year} {2021})},\ \Eprint
  {https://arxiv.org/abs/2005.02228} {arXiv:2005.02228 [gr-qc]} \BibitemShut
  {NoStop}%
\bibitem [{\citenamefont {Barker}\ and\ \citenamefont
  {Zell}(2024)}]{Barker:2023fem}%
  \BibitemOpen
  \bibfield  {author} {\bibinfo {author} {\bibfnamefont {W.}~\bibnamefont
  {Barker}}\ and\ \bibinfo {author} {\bibfnamefont {S.}~\bibnamefont {Zell}},\
  }\bibfield  {title} {\bibinfo {title} {{Einstein-Proca theory from the
  Einstein-Cartan formulation}},\ }\href
  {https://doi.org/10.1103/PhysRevD.109.024007} {\bibfield  {journal} {\bibinfo
   {journal} {Phys. Rev. D}\ }\textbf {\bibinfo {volume} {109}},\ \bibinfo
  {pages} {024007} (\bibinfo {year} {2024})},\ \Eprint
  {https://arxiv.org/abs/2306.14953} {arXiv:2306.14953 [hep-th]} \BibitemShut
  {NoStop}%
\bibitem [{\citenamefont {Vitagliano}\ \emph {et~al.}(2010)\citenamefont
  {Vitagliano}, \citenamefont {Sotiriou},\ and\ \citenamefont
  {Liberati}}]{Vitagliano:2010pq}%
  \BibitemOpen
  \bibfield  {author} {\bibinfo {author} {\bibfnamefont {V.}~\bibnamefont
  {Vitagliano}}, \bibinfo {author} {\bibfnamefont {T.~P.}\ \bibnamefont
  {Sotiriou}},\ and\ \bibinfo {author} {\bibfnamefont {S.}~\bibnamefont
  {Liberati}},\ }\bibfield  {title} {\bibinfo {title} {{The dynamics of
  generalized Palatini Theories of Gravity}},\ }\href
  {https://doi.org/10.1103/PhysRevD.82.084007} {\bibfield  {journal} {\bibinfo
  {journal} {Phys. Rev. D}\ }\textbf {\bibinfo {volume} {82}},\ \bibinfo
  {pages} {084007} (\bibinfo {year} {2010})},\ \Eprint
  {https://arxiv.org/abs/1007.3937} {arXiv:1007.3937 [gr-qc]} \BibitemShut
  {NoStop}%
\bibitem [{\citenamefont {Allemandi}\ \emph {et~al.}(2004)\citenamefont
  {Allemandi}, \citenamefont {Borowiec},\ and\ \citenamefont
  {Francaviglia}}]{Allemandi:2004wn}%
  \BibitemOpen
  \bibfield  {author} {\bibinfo {author} {\bibfnamefont {G.}~\bibnamefont
  {Allemandi}}, \bibinfo {author} {\bibfnamefont {A.}~\bibnamefont
  {Borowiec}},\ and\ \bibinfo {author} {\bibfnamefont {M.}~\bibnamefont
  {Francaviglia}},\ }\bibfield  {title} {\bibinfo {title} {{Accelerated
  cosmological models in Ricci squared gravity}},\ }\href
  {https://doi.org/10.1103/PhysRevD.70.103503} {\bibfield  {journal} {\bibinfo
  {journal} {Phys. Rev. D}\ }\textbf {\bibinfo {volume} {70}},\ \bibinfo
  {pages} {103503} (\bibinfo {year} {2004})},\ \Eprint
  {https://arxiv.org/abs/hep-th/0407090} {arXiv:hep-th/0407090} \BibitemShut
  {NoStop}%
\bibitem [{\citenamefont {Li}\ \emph {et~al.}(2007)\citenamefont {Li},
  \citenamefont {Barrow},\ and\ \citenamefont {Mota}}]{Li:2007xw}%
  \BibitemOpen
  \bibfield  {author} {\bibinfo {author} {\bibfnamefont {B.}~\bibnamefont
  {Li}}, \bibinfo {author} {\bibfnamefont {J.~D.}\ \bibnamefont {Barrow}},\
  and\ \bibinfo {author} {\bibfnamefont {D.~F.}\ \bibnamefont {Mota}},\
  }\bibfield  {title} {\bibinfo {title} {{The Cosmology of Ricci-Tensor-Squared
  gravity in the Palatini variational approach}},\ }\href
  {https://doi.org/10.1103/PhysRevD.76.104047} {\bibfield  {journal} {\bibinfo
  {journal} {Phys. Rev. D}\ }\textbf {\bibinfo {volume} {76}},\ \bibinfo
  {pages} {104047} (\bibinfo {year} {2007})},\ \Eprint
  {https://arxiv.org/abs/0707.2664} {arXiv:0707.2664 [gr-qc]} \BibitemShut
  {NoStop}%
\bibitem [{\citenamefont {Olmo}\ \emph {et~al.}(2009)\citenamefont {Olmo},
  \citenamefont {Sanchis-Alepuz},\ and\ \citenamefont
  {Tripathi}}]{Olmo:2009xy}%
  \BibitemOpen
  \bibfield  {author} {\bibinfo {author} {\bibfnamefont {G.~J.}\ \bibnamefont
  {Olmo}}, \bibinfo {author} {\bibfnamefont {H.}~\bibnamefont
  {Sanchis-Alepuz}},\ and\ \bibinfo {author} {\bibfnamefont {S.}~\bibnamefont
  {Tripathi}},\ }\bibfield  {title} {\bibinfo {title} {{Dynamical Aspects of
  Generalized Palatini Theories of Gravity}},\ }\href
  {https://doi.org/10.1103/PhysRevD.80.024013} {\bibfield  {journal} {\bibinfo
  {journal} {Phys. Rev. D}\ }\textbf {\bibinfo {volume} {80}},\ \bibinfo
  {pages} {024013} (\bibinfo {year} {2009})},\ \Eprint
  {https://arxiv.org/abs/0907.2787} {arXiv:0907.2787 [gr-qc]} \BibitemShut
  {NoStop}%
\bibitem [{\citenamefont {Barragan}\ and\ \citenamefont
  {Olmo}(2010)}]{Barragan:2010qb}%
  \BibitemOpen
  \bibfield  {author} {\bibinfo {author} {\bibfnamefont {C.}~\bibnamefont
  {Barragan}}\ and\ \bibinfo {author} {\bibfnamefont {G.~J.}\ \bibnamefont
  {Olmo}},\ }\bibfield  {title} {\bibinfo {title} {{Isotropic and Anisotropic
  Bouncing Cosmologies in Palatini Gravity}},\ }\href
  {https://doi.org/10.1103/PhysRevD.82.084015} {\bibfield  {journal} {\bibinfo
  {journal} {Phys. Rev. D}\ }\textbf {\bibinfo {volume} {82}},\ \bibinfo
  {pages} {084015} (\bibinfo {year} {2010})},\ \Eprint
  {https://arxiv.org/abs/1005.4136} {arXiv:1005.4136 [gr-qc]} \BibitemShut
  {NoStop}%
\bibitem [{\citenamefont {Bauer}(2011)}]{Bauer:2011sft}%
  \BibitemOpen
  \bibfield  {author} {\bibinfo {author} {\bibfnamefont {F.}~\bibnamefont
  {Bauer}},\ }\bibfield  {title} {\bibinfo {title} {{Filtering out the
  cosmological constant in the Palatini formalism of modified gravity}},\
  }\href {https://doi.org/10.1007/s10714-011-1153-2} {\bibfield  {journal}
  {\bibinfo  {journal} {Gen. Rel. Grav.}\ }\textbf {\bibinfo {volume} {43}},\
  \bibinfo {pages} {1733} (\bibinfo {year} {2011})},\ \Eprint
  {https://arxiv.org/abs/1007.2546} {arXiv:1007.2546 [gr-qc]} \BibitemShut
  {NoStop}%
\bibitem [{\citenamefont {{Barker}}\ and\ \citenamefont
  {{Marzo}}(2024)}]{Barker:2024}%
  \BibitemOpen
  \bibfield  {author} {\bibinfo {author} {\bibfnamefont {W.}~\bibnamefont
  {{Barker}}}\ and\ \bibinfo {author} {\bibfnamefont {C.}~\bibnamefont
  {{Marzo}}},\ }\bibfield  {title} {\bibinfo {title} {{Particle Spectrum for
  any tensor Lagrangian (PSALTer) --- in prep}},\ }\href@noop {} {\  (\bibinfo
  {year} {2024})}\BibitemShut {NoStop}%
\bibitem [{\citenamefont {Barker}(2023{\natexlab{a}})}]{Barker:2023bmr}%
  \BibitemOpen
  \bibfield  {author} {\bibinfo {author} {\bibfnamefont {W.}~\bibnamefont
  {Barker}},\ }\bibfield  {title} {\bibinfo {title} {{Particle spectra of
  gravity based on internal symmetry of quantum fields}},\ }\href@noop {} {\
  (\bibinfo {year} {2023}{\natexlab{a}})},\ \Eprint
  {https://arxiv.org/abs/2311.11790} {arXiv:2311.11790 [hep-th]} \BibitemShut
  {NoStop}%
\bibitem [{\citenamefont {Partanen}\ and\ \citenamefont
  {Tulkki}(2023{\natexlab{a}})}]{Partanen:2023dkt}%
  \BibitemOpen
  \bibfield  {author} {\bibinfo {author} {\bibfnamefont {M.}~\bibnamefont
  {Partanen}}\ and\ \bibinfo {author} {\bibfnamefont {J.}~\bibnamefont
  {Tulkki}},\ }\bibfield  {title} {\bibinfo {title} {{Gravity based on internal
  symmetry of quantum fields}},\ }\href@noop {} {\  (\bibinfo {year}
  {2023}{\natexlab{a}})},\ \Eprint {https://arxiv.org/abs/2310.01460}
  {arXiv:2310.01460 [gr-qc]} \BibitemShut {NoStop}%
\bibitem [{\citenamefont {Partanen}\ and\ \citenamefont
  {Tulkki}(2023{\natexlab{b}})}]{Partanen:2023sjn}%
  \BibitemOpen
  \bibfield  {author} {\bibinfo {author} {\bibfnamefont {M.}~\bibnamefont
  {Partanen}}\ and\ \bibinfo {author} {\bibfnamefont {J.}~\bibnamefont
  {Tulkki}},\ }\bibfield  {title} {\bibinfo {title} {{QED based on
  eight-dimensional spinorial wave equation of the electromagnetic field and
  the emergence of quantum gravity}},\ }\href@noop {} {\  (\bibinfo {year}
  {2023}{\natexlab{b}})},\ \Eprint {https://arxiv.org/abs/2310.02285}
  {arXiv:2310.02285 [physics.gen-ph]} \BibitemShut {NoStop}%
\bibitem [{\citenamefont {Gupta}(1954)}]{Gupta:1954zz}%
  \BibitemOpen
  \bibfield  {author} {\bibinfo {author} {\bibfnamefont {S.~N.}\ \bibnamefont
  {Gupta}},\ }\bibfield  {title} {\bibinfo {title} {{Gravitation and
  Electromagnetism}},\ }\href {https://doi.org/10.1103/PhysRev.96.1683}
  {\bibfield  {journal} {\bibinfo  {journal} {Phys. Rev.}\ }\textbf {\bibinfo
  {volume} {96}},\ \bibinfo {pages} {1683} (\bibinfo {year}
  {1954})}\BibitemShut {NoStop}%
\bibitem [{\citenamefont {Deser}\ and\ \citenamefont
  {Arnowitt}(1963)}]{Deser:1963zzc}%
  \BibitemOpen
  \bibfield  {author} {\bibinfo {author} {\bibfnamefont {S.}~\bibnamefont
  {Deser}}\ and\ \bibinfo {author} {\bibfnamefont {R.}~\bibnamefont
  {Arnowitt}},\ }\bibfield  {title} {\bibinfo {title} {{Interaction Among Gauge
  Vector Fields}},\ }\href {https://doi.org/10.1016/0029-5582(63)90081-6}
  {\bibfield  {journal} {\bibinfo  {journal} {Nucl. Phys.}\ }\textbf {\bibinfo
  {volume} {49}},\ \bibinfo {pages} {133} (\bibinfo {year} {1963})}\BibitemShut
  {NoStop}%
\bibitem [{\citenamefont {Deser}(1970)}]{Deser:1969wk}%
  \BibitemOpen
  \bibfield  {author} {\bibinfo {author} {\bibfnamefont {S.}~\bibnamefont
  {Deser}},\ }\bibfield  {title} {\bibinfo {title} {{Selfinteraction and gauge
  invariance}},\ }\href {https://doi.org/10.1007/BF00759198} {\bibfield
  {journal} {\bibinfo  {journal} {Gen. Rel. Grav.}\ }\textbf {\bibinfo {volume}
  {1}},\ \bibinfo {pages} {9} (\bibinfo {year} {1970})},\ \Eprint
  {https://arxiv.org/abs/gr-qc/0411023} {arXiv:gr-qc/0411023} \BibitemShut
  {NoStop}%
\bibitem [{\citenamefont {Fang}\ and\ \citenamefont
  {Fronsdal}(1979)}]{Fang:1978rc}%
  \BibitemOpen
  \bibfield  {author} {\bibinfo {author} {\bibfnamefont {J.}~\bibnamefont
  {Fang}}\ and\ \bibinfo {author} {\bibfnamefont {C.}~\bibnamefont
  {Fronsdal}},\ }\bibfield  {title} {\bibinfo {title} {{Deformation of Gauge
  Groups. Gravitation}},\ }\href {https://doi.org/10.1063/1.524007} {\bibfield
  {journal} {\bibinfo  {journal} {J. Math. Phys.}\ }\textbf {\bibinfo {volume}
  {20}},\ \bibinfo {pages} {2264} (\bibinfo {year} {1979})}\BibitemShut
  {NoStop}%
\bibitem [{\citenamefont {Lin}\ \emph {et~al.}(2020)\citenamefont {Lin},
  \citenamefont {Hobson},\ and\ \citenamefont {Lasenby}}]{Lin:2019ugq}%
  \BibitemOpen
  \bibfield  {author} {\bibinfo {author} {\bibfnamefont {Y.-C.}\ \bibnamefont
  {Lin}}, \bibinfo {author} {\bibfnamefont {M.~P.}\ \bibnamefont {Hobson}},\
  and\ \bibinfo {author} {\bibfnamefont {A.~N.}\ \bibnamefont {Lasenby}},\
  }\bibfield  {title} {\bibinfo {title} {{Power-counting renormalizable,
  ghost-and-tachyon-free Poincar\'e gauge theories}},\ }\href
  {https://doi.org/10.1103/PhysRevD.101.064038} {\bibfield  {journal} {\bibinfo
   {journal} {Phys. Rev. D}\ }\textbf {\bibinfo {volume} {101}},\ \bibinfo
  {pages} {064038} (\bibinfo {year} {2020})},\ \Eprint
  {https://arxiv.org/abs/1910.14197} {arXiv:1910.14197 [gr-qc]} \BibitemShut
  {NoStop}%
\bibitem [{\citenamefont {Kuhfuss}\ and\ \citenamefont
  {Nitsch}(1986)}]{Kuhfuss:1986rb}%
  \BibitemOpen
  \bibfield  {author} {\bibinfo {author} {\bibfnamefont {R.}~\bibnamefont
  {Kuhfuss}}\ and\ \bibinfo {author} {\bibfnamefont {J.}~\bibnamefont
  {Nitsch}},\ }\bibfield  {title} {\bibinfo {title} {{Propagating Modes in
  Gauge Field Theories of Gravity}},\ }\href
  {https://doi.org/10.1007/BF00763447} {\bibfield  {journal} {\bibinfo
  {journal} {Gen. Rel. Grav.}\ }\textbf {\bibinfo {volume} {18}},\ \bibinfo
  {pages} {1207} (\bibinfo {year} {1986})}\BibitemShut {NoStop}%
\bibitem [{\citenamefont {Mendon\c{c}a}\ and\ \citenamefont
  {Schimidt~Bittencourt}(2020)}]{Mendonca:2019gco}%
  \BibitemOpen
  \bibfield  {author} {\bibinfo {author} {\bibfnamefont {E.~L.}\ \bibnamefont
  {Mendon\c{c}a}}\ and\ \bibinfo {author} {\bibfnamefont {R.}~\bibnamefont
  {Schimidt~Bittencourt}},\ }\bibfield  {title} {\bibinfo {title} {{Unitarity
  of Singh-Hagen model in $D$ dimensions}},\ }\href
  {https://doi.org/10.1155/2020/8425745} {\bibfield  {journal} {\bibinfo
  {journal} {Adv. High Energy Phys.}\ }\textbf {\bibinfo {volume} {2020}},\
  \bibinfo {pages} {8425745} (\bibinfo {year} {2020})},\ \Eprint
  {https://arxiv.org/abs/1902.05118} {arXiv:1902.05118 [hep-th]} \BibitemShut
  {NoStop}%
\bibitem [{\citenamefont {{Rivers}}(1964)}]{Rivers1964}%
  \BibitemOpen
  \bibfield  {author} {\bibinfo {author} {\bibfnamefont {R.~J.}\ \bibnamefont
  {{Rivers}}},\ }\bibfield  {title} {\bibinfo {title} {{Lagrangian theory for
  neutral massive spin-2 fields}},\ }\href {https://doi.org/10.1007/BF02734585}
  {\bibfield  {journal} {\bibinfo  {journal} {Il Nuovo Cimento}\ }\textbf
  {\bibinfo {volume} {34}},\ \bibinfo {pages} {386} (\bibinfo {year}
  {1964})}\BibitemShut {NoStop}%
\bibitem [{\citenamefont {Marzo}(2022{\natexlab{b}})}]{Marzo:2021esg}%
  \BibitemOpen
  \bibfield  {author} {\bibinfo {author} {\bibfnamefont {C.}~\bibnamefont
  {Marzo}},\ }\bibfield  {title} {\bibinfo {title} {{Ghost and tachyon free
  propagation up to spin 3 in Lorentz invariant field theories}},\ }\href
  {https://doi.org/10.1103/PhysRevD.105.065017} {\bibfield  {journal} {\bibinfo
   {journal} {Phys. Rev. D}\ }\textbf {\bibinfo {volume} {105}},\ \bibinfo
  {pages} {065017} (\bibinfo {year} {2022}{\natexlab{b}})},\ \Eprint
  {https://arxiv.org/abs/2108.11982} {arXiv:2108.11982 [hep-ph]} \BibitemShut
  {NoStop}%
\bibitem [{\citenamefont {Beltran~Jimenez}\ \emph {et~al.}(2016)\citenamefont
  {Beltran~Jimenez}, \citenamefont {Heisenberg},\ and\ \citenamefont
  {Koivisto}}]{BeltranJimenez:2016wxw}%
  \BibitemOpen
  \bibfield  {author} {\bibinfo {author} {\bibfnamefont {J.}~\bibnamefont
  {Beltran~Jimenez}}, \bibinfo {author} {\bibfnamefont {L.}~\bibnamefont
  {Heisenberg}},\ and\ \bibinfo {author} {\bibfnamefont {T.~S.}\ \bibnamefont
  {Koivisto}},\ }\bibfield  {title} {\bibinfo {title} {{Cosmology for quadratic
  gravity in generalized Weyl geometry}},\ }\href
  {https://doi.org/10.1088/1475-7516/2016/04/046} {\bibfield  {journal}
  {\bibinfo  {journal} {JCAP}\ }\textbf {\bibinfo {volume} {04}},\ \bibinfo
  {pages} {046}},\ \Eprint {https://arxiv.org/abs/1602.07287} {arXiv:1602.07287
  [hep-th]} \BibitemShut {NoStop}%
\bibitem [{\citenamefont {Iosifidis}(2019)}]{Iosifidis:2018jwu}%
  \BibitemOpen
  \bibfield  {author} {\bibinfo {author} {\bibfnamefont {D.}~\bibnamefont
  {Iosifidis}},\ }\bibfield  {title} {\bibinfo {title} {{Exactly Solvable
  Connections in Metric-Affine Gravity}},\ }\href
  {https://doi.org/10.1088/1361-6382/ab0be2} {\bibfield  {journal} {\bibinfo
  {journal} {Class. Quant. Grav.}\ }\textbf {\bibinfo {volume} {36}},\ \bibinfo
  {pages} {085001} (\bibinfo {year} {2019})},\ \Eprint
  {https://arxiv.org/abs/1812.04031} {arXiv:1812.04031 [gr-qc]} \BibitemShut
  {NoStop}%
\bibitem [{\citenamefont {Hehl}\ \emph {et~al.}(1977)\citenamefont {Hehl},
  \citenamefont {Kerlick}, \citenamefont {Lord},\ and\ \citenamefont
  {Smalley}}]{Hehl:1977fj}%
  \BibitemOpen
  \bibfield  {author} {\bibinfo {author} {\bibfnamefont {F.~W.}\ \bibnamefont
  {Hehl}}, \bibinfo {author} {\bibfnamefont {G.~D.}\ \bibnamefont {Kerlick}},
  \bibinfo {author} {\bibfnamefont {E.~A.}\ \bibnamefont {Lord}},\ and\
  \bibinfo {author} {\bibfnamefont {L.~L.}\ \bibnamefont {Smalley}},\
  }\bibfield  {title} {\bibinfo {title} {{Hypermomentum and the Microscopic
  Violation of the Riemannian Constraint in General Relativity}},\ }\href
  {https://doi.org/10.1016/0370-2693(77)90347-1} {\bibfield  {journal}
  {\bibinfo  {journal} {Phys. Lett. B}\ }\textbf {\bibinfo {volume} {70}},\
  \bibinfo {pages} {70} (\bibinfo {year} {1977})}\BibitemShut {NoStop}%
\bibitem [{\citenamefont {Iosifidis}\ and\ \citenamefont
  {Koivisto}(2023)}]{Iosifidis:2023kyf}%
  \BibitemOpen
  \bibfield  {author} {\bibinfo {author} {\bibfnamefont {D.}~\bibnamefont
  {Iosifidis}}\ and\ \bibinfo {author} {\bibfnamefont {T.~S.}\ \bibnamefont
  {Koivisto}},\ }\bibfield  {title} {\bibinfo {title} {{Hyperhydrodynamics:
  Relativistic Viscous Fluids Emerging from Hypermomentum}},\ }\href@noop {} {\
   (\bibinfo {year} {2023})},\ \Eprint {https://arxiv.org/abs/2312.06780}
  {arXiv:2312.06780 [gr-qc]} \BibitemShut {NoStop}%
\bibitem [{\citenamefont {Iosifidis}(2021{\natexlab{a}})}]{Iosifidis:2021nra}%
  \BibitemOpen
  \bibfield  {author} {\bibinfo {author} {\bibfnamefont {D.}~\bibnamefont
  {Iosifidis}},\ }\bibfield  {title} {\bibinfo {title} {{The Perfect Hyperfluid
  of Metric-Affine Gravity: The Foundation}},\ }\href
  {https://doi.org/10.1088/1475-7516/2021/04/072} {\bibfield  {journal}
  {\bibinfo  {journal} {JCAP}\ }\textbf {\bibinfo {volume} {04}},\ \bibinfo
  {pages} {072}},\ \Eprint {https://arxiv.org/abs/2101.07289} {arXiv:2101.07289
  [gr-qc]} \BibitemShut {NoStop}%
\bibitem [{\citenamefont {Iosifidis}(2021{\natexlab{b}})}]{Iosifidis:2020upr}%
  \BibitemOpen
  \bibfield  {author} {\bibinfo {author} {\bibfnamefont {D.}~\bibnamefont
  {Iosifidis}},\ }\bibfield  {title} {\bibinfo {title} {{Non-Riemannian
  cosmology: The role of shear hypermomentum}},\ }\href
  {https://doi.org/10.1142/S0219887821501292} {\bibfield  {journal} {\bibinfo
  {journal} {Int. J. Geom. Meth. Mod. Phys.}\ }\textbf {\bibinfo {volume}
  {18}},\ \bibinfo {pages} {2150129} (\bibinfo {year} {2021}{\natexlab{b}})},\
  \Eprint {https://arxiv.org/abs/2010.00875} {arXiv:2010.00875 [gr-qc]}
  \BibitemShut {NoStop}%
\bibitem [{\citenamefont {Iosifidis}(2020)}]{Iosifidis:2020gth}%
  \BibitemOpen
  \bibfield  {author} {\bibinfo {author} {\bibfnamefont {D.}~\bibnamefont
  {Iosifidis}},\ }\bibfield  {title} {\bibinfo {title} {{Cosmological
  Hyperfluids, Torsion and Non-metricity}},\ }\href
  {https://doi.org/10.1140/epjc/s10052-020-08634-z} {\bibfield  {journal}
  {\bibinfo  {journal} {Eur. Phys. J. C}\ }\textbf {\bibinfo {volume} {80}},\
  \bibinfo {pages} {1042} (\bibinfo {year} {2020})},\ \Eprint
  {https://arxiv.org/abs/2003.07384} {arXiv:2003.07384 [gr-qc]} \BibitemShut
  {NoStop}%
\bibitem [{\citenamefont {{Penrose}}(1955)}]{1955PCPS...51..406P}%
  \BibitemOpen
  \bibfield  {author} {\bibinfo {author} {\bibfnamefont {R.}~\bibnamefont
  {{Penrose}}},\ }\bibfield  {title} {\bibinfo {title} {{A generalized inverse
  for matrices}},\ }\href {https://doi.org/10.1017/S0305004100030401}
  {\bibfield  {journal} {\bibinfo  {journal} {Proceedings of the Cambridge
  Philosophical Society}\ }\textbf {\bibinfo {volume} {51}},\ \bibinfo {pages}
  {406} (\bibinfo {year} {1955})}\BibitemShut {NoStop}%
\bibitem [{\citenamefont {{Penrose}}(1956)}]{1956PCPS...52...17P}%
  \BibitemOpen
  \bibfield  {author} {\bibinfo {author} {\bibfnamefont {R.}~\bibnamefont
  {{Penrose}}},\ }\bibfield  {title} {\bibinfo {title} {{On best approximate
  solutions of linear matrix equations}},\ }\href
  {https://doi.org/10.1017/S0305004100030929} {\bibfield  {journal} {\bibinfo
  {journal} {Proceedings of the Cambridge Philosophical Society}\ }\textbf
  {\bibinfo {volume} {52}},\ \bibinfo {pages} {17} (\bibinfo {year}
  {1956})}\BibitemShut {NoStop}%
\bibitem [{\citenamefont {{Barker}}\ and\ \citenamefont
  {{Marzo}}()}]{SupplementalMaterials}%
  \BibitemOpen
  \bibfield  {author} {\bibinfo {author} {\bibfnamefont {W.}~\bibnamefont
  {{Barker}}}\ and\ \bibinfo {author} {\bibfnamefont {C.}~\bibnamefont
  {{Marzo}}},\ }\href@noop {} {\emph {\bibinfo {title} {{Supplemental materials
  hosted at
  \href{https://github.com/wevbarker/SupplementalMaterials-2402}{github.com/wevbarker/SupplementalMaterials-2402}}}}}\BibitemShut
  {NoStop}%
\bibitem [{\citenamefont {Iosifidis}\ and\ \citenamefont
  {Koivisto}(2019)}]{Iosifidis:2018zwo}%
  \BibitemOpen
  \bibfield  {author} {\bibinfo {author} {\bibfnamefont {D.}~\bibnamefont
  {Iosifidis}}\ and\ \bibinfo {author} {\bibfnamefont {T.}~\bibnamefont
  {Koivisto}},\ }\bibfield  {title} {\bibinfo {title} {{Scale transformations
  in metric-affine geometry}},\ }\href
  {https://doi.org/10.3390/universe5030082} {\bibfield  {journal} {\bibinfo
  {journal} {Universe}\ }\textbf {\bibinfo {volume} {5}},\ \bibinfo {pages}
  {82} (\bibinfo {year} {2019})},\ \Eprint {https://arxiv.org/abs/1810.12276}
  {arXiv:1810.12276 [gr-qc]} \BibitemShut {NoStop}%
\bibitem [{\citenamefont {Hehl}\ and\ \citenamefont
  {Macias}(1999)}]{Hehl:1999sb}%
  \BibitemOpen
  \bibfield  {author} {\bibinfo {author} {\bibfnamefont {F.~W.}\ \bibnamefont
  {Hehl}}\ and\ \bibinfo {author} {\bibfnamefont {A.}~\bibnamefont {Macias}},\
  }\bibfield  {title} {\bibinfo {title} {{Metric affine gauge theory of
  gravity. 2. Exact solutions}},\ }\href
  {https://doi.org/10.1142/S0218271899000316} {\bibfield  {journal} {\bibinfo
  {journal} {Int. J. Mod. Phys. D}\ }\textbf {\bibinfo {volume} {8}},\ \bibinfo
  {pages} {399} (\bibinfo {year} {1999})},\ \Eprint
  {https://arxiv.org/abs/gr-qc/9902076} {arXiv:gr-qc/9902076} \BibitemShut
  {NoStop}%
\bibitem [{\citenamefont {Obukhov}\ \emph {et~al.}(1997)\citenamefont
  {Obukhov}, \citenamefont {Vlachynsky}, \citenamefont {Esser},\ and\
  \citenamefont {Hehl}}]{PhysRevD.56.7769}%
  \BibitemOpen
  \bibfield  {author} {\bibinfo {author} {\bibfnamefont {Y.~N.}\ \bibnamefont
  {Obukhov}}, \bibinfo {author} {\bibfnamefont {E.~J.}\ \bibnamefont
  {Vlachynsky}}, \bibinfo {author} {\bibfnamefont {W.}~\bibnamefont {Esser}},\
  and\ \bibinfo {author} {\bibfnamefont {F.~W.}\ \bibnamefont {Hehl}},\
  }\bibfield  {title} {\bibinfo {title} {Effective einstein theory from
  metric-affine gravity models via irreducible decompositions},\ }\href
  {https://doi.org/10.1103/PhysRevD.56.7769} {\bibfield  {journal} {\bibinfo
  {journal} {Phys. Rev. D}\ }\textbf {\bibinfo {volume} {56}},\ \bibinfo
  {pages} {7769} (\bibinfo {year} {1997})}\BibitemShut {NoStop}%
\bibitem [{\citenamefont {\c{S}eng\"or}(2023)}]{Sengor:2022kji}%
  \BibitemOpen
  \bibfield  {author} {\bibinfo {author} {\bibfnamefont {G.}~\bibnamefont
  {\c{S}eng\"or}},\ }\bibfield  {title} {\bibinfo {title} {{Particles of a de
  Sitter Universe}},\ }\href {https://doi.org/10.3390/universe9020059}
  {\bibfield  {journal} {\bibinfo  {journal} {Universe}\ }\textbf {\bibinfo
  {volume} {9}},\ \bibinfo {pages} {59} (\bibinfo {year} {2023})},\ \Eprint
  {https://arxiv.org/abs/2212.10626} {arXiv:2212.10626 [hep-th]} \BibitemShut
  {NoStop}%
\bibitem [{\citenamefont {Tekin}(2016)}]{Tekin:2016vli}%
  \BibitemOpen
  \bibfield  {author} {\bibinfo {author} {\bibfnamefont {B.}~\bibnamefont
  {Tekin}},\ }\bibfield  {title} {\bibinfo {title} {{Particle Content of
  Quadratic and $f(R_{\mu\nu\sigma \rho})$ Theories in $(A)dS$}},\ }\href
  {https://doi.org/10.1103/PhysRevD.93.101502} {\bibfield  {journal} {\bibinfo
  {journal} {Phys. Rev. D}\ }\textbf {\bibinfo {volume} {93}},\ \bibinfo
  {pages} {101502} (\bibinfo {year} {2016})},\ \Eprint
  {https://arxiv.org/abs/1604.00891} {arXiv:1604.00891 [hep-th]} \BibitemShut
  {NoStop}%
\bibitem [{\citenamefont {Barker}\ \emph {et~al.}(2021)\citenamefont {Barker},
  \citenamefont {Lasenby}, \citenamefont {Hobson},\ and\ \citenamefont
  {Handley}}]{Barker:2021oez}%
  \BibitemOpen
  \bibfield  {author} {\bibinfo {author} {\bibfnamefont {W.~E.~V.}\
  \bibnamefont {Barker}}, \bibinfo {author} {\bibfnamefont {A.~N.}\
  \bibnamefont {Lasenby}}, \bibinfo {author} {\bibfnamefont {M.~P.}\
  \bibnamefont {Hobson}},\ and\ \bibinfo {author} {\bibfnamefont {W.~J.}\
  \bibnamefont {Handley}},\ }\bibfield  {title} {\bibinfo {title} {{Nonlinear
  Hamiltonian analysis of new quadratic torsion theories: Cases with
  curvature-free constraints}},\ }\href
  {https://doi.org/10.1103/PhysRevD.104.084036} {\bibfield  {journal} {\bibinfo
   {journal} {Phys. Rev. D}\ }\textbf {\bibinfo {volume} {104}},\ \bibinfo
  {pages} {084036} (\bibinfo {year} {2021})},\ \Eprint
  {https://arxiv.org/abs/2101.02645} {arXiv:2101.02645 [gr-qc]} \BibitemShut
  {NoStop}%
\bibitem [{\citenamefont {Barker}(2023{\natexlab{b}})}]{Barker:2022kdk}%
  \BibitemOpen
  \bibfield  {author} {\bibinfo {author} {\bibfnamefont {W.~E.~V.}\
  \bibnamefont {Barker}},\ }\bibfield  {title} {\bibinfo {title}
  {{Supercomputers against strong coupling in gravity with curvature and
  torsion}},\ }\href {https://doi.org/10.1140/epjc/s10052-023-11179-6}
  {\bibfield  {journal} {\bibinfo  {journal} {Eur. Phys. J. C}\ }\textbf
  {\bibinfo {volume} {83}},\ \bibinfo {pages} {228} (\bibinfo {year}
  {2023}{\natexlab{b}})},\ \Eprint {https://arxiv.org/abs/2206.00658}
  {arXiv:2206.00658 [gr-qc]} \BibitemShut {NoStop}%
\bibitem [{\citenamefont {Martin-Garcia}\ \emph {et~al.}(2007)\citenamefont
  {Martin-Garcia}, \citenamefont {Portugal},\ and\ \citenamefont
  {Manssur}}]{Martin-Garcia:2007bqa}%
  \BibitemOpen
  \bibfield  {author} {\bibinfo {author} {\bibfnamefont {J.~M.}\ \bibnamefont
  {Martin-Garcia}}, \bibinfo {author} {\bibfnamefont {R.}~\bibnamefont
  {Portugal}},\ and\ \bibinfo {author} {\bibfnamefont {L.~R.~U.}\ \bibnamefont
  {Manssur}},\ }\bibfield  {title} {\bibinfo {title} {{The Invar Tensor
  Package}},\ }\href {https://doi.org/10.1016/j.cpc.2007.05.015} {\bibfield
  {journal} {\bibinfo  {journal} {Comput. Phys. Commun.}\ }\textbf {\bibinfo
  {volume} {177}},\ \bibinfo {pages} {640} (\bibinfo {year} {2007})},\ \Eprint
  {https://arxiv.org/abs/0704.1756} {arXiv:0704.1756 [cs.SC]} \BibitemShut
  {NoStop}%
\bibitem [{\citenamefont {Martin-Garcia}\ \emph {et~al.}(2008)\citenamefont
  {Martin-Garcia}, \citenamefont {Yllanes},\ and\ \citenamefont
  {Portugal}}]{Martin-Garcia:2008yei}%
  \BibitemOpen
  \bibfield  {author} {\bibinfo {author} {\bibfnamefont {J.~M.}\ \bibnamefont
  {Martin-Garcia}}, \bibinfo {author} {\bibfnamefont {D.}~\bibnamefont
  {Yllanes}},\ and\ \bibinfo {author} {\bibfnamefont {R.}~\bibnamefont
  {Portugal}},\ }\bibfield  {title} {\bibinfo {title} {{The Invar tensor
  package: Differential invariants of Riemann}},\ }\href
  {https://doi.org/10.1016/j.cpc.2008.04.018} {\bibfield  {journal} {\bibinfo
  {journal} {Comput. Phys. Commun.}\ }\textbf {\bibinfo {volume} {179}},\
  \bibinfo {pages} {586} (\bibinfo {year} {2008})},\ \Eprint
  {https://arxiv.org/abs/0802.1274} {arXiv:0802.1274 [cs.SC]} \BibitemShut
  {NoStop}%
\bibitem [{\citenamefont
  {Mart\'\i{}n-Garc\'\i{}a}(2008)}]{Martin-Garcia:2008ysv}%
  \BibitemOpen
  \bibfield  {author} {\bibinfo {author} {\bibfnamefont {J.~M.}\ \bibnamefont
  {Mart\'\i{}n-Garc\'\i{}a}},\ }\bibfield  {title} {\bibinfo {title} {{xPerm:
  fast index canonicalization for tensor computer algebra}},\ }\href
  {https://doi.org/10.1016/j.cpc.2008.05.009} {\bibfield  {journal} {\bibinfo
  {journal} {Comput. Phys. Commun.}\ }\textbf {\bibinfo {volume} {179}},\
  \bibinfo {pages} {597} (\bibinfo {year} {2008})},\ \Eprint
  {https://arxiv.org/abs/0803.0862} {arXiv:0803.0862 [cs.SC]} \BibitemShut
  {NoStop}%
\bibitem [{\citenamefont {Brizuela}\ \emph {et~al.}(2009)\citenamefont
  {Brizuela}, \citenamefont {Martin-Garcia},\ and\ \citenamefont
  {Mena~Marugan}}]{Brizuela:2008ra}%
  \BibitemOpen
  \bibfield  {author} {\bibinfo {author} {\bibfnamefont {D.}~\bibnamefont
  {Brizuela}}, \bibinfo {author} {\bibfnamefont {J.~M.}\ \bibnamefont
  {Martin-Garcia}},\ and\ \bibinfo {author} {\bibfnamefont {G.~A.}\
  \bibnamefont {Mena~Marugan}},\ }\bibfield  {title} {\bibinfo {title} {{xPert:
  Computer algebra for metric perturbation theory}},\ }\href
  {https://doi.org/10.1007/s10714-009-0773-2} {\bibfield  {journal} {\bibinfo
  {journal} {Gen. Rel. Grav.}\ }\textbf {\bibinfo {volume} {41}},\ \bibinfo
  {pages} {2415} (\bibinfo {year} {2009})},\ \Eprint
  {https://arxiv.org/abs/0807.0824} {arXiv:0807.0824 [gr-qc]} \BibitemShut
  {NoStop}%
\bibitem [{\citenamefont {Nutma}(2014)}]{Nutma:2013zea}%
  \BibitemOpen
  \bibfield  {author} {\bibinfo {author} {\bibfnamefont {T.}~\bibnamefont
  {Nutma}},\ }\bibfield  {title} {\bibinfo {title} {{xTras : A field-theory
  inspired xAct package for mathematica}},\ }\href
  {https://doi.org/10.1016/j.cpc.2014.02.006} {\bibfield  {journal} {\bibinfo
  {journal} {Comput. Phys. Commun.}\ }\textbf {\bibinfo {volume} {185}},\
  \bibinfo {pages} {1719} (\bibinfo {year} {2014})},\ \Eprint
  {https://arxiv.org/abs/1308.3493} {arXiv:1308.3493 [cs.SC]} \BibitemShut
  {NoStop}%
\bibitem [{\citenamefont {Melichev}\ and\ \citenamefont
  {Percacci}(2023)}]{Melichev:2023lwj}%
  \BibitemOpen
  \bibfield  {author} {\bibinfo {author} {\bibfnamefont {O.}~\bibnamefont
  {Melichev}}\ and\ \bibinfo {author} {\bibfnamefont {R.}~\bibnamefont
  {Percacci}},\ }\bibfield  {title} {\bibinfo {title} {{On the renormalization
  of Poincar\'e gauge theories}},\ }\href@noop {} {\  (\bibinfo {year}
  {2023})},\ \Eprint {https://arxiv.org/abs/2307.02336} {arXiv:2307.02336
  [hep-th]} \BibitemShut {NoStop}%
\bibitem [{\citenamefont {Vandepeer~Barker}(2022)}]{VandepeerBarker:2022xnp}%
  \BibitemOpen
  \bibfield  {author} {\bibinfo {author} {\bibfnamefont {W.~E.}\ \bibnamefont
  {Vandepeer~Barker}},\ }\emph {\bibinfo {title} {{Gauge theories of
  gravity}}},\ \href {https://doi.org/10.17863/CAM.86972} {Ph.D. thesis},\
  \bibinfo  {school} {Cambridge U.} (\bibinfo {year} {2022})\BibitemShut
  {NoStop}%
\bibitem [{\citenamefont {Barker}(2023{\natexlab{c}})}]{Barker:2022jsh}%
  \BibitemOpen
  \bibfield  {author} {\bibinfo {author} {\bibfnamefont {W.~E.~V.}\
  \bibnamefont {Barker}},\ }\bibfield  {title} {\bibinfo {title} {{Geometric
  multipliers and partial teleparallelism in Poincar\'e gauge theory}},\ }\href
  {https://doi.org/10.1103/PhysRevD.108.024053} {\bibfield  {journal} {\bibinfo
   {journal} {Phys. Rev. D}\ }\textbf {\bibinfo {volume} {108}},\ \bibinfo
  {pages} {024053} (\bibinfo {year} {2023}{\natexlab{c}})},\ \Eprint
  {https://arxiv.org/abs/2205.13534} {arXiv:2205.13534 [gr-qc]} \BibitemShut
  {NoStop}%
\bibitem [{\citenamefont {Barker}\ \emph
  {et~al.}(2020{\natexlab{a}})\citenamefont {Barker}, \citenamefont {Lasenby},
  \citenamefont {Hobson},\ and\ \citenamefont {Handley}}]{Barker:2020elg}%
  \BibitemOpen
  \bibfield  {author} {\bibinfo {author} {\bibfnamefont {W.~E.~V.}\
  \bibnamefont {Barker}}, \bibinfo {author} {\bibfnamefont {A.~N.}\
  \bibnamefont {Lasenby}}, \bibinfo {author} {\bibfnamefont {M.~P.}\
  \bibnamefont {Hobson}},\ and\ \bibinfo {author} {\bibfnamefont {W.~J.}\
  \bibnamefont {Handley}},\ }\bibfield  {title} {\bibinfo {title} {{Mapping
  Poincar\'e gauge cosmology to Horndeski theory for emergent dark energy}},\
  }\href {https://doi.org/10.1103/PhysRevD.102.084002} {\bibfield  {journal}
  {\bibinfo  {journal} {Phys. Rev. D}\ }\textbf {\bibinfo {volume} {102}},\
  \bibinfo {pages} {084002} (\bibinfo {year} {2020}{\natexlab{a}})},\ \Eprint
  {https://arxiv.org/abs/2006.03581} {arXiv:2006.03581 [gr-qc]} \BibitemShut
  {NoStop}%
\bibitem [{\citenamefont {Barker}\ \emph
  {et~al.}(2020{\natexlab{b}})\citenamefont {Barker}, \citenamefont {Lasenby},
  \citenamefont {Hobson},\ and\ \citenamefont {Handley}}]{Barker:2020gcp}%
  \BibitemOpen
  \bibfield  {author} {\bibinfo {author} {\bibfnamefont {W.~E.~V.}\
  \bibnamefont {Barker}}, \bibinfo {author} {\bibfnamefont {A.~N.}\
  \bibnamefont {Lasenby}}, \bibinfo {author} {\bibfnamefont {M.~P.}\
  \bibnamefont {Hobson}},\ and\ \bibinfo {author} {\bibfnamefont {W.~J.}\
  \bibnamefont {Handley}},\ }\bibfield  {title} {\bibinfo {title} {{Systematic
  study of background cosmology in unitary Poincar\'e gauge theories with
  application to emergent dark radiation and $H_0$ tension}},\ }\href
  {https://doi.org/10.1103/PhysRevD.102.024048} {\bibfield  {journal} {\bibinfo
   {journal} {Phys. Rev. D}\ }\textbf {\bibinfo {volume} {102}},\ \bibinfo
  {pages} {024048} (\bibinfo {year} {2020}{\natexlab{b}})},\ \Eprint
  {https://arxiv.org/abs/2003.02690} {arXiv:2003.02690 [gr-qc]} \BibitemShut
  {NoStop}%
\bibitem [{\citenamefont {Blixt}\ \emph {et~al.}(2022)\citenamefont {Blixt},
  \citenamefont {Ferraro}, \citenamefont {Golovnev},\ and\ \citenamefont
  {Guzm\'an}}]{Blixt:2022rpl}%
  \BibitemOpen
  \bibfield  {author} {\bibinfo {author} {\bibfnamefont {D.}~\bibnamefont
  {Blixt}}, \bibinfo {author} {\bibfnamefont {R.}~\bibnamefont {Ferraro}},
  \bibinfo {author} {\bibfnamefont {A.}~\bibnamefont {Golovnev}},\ and\
  \bibinfo {author} {\bibfnamefont {M.-J.}\ \bibnamefont {Guzm\'an}},\
  }\bibfield  {title} {\bibinfo {title} {{Lorentz gauge-invariant variables in
  torsion-based theories of gravity}},\ }\href
  {https://doi.org/10.1103/PhysRevD.105.084029} {\bibfield  {journal} {\bibinfo
   {journal} {Phys. Rev. D}\ }\textbf {\bibinfo {volume} {105}},\ \bibinfo
  {pages} {084029} (\bibinfo {year} {2022})},\ \Eprint
  {https://arxiv.org/abs/2201.11102} {arXiv:2201.11102 [gr-qc]} \BibitemShut
  {NoStop}%
\bibitem [{\citenamefont {Blixt}\ \emph {et~al.}(2023)\citenamefont {Blixt},
  \citenamefont {Hohmann}, \citenamefont {Koivisto},\ and\ \citenamefont
  {Marzola}}]{Blixt:2023qbg}%
  \BibitemOpen
  \bibfield  {author} {\bibinfo {author} {\bibfnamefont {D.}~\bibnamefont
  {Blixt}}, \bibinfo {author} {\bibfnamefont {M.}~\bibnamefont {Hohmann}},
  \bibinfo {author} {\bibfnamefont {T.}~\bibnamefont {Koivisto}},\ and\
  \bibinfo {author} {\bibfnamefont {L.}~\bibnamefont {Marzola}},\ }\bibfield
  {title} {\bibinfo {title} {{Teleparallel bigravity}},\ }\href
  {https://doi.org/10.1140/epjc/s10052-023-12247-7} {\bibfield  {journal}
  {\bibinfo  {journal} {Eur. Phys. J. C}\ }\textbf {\bibinfo {volume} {83}},\
  \bibinfo {pages} {1120} (\bibinfo {year} {2023})},\ \Eprint
  {https://arxiv.org/abs/2305.03504} {arXiv:2305.03504 [gr-qc]} \BibitemShut
  {NoStop}%
\bibitem [{\citenamefont {Rigouzzo}\ and\ \citenamefont
  {Zell}(2022)}]{Rigouzzo:2022yan}%
  \BibitemOpen
  \bibfield  {author} {\bibinfo {author} {\bibfnamefont {C.}~\bibnamefont
  {Rigouzzo}}\ and\ \bibinfo {author} {\bibfnamefont {S.}~\bibnamefont
  {Zell}},\ }\bibfield  {title} {\bibinfo {title} {{Coupling metric-affine
  gravity to a Higgs-like scalar field}},\ }\href
  {https://doi.org/10.1103/PhysRevD.106.024015} {\bibfield  {journal} {\bibinfo
   {journal} {Phys. Rev. D}\ }\textbf {\bibinfo {volume} {106}},\ \bibinfo
  {pages} {024015} (\bibinfo {year} {2022})},\ \Eprint
  {https://arxiv.org/abs/2204.03003} {arXiv:2204.03003 [hep-th]} \BibitemShut
  {NoStop}%
\end{thebibliography}%

\appendix
\counterwithin{figure}{section}

\section{Zero non-metricity with~\PSALTer{}}\label{SecMetApp}
\begin{figure*}[ht]
\includegraphics[width=\textwidth]{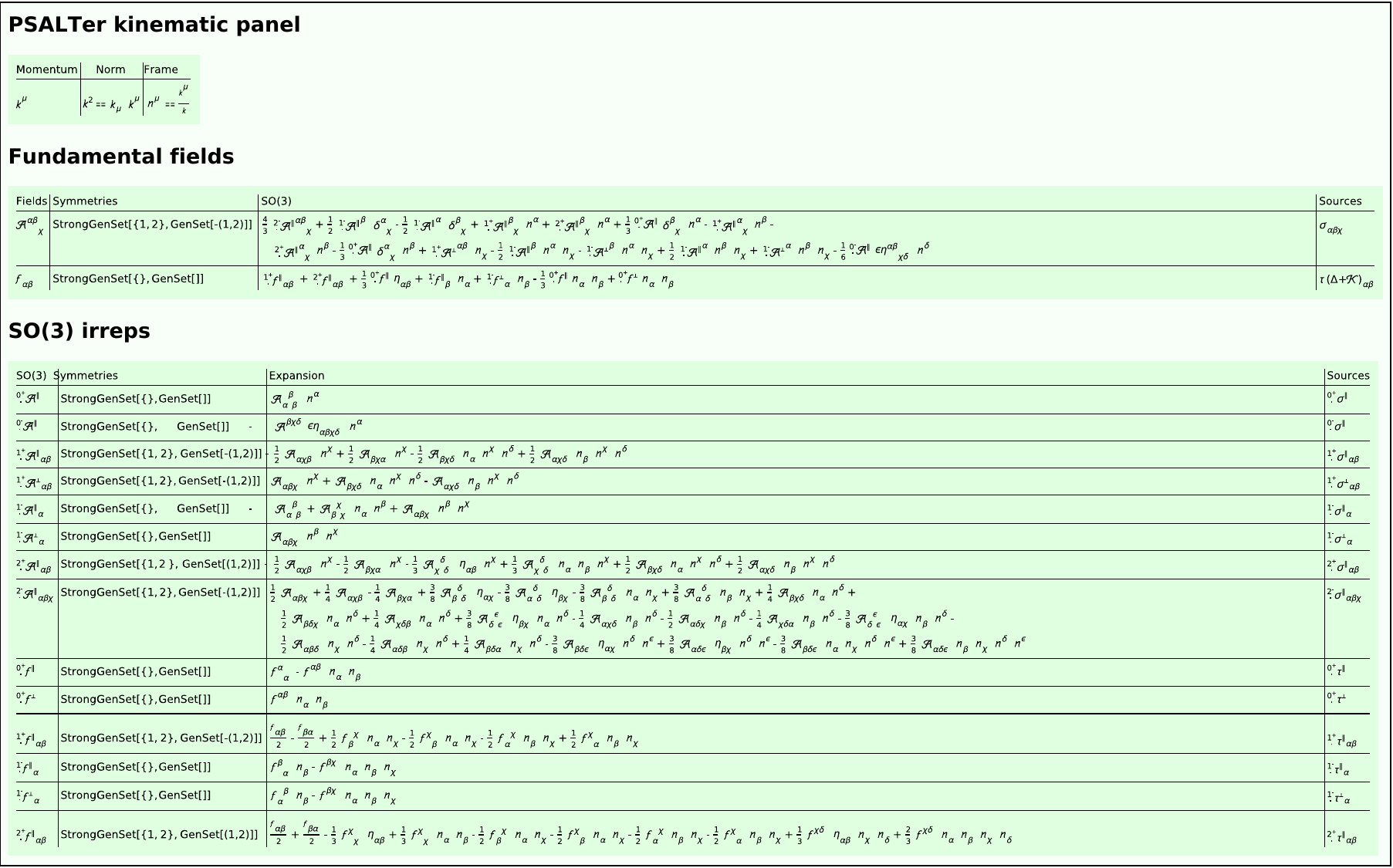}
	\caption{\label{PoincareGaugeTheory} Kinematic structure of Poincar\'e gauge theory (PGT), as used in~\cref{AnnalaRasanenColumn4}. We re-purpose the PGT kinematic module in \PSALTer{} for the study of zero-non-metricity MAG in~\cref{SecMet}. Because of kinematic differences between the PGT and MAG (which are nullified by the extra Lorentz symmetry in PGT), the irreps displayed here do not completely map to those in~\cref{ParticleDec2,ParticleDecNonMet}. The key point is that the spin connection $\tensor{\mathscr{A}}{^{ij}_\mu}\equiv\tensor{\mathscr{A}}{^{[ij]}_\mu}$ maps to the antisymmetric distortion $\MAGD{_{\lambda\mu\nu}}\equiv\MAGD{_{\lambda[\mu\nu]}}$, and the asymmetric tetrad perturbation $\tensor{f}{_i^\mu}$ contains at least the d.o.f in the symmetric metric perturbation $\tensor{h}{_{\mu\nu}}\equiv\tensor{h}{_{(\mu\nu)}}$. Note that the $2^-$ state has a hidden multi-term cyclic symmetry on all its indices, which is not accommodated by the C language implementation of the Butler--Portugal algorithm~\cite{Martin-Garcia:2008ysv}.}
\end{figure*}

\begin{figure*}[ht]
\includegraphics[height=\textwidth,angle=90]{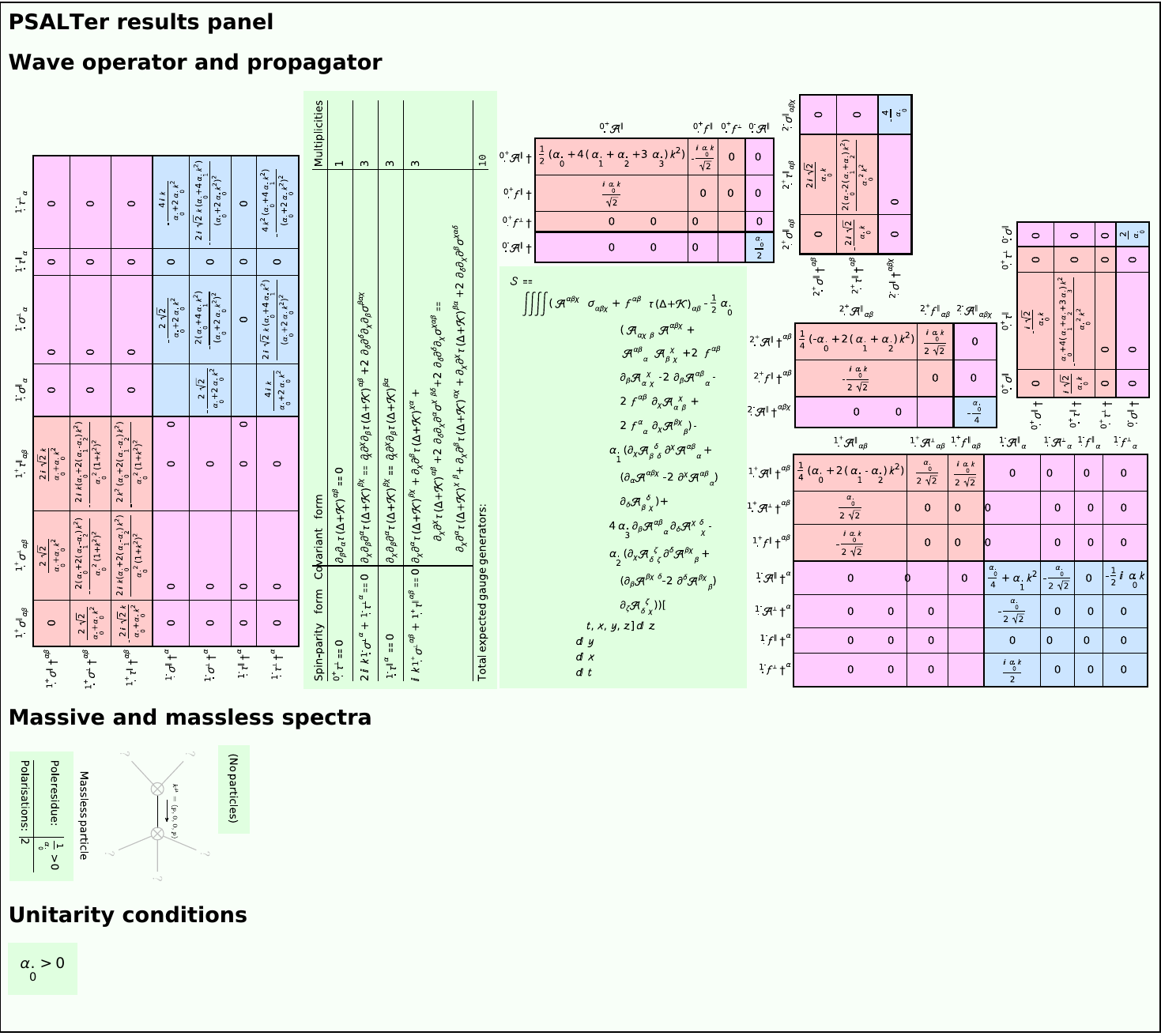}
	\caption{\label{AnnalaRasanenColumn4} The full spectrum of the general theory in~\cref{ActionMetricMAG}, but interpreted as a Poincar\'e gauge theory (PGT) in~\cref{PGTVersion}. Kinematically, the 10 d.o.f of the metric are replaced with the 16 d.o.f of the tetrad field. Consequently however, the additional gauging of the Lorentz group resusts in six extra gauge generatrs on top of the diffeomorphism (translation) generators, so the formulations are not physically distinguishable. All the quantities in this output are defined in~\cref{PoincareGaugeTheory}.}
\end{figure*}

The full spectrum of the general theory in~\cref{ActionMetricMAG} is given in~\cref{AnnalaRasanenColumn4}. Whilst the formulation in~\cref{SecMet} is consistent with the MAG conventions set out in~\cref{MAGConv}, we take the unusual approach of re-formulating the theory as a Poincar\'e gauge theory (PGT)~\cite{Melichev:2023lwj} when presenting the \PSALTer{} analysis. This allows us to recycle a pre-existing PGT kinematic module within \PSALTer{}, and meanwhile the spectral analysis of~\cref{AnnalaRasanenColumn4} is already well-known anyway~\cite{Barker:2023fem}. The PGT kinematic module is displayed in~\cref{PoincareGaugeTheory}.

Whilst our MAG conventions are designed to be identical to~\cite{Percacci:2020ddy}, our PGT conventions will be identical to those in~\cite{Barker:2023bmr,Barker:2023bmr,VandepeerBarker:2022xnp,Barker:2022jsh,Barker:2021oez,Barker:2020elg,Barker:2020gcp}. To define the PGT, we introduce $\tensor{e}{^i_\mu}$ and $\tensor{e}{_i^\mu}$ as the co-tetrad and tetrad components, which are associated with Roman Lorentz (i.e. anholonomic) indices, so that we can compare with the MAG metric $\tensor{e}{^i_\mu}\tensor{e}{^j_\nu}\tensor{\eta}{_{ij}}\cong\tensor{g}{_{\mu\nu}}$ and inverse $\tensor{e}{_i^\mu}\tensor{e}{_j^\nu}\tensor{\eta}{^{ij}}\cong\tensor{g}{^{\mu\nu}}$ with identities $\tensor{e}{^i_\mu}\tensor{e}{_i^\nu}\equiv\tensor*{\delta}{_\mu^\nu}$ and $\tensor{e}{^i_\mu}\tensor{e}{_j^\mu}\equiv\tensor*{\delta}{_j^i}$ as kinematic restrictions. There is also a spin connection $\tensor{\mathscr{A}}{^{ij}_\mu}\equiv\tensor{\mathscr{A}}{^{[ij]}_\mu}$, so that the PGT torsion and PGT curvature are
\begin{subequations}
	\begin{align}
		\tensor{\mathcal{T}}{^k_{ij}}&\equiv 2\tensor{e}{_i^\mu}\tensor{e}{_j^\nu}\big(\PD{_{[\mu|}}\tensor{e}{^k_{|\nu]}}+\tensor{\mathscr{A}}{^k_{m[\mu|}}\tensor{e}{^m_{|\nu]}}\big),\label{PGTTorsion}
	\\
		 \tensor{\mathcal{R}}{^{kl}_{ij}}&\equiv 2\tensor{e}{_i^\mu}\tensor{e}{_j^\nu}\big(\PD{_{[\mu|}}\tensor{\mathscr{A}}{^{kl}_{|\nu]}}+\tensor{\mathscr{A}}{^k_{m[\mu|}}\tensor{\mathscr{A}}{^{ml}_{|\nu]}}\big).\label{PGTCurvature}
	\end{align}
\end{subequations}
The MAG torsion and curvature in~\cref{tqdef,FDef} are precisely analogous to the PGT counterparts in~\cref{PGTTorsion,PGTCurvature} respectively, through the relations $\MAGT{_{\mu}^\alpha_{\nu}}\cong\tensor{e}{^i_\mu}\tensor{e}{_k^\alpha}\tensor{e}{^j_\nu}\tensor{\mathcal{T}}{^k_{ij}}$ and $\MAGF{_{\mu\nu}^\rho_\sigma}\cong\tensor{e}{^i_\mu}\tensor{e}{^j_\nu}\tensor{e}{_k^\rho}\tensor{e}{_l_\sigma}\tensor{\mathcal{R}}{^{k}_{lij}}$, where we pay attention to the different ordering of the indices according to the two conventions. In terms of the PGT field strength tensors, the action in~\cref{ActionMetricMAG} corresponds to
\begin{equation}\label{PGTVersion}
	S[e,\mathscr{A}]=\int\mathrm{d}^4x e\bigg[\alpha_0\mathcal{R}+\tensor{\mathcal{R}}{^{ij}}\big(\alpha_1\tensor{\mathcal{R}}{_{ij}}+\alpha_2\tensor{\mathcal{R}}{_{ji}}\big)\bigg],
\end{equation}
where $S[e,\mathscr{A}]\cong S[g,A]$ and we adopt the dimensionful coupling $\alpha_0$ and dimensionless couplings $\alpha_1$ and $\alpha_2$ in place of $a_0$, $g_1$ and $g_2$. The contractions are defined $\tensor{\mathcal{R}}{_{ij}}\equiv\tensor{\mathcal{R}}{^l_{ilj}}$ and $\mathcal{R}\equiv\tensor{\mathcal{R}}{^i_i}$ with the measure $e\equiv\det\big(\tensor{e}{^i_\mu}\big)\cong\sqrt{-g}$. In the weak-field regime, we take $\tensor{\mathscr{A}}{^{ij}_\mu}$ to be inherently perturbative, and we define the exact tetrad perturbation $\tensor{e}{_i^\mu}\equiv\tensor*{\delta}{_i^\mu}+\tensor{f}{_i^\mu}$, i.e. the `Kronecker' choice of Minkowski vacuum~\cite{Barker:2023bmr,Blixt:2022rpl,Blixt:2023qbg}. 

To lowest order in the quadratic action, the Greek and Roman indices are then interchangeable --- indeed \PSALTer{} only knows about one set of Lorentz indices on the Minkowski background, and these are strictly associated with Greek indices which represent Cartesian coordinates. There are 16 d.o.f in $\tensor{f}{_i^\mu}$ and 24 d.o.f in $\tensor{\mathscr{A}}{^{ij}_\mu}$. The latter can clearly be accounted for, in the second-order formulation, by the d.o.f in~\cref{ParticleDecNonMet}. Most of the former are accounted for by the metric d.o.f in~\cref{ParticleDec2}, but there are six further d.o.f in the antisymmetric part of the tetrad which do not appear in MAG. This is not a problem, because, these six d.o.f are immediately eliminated by the six gauge generators of the Lorentz symmetry, part of the Poincar\'e symmetry, which also is not visible in the MAG. As a consequence, the spin-one matrices in~\cref{AnnalaRasanenColumn4} have two rows and two columns more than they would do in the MAG formulation, but the dimension of their null space also increases by two. Kinematic extensions of the theory which are cancelled by symmetries in this way do not alter the physics, and in this sense we understand the zero-non-metricity MAG and the PGT to be equivalent theories.

Conjugate to the tetrad perturbation $\tensor{f}{_i^\mu}$ and the spin connection $\tensor{\mathscr{A}}{^{ij}_\mu}$ are the translational source (asymmetric stress-energy tensor) $\tensor{\tau}{^i_\mu}$ and matter spin current $\tensor{\sigma}{_{ij}^\mu}$~\cite{Rigouzzo:2023sbb,Rigouzzo:2022yan,Karananas:2021zkl}. The reduced-index $\mathrm{SO}(3)$ irreducible parts of these fields and sources label the rows and columns of the matrices in~\cref{AnnalaRasanenColumn4}, and have spin-parity ($J^P$) labels to identify them. Duplicate $J^P$ states are distinguished by additional parallel ($\parallel$) and perpendicular ($\perp$) labels --- but there is no significant meaning behind these auxiliary labels.

\section{Zero torsion with~\PSALTer{}}\label{SecTorApp}
\begin{figure*}[ht]
\includegraphics[width=\textwidth]{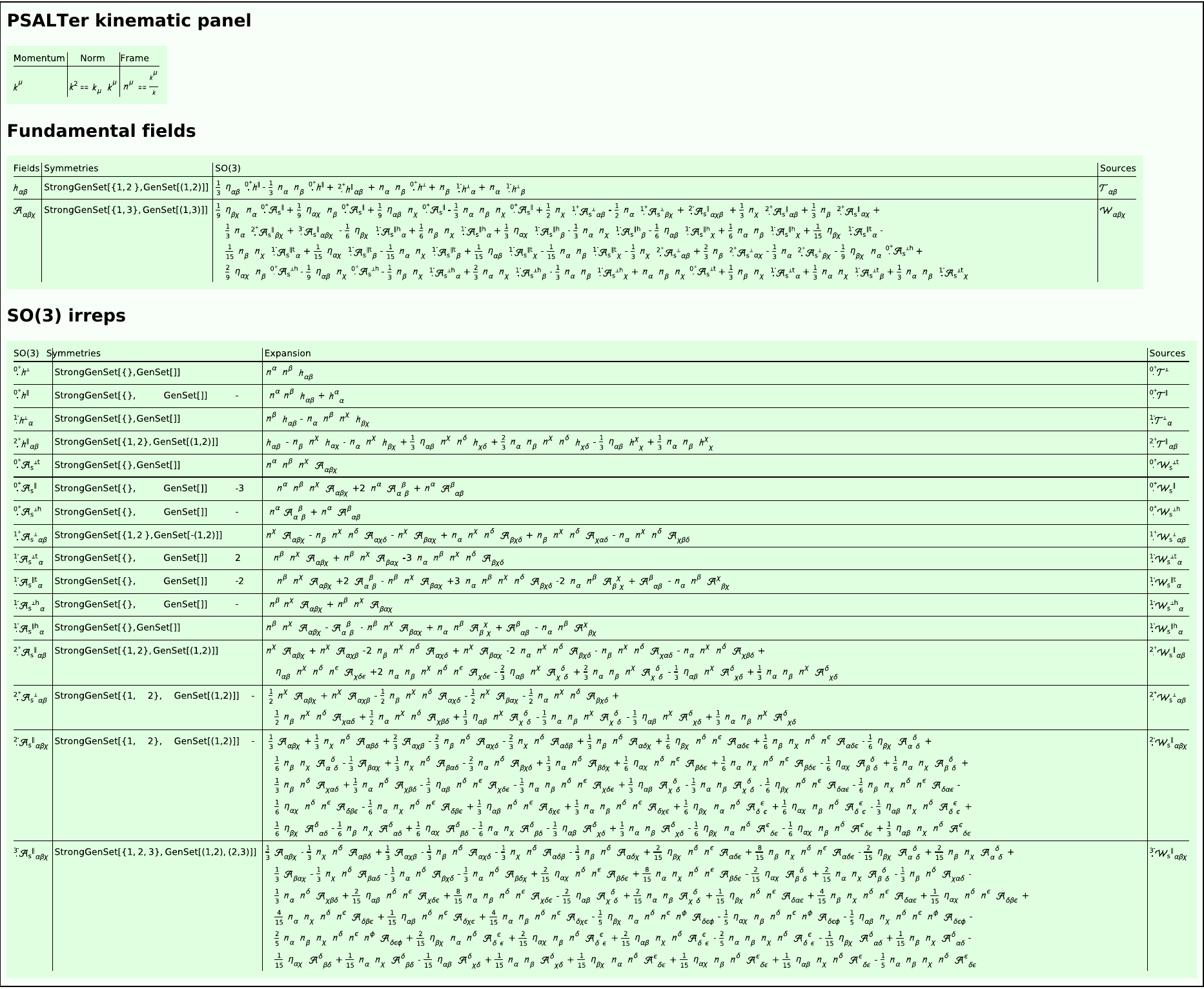}
	\caption{\label{ZeroTorsionPalatini} Kinematic structure of zero-torsion MAG, as studied throughout~\cref{SecTor}. The $\mathrm{SO}(3)$ irreps precisely correspond to those in~\cref{ParticleDecT,ParticleDec2}, though the labelling of duplicate $J^P$ states is different from that in~\cite{Percacci:2020ddy}. The key point is that the connection field carries an extra symmetry restriction $\MAGA{_\mu^\rho_\nu}\equiv\MAGA{_{(\mu|}^\rho_{|\nu)}}$, and by referring to~\cref{tqdef} we see that this kills off the torsion in the first-order formulation. In moving to the second-order formulation, we make a slight notational abuse in~\cref{Reparam}, but from~\cref{TorNonMe} we see that the effect will still be as desired if we keep using this kinematic module. As with~\cref{PoincareGaugeTheory}, the $2^-$ and $3^-$ states have extra cyclic symmetries which are hidden. These definitions are used in~\cref{ZeroTorsionGeneralFirstOrder,ZeroTorsionGeneralSecondOrder,ZeroTorsionNo2pNo1pSimpleFirstOrder,ZeroTorsionNo2pNo1pSimpleSecondOrder,ZeroTorsionNo2pNo1pComplicatedFirstOrder,ZeroTorsionNo2pNo1pComplicatedSecondOrder}.}
\end{figure*}

\begin{figure*}[ht]
\includegraphics[width=\textwidth]{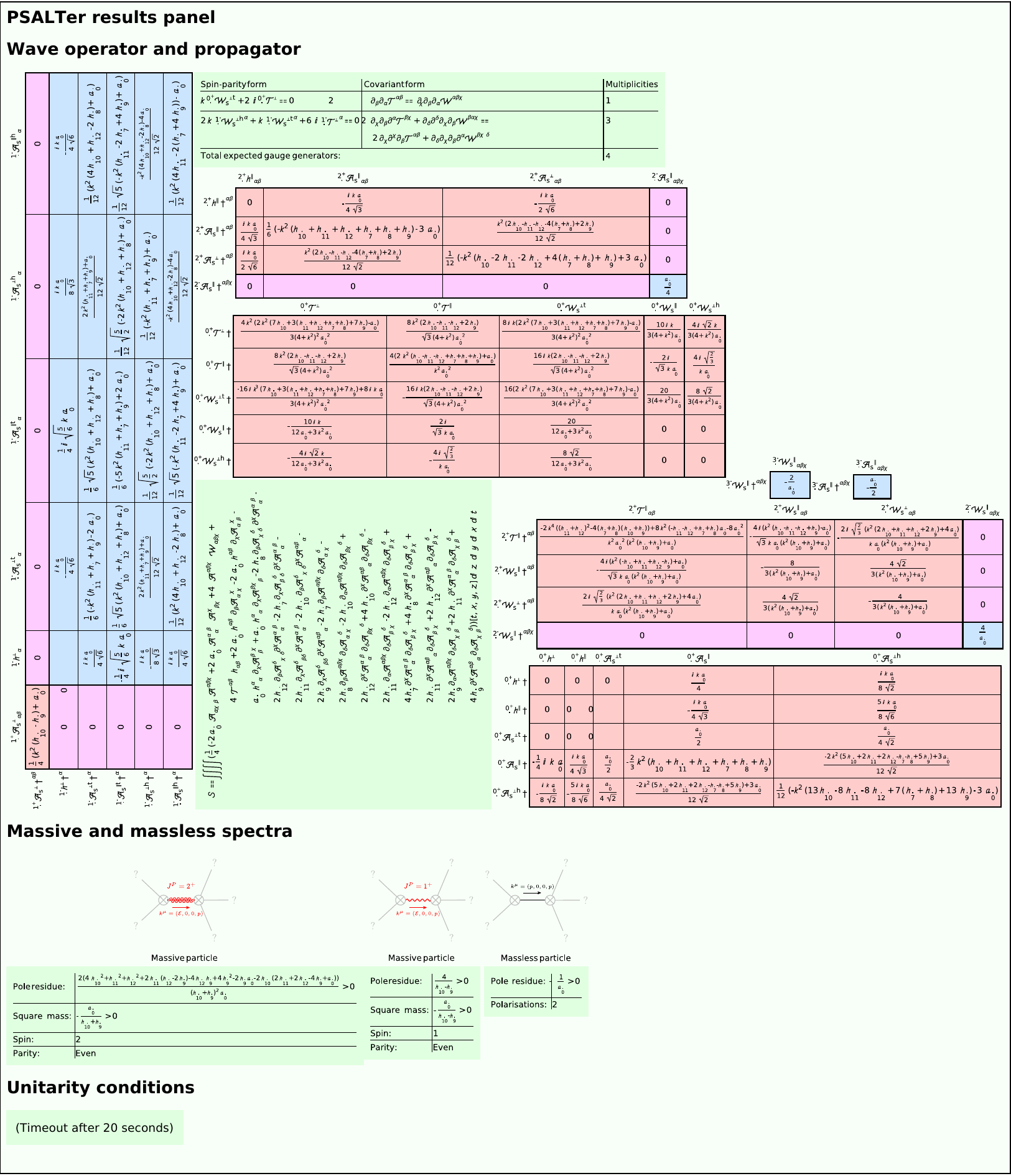}
	\caption{\label{ZeroTorsionGeneralFirstOrder} The full spectrum of the general theory in~\cref{ActionTLessMAG}. No general unitarity conditions are obtained, without further tuning. All the quantities in this output are defined in~\cref{ZeroTorsionPalatini}.}
\end{figure*}

Unlike in~\cref{SecMetApp}, our \PSALTer{} analysis corresponding to~\cref{SecTor} is fully grounded in the MAG formulation. The zero-torsion MAG kinematic module is displayed in~\cref{ZeroTorsionPalatini}. The first-order analyses in~\cref{ZeroTorsionGeneralFirstOrder,ZeroTorsionNo2pNo1pSimpleFirstOrder,ZeroTorsionNo2pNo1pComplicatedFirstOrder} share all our notational conventions above: the fields $\tensor{h}{_{\mu\nu}}$ and $\MAGA{_\mu^\rho_\nu}$ are perturbative. To reach the second-order formulation, we only have to edit the quadratic action before substituting into the \texttt{ParticleSpectrum} function (which is the main function provided by the \PSALTer{} package). The reparameterisation used to transform the quadratic action is
\begin{equation}\label{Reparam}
	\MAGA{_\mu^\rho_\nu}\mapsto\MAGA{_\mu^\rho_\nu}+\frac{1}{2}\big(2\PD{_{(\mu|}}\tensor{h}{_{\lambda|\nu)}}-\PD{_\lambda}\tensor{h}{_{\mu\nu}}\big).
\end{equation}
To lowest order in perturbative fields,~\cref{Reparam} captures the transition from $\MAGA{_\mu^\rho_\nu}$ to $\MAGD{_\mu^\rho_\nu}$ set out in~\cref{MAGConv}. The notation is slightly abusive, because $\MAGA{_\mu^\rho_\nu}$ on the RHS of~\cref{Reparam} is really $\MAGD{_\mu^\rho_\nu}$. But since there is no advantage in defining a new kinematic module for \PSALTer{} just to avoid the notational conflict, we therefore lazily recycle the first-order zero-torsion MAG module for all our second-order calculations.

Conjugate to the metric perturbation $\tensor{h}{_{\mu\nu}}$ and the affine connection $\MAGA{_\mu^\rho_\nu}$ are the (symmetric) stress-energy tensor $\tensor{T}{^{\mu\nu}}$ and the current $\tensor{W}{^\mu_\rho^\nu}$ which in MAG has become known as the \emph{hypermomentum}. As with the PGT notation in~\cref{AnnalaRasanenColumn4}, the $J^P$ states are labelled as such. To distinguish the duplicate $J^P$ states, apart from the ($\parallel$) and ($\perp$) symbols, we use the letters ($\mathsf{s}$), ($\mathsf{h}$) and ($\mathsf{t}$) --- once again there is no significant meaning behind these labels. Different labels (numerical subscripts) are used in~\cref{ParticleDecT}.

\begin{figure*}[ht]
\includegraphics[width=\textwidth]{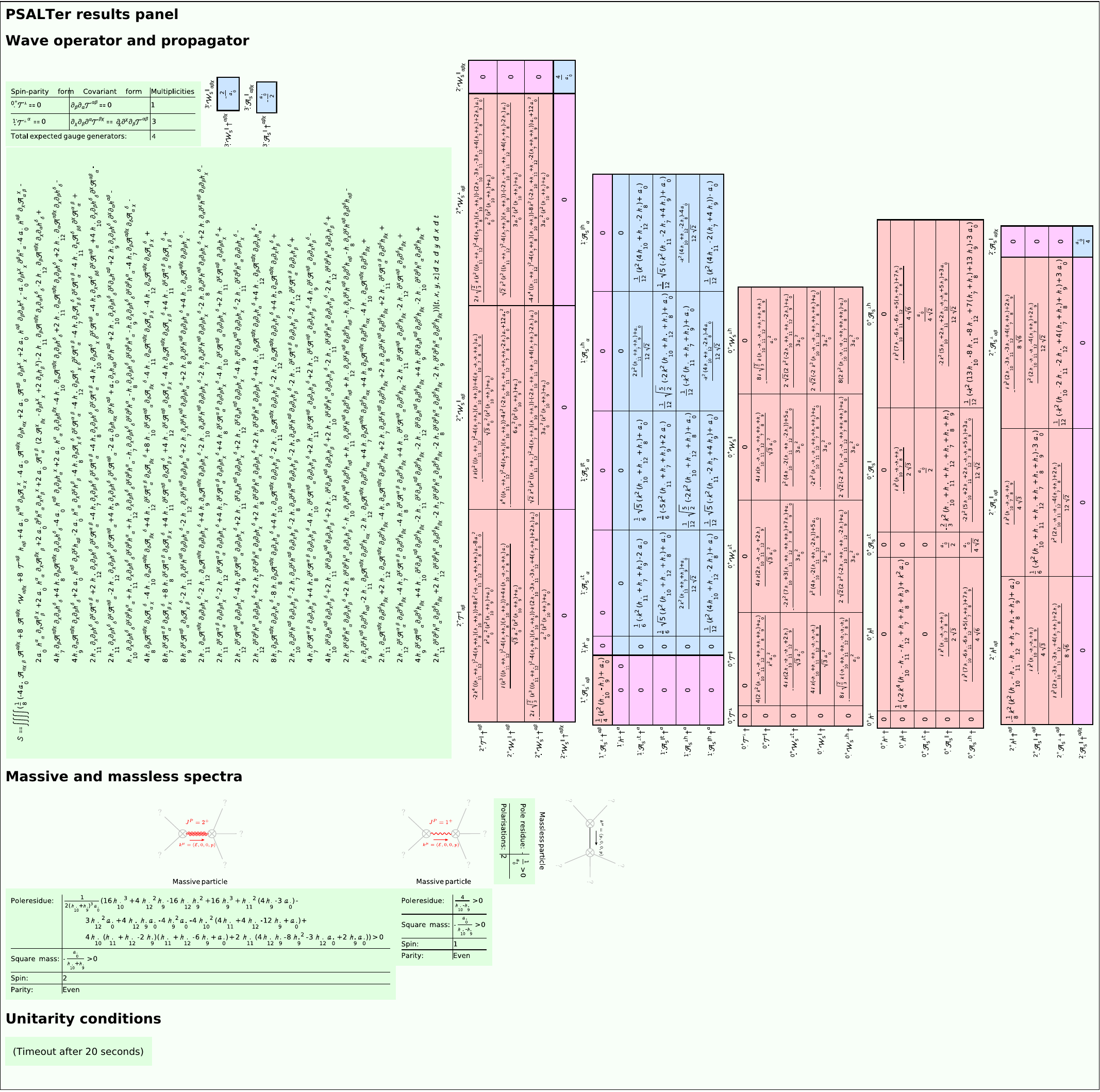}
	\caption{\label{ZeroTorsionGeneralSecondOrder} The results in~\cref{ZeroTorsionGeneralFirstOrder} repeated in the second-order formulation. Note that the quadratic action in this case contains very many more operators than does~\cref{ZeroTorsionGeneralFirstOrder}. The matrix elements and the forms of the pole residues are expected to change, but the mass spectrum is the same. Once again, the unitarity conditions are not obtained for the full theory: despite the apparent changes to the residues, such conditions should be invariant under re-parameterisations. All the quantities in this output are defined in~\cref{ZeroTorsionPalatini}.}
\end{figure*}

\begin{figure*}[ht]
\includegraphics[width=\textwidth]{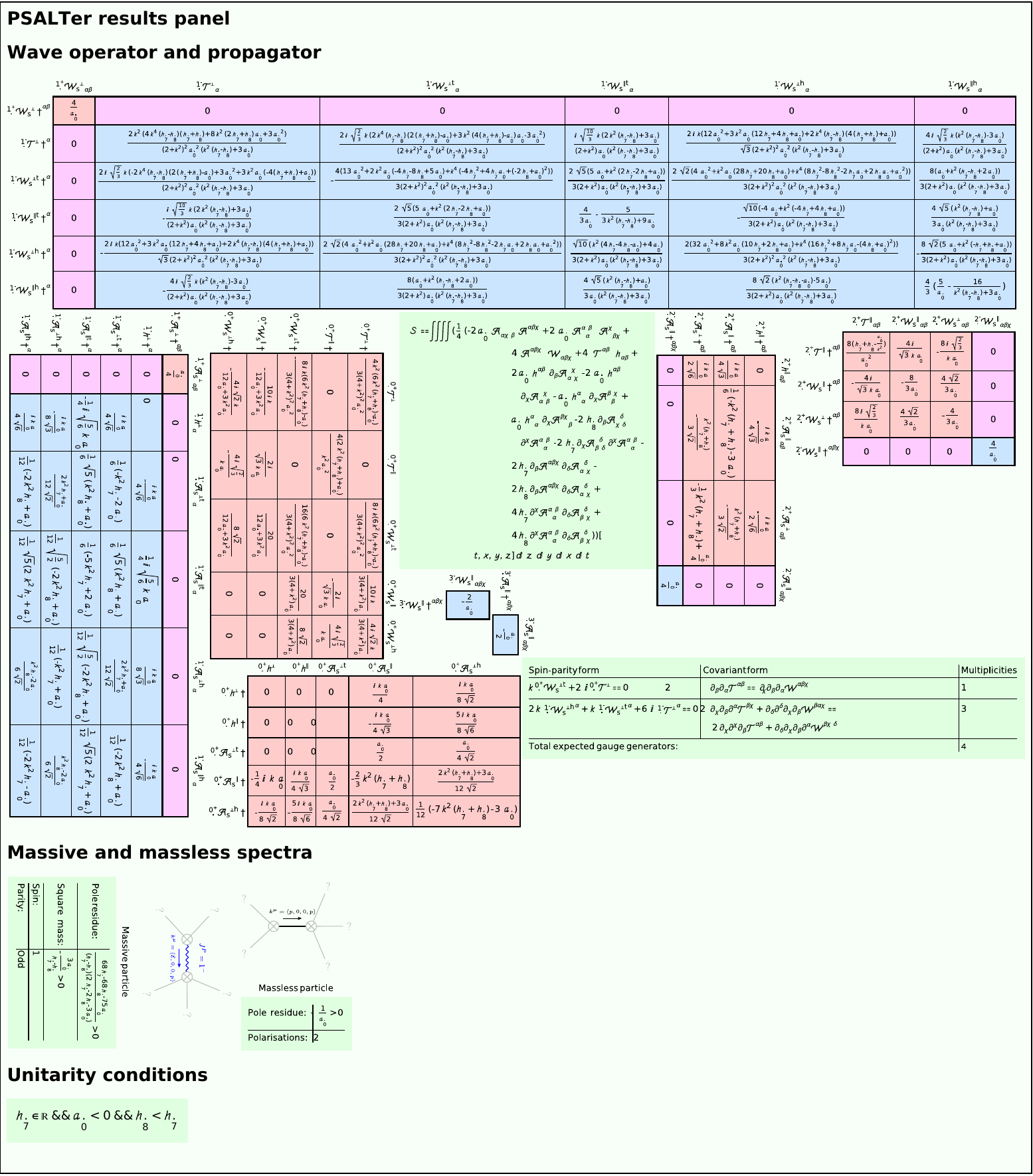}
	\caption{\label{ZeroTorsionNo2pNo1pSimpleFirstOrder} The spectrum of the theory in~\cref{ActionTLessMAG} with the additional constraints $c_9=c_{10}=c_{11}=c_{12}=0$. These are sufficient to eliminate the $2^+$ and $1^+$ massive states, leaving only the massive $1^-$ state in~\cref{sp1meno}. The overall theory is clearly unitary. All the quantities in this output are defined in~\cref{ZeroTorsionPalatini}.}
\end{figure*}

\begin{figure*}[ht]
\includegraphics[height=\textwidth,angle=90]{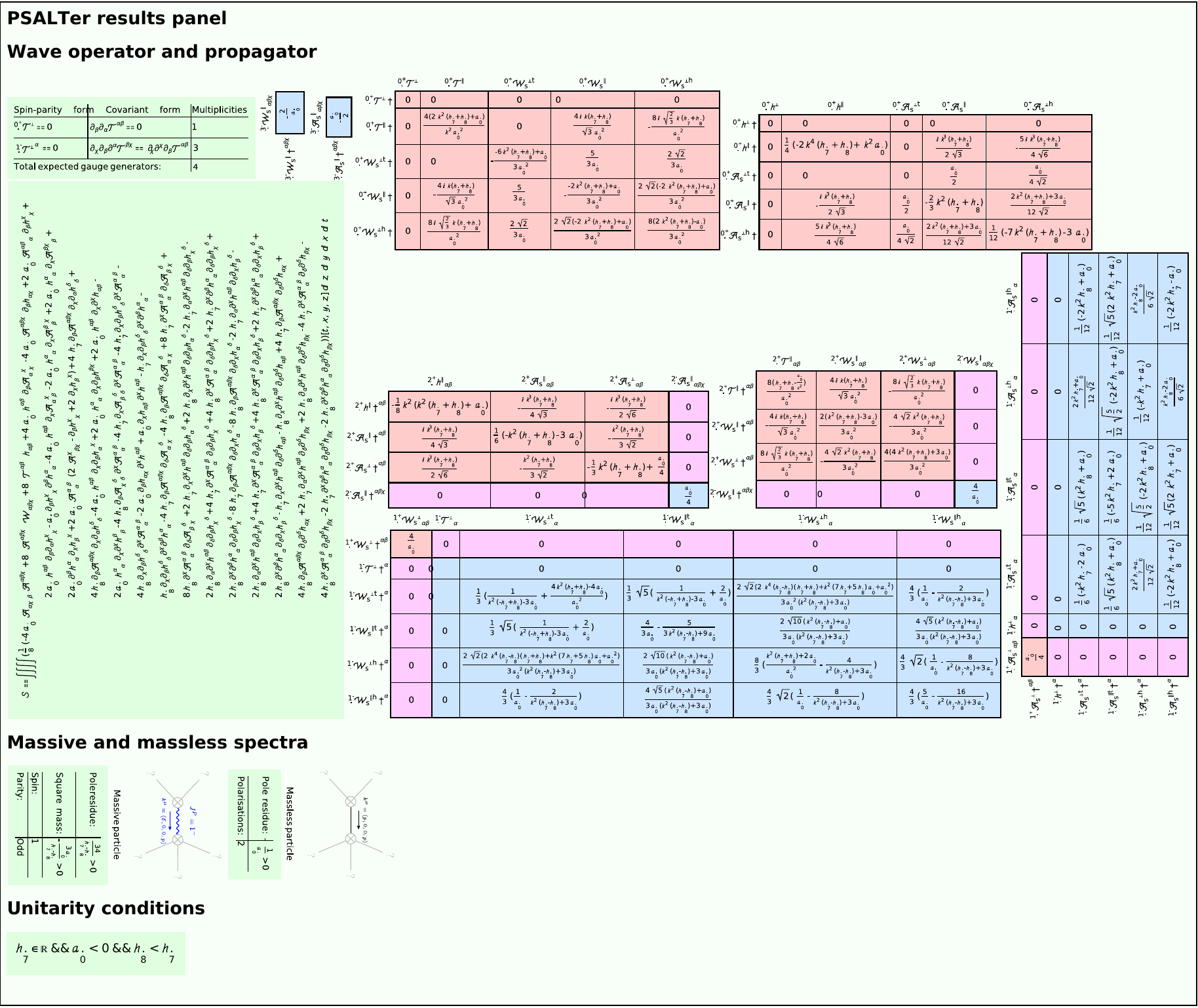}
	\caption{\label{ZeroTorsionNo2pNo1pSimpleSecondOrder} The results in~\cref{ZeroTorsionNo2pNo1pSimpleFirstOrder} repeated in the second-order formulation. Note the much larger quadratic expansion, but the consistent mass spectrum and unitarity conditions. The results are used in~\cref{sp1meno}. All the quantities in this output are defined in~\cref{ZeroTorsionPalatini}.}
\end{figure*}

\begin{figure*}[ht]
\includegraphics[width=\textwidth]{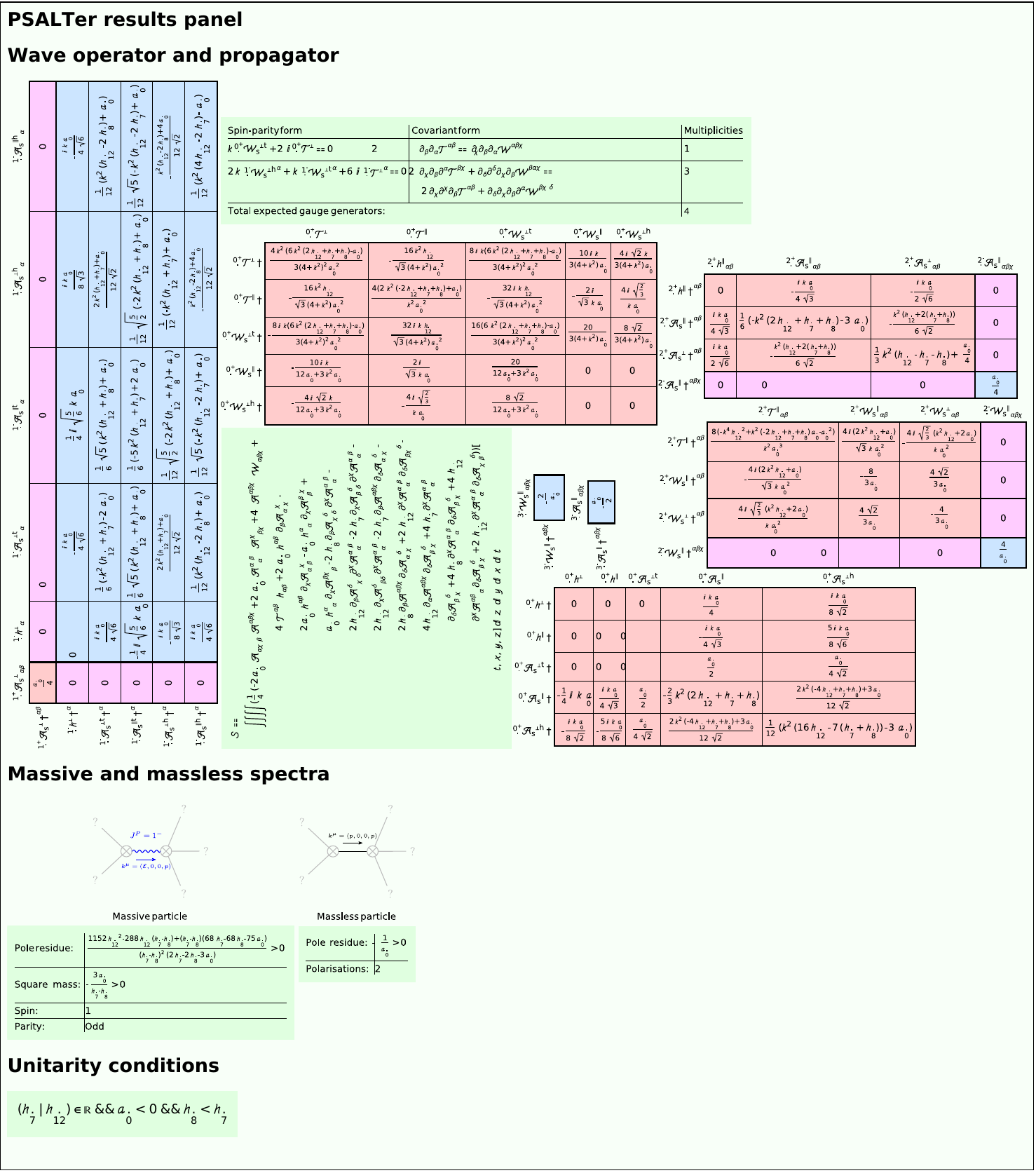}
	\caption{\label{ZeroTorsionNo2pNo1pComplicatedFirstOrder} The spectrum of the theory in~\cref{ActionTLessMAG} with the additional constraints $c_9=c_{10}=c_{11}-c_{12}=0$. As with~\cref{ZeroTorsionNo2pNo1pSimpleFirstOrder} these are sufficient to eliminate the $2^+$ and $1^+$ massive states, leaving only the massive $1^-$ state in~\cref{sp1meno}. The overall theory is clearly unitary. All the quantities in this output are defined in~\cref{ZeroTorsionPalatini}.}
\end{figure*}

\begin{figure*}[ht]
\includegraphics[height=0.9\textheight]{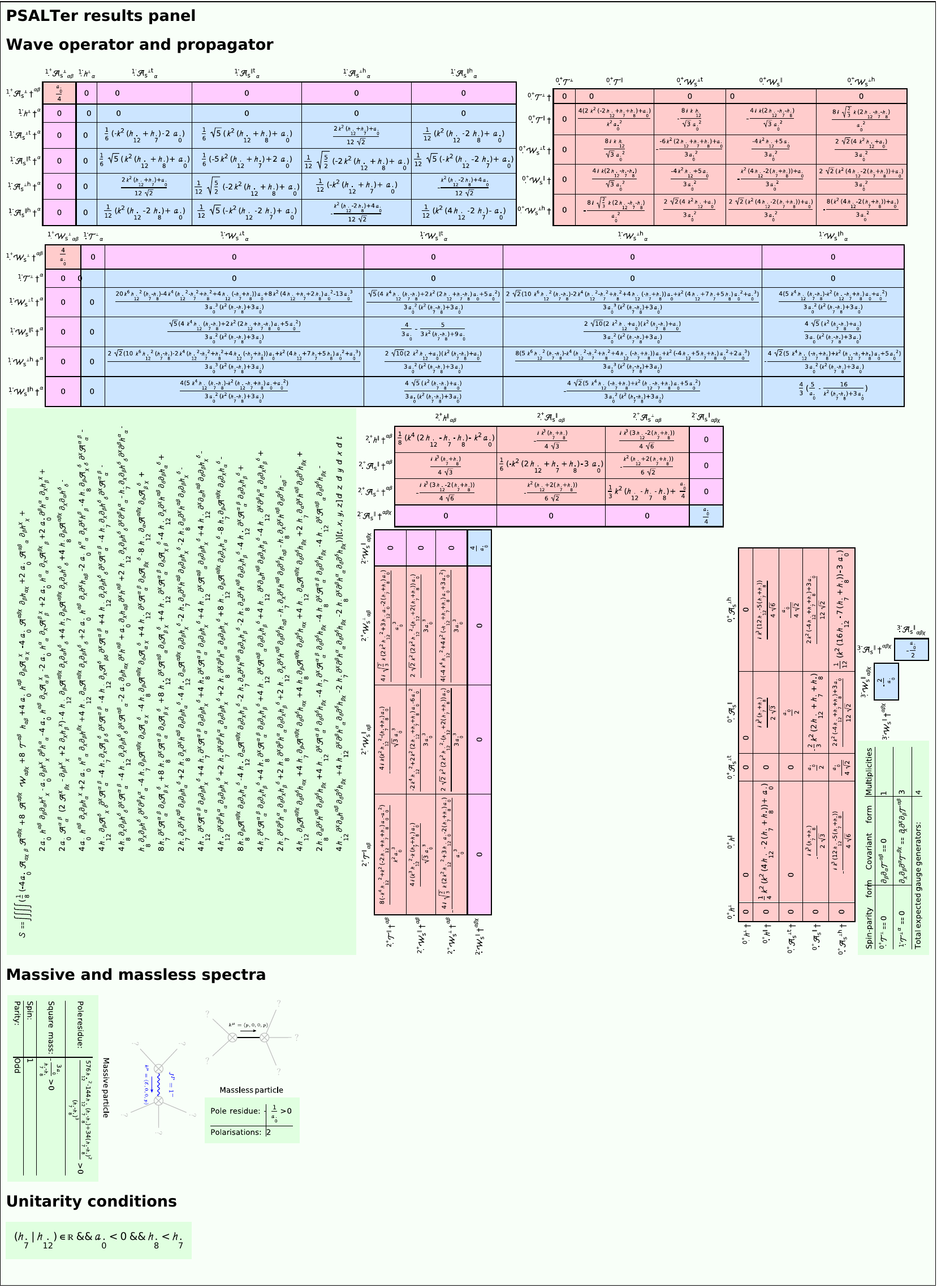}
	\caption{\label{ZeroTorsionNo2pNo1pComplicatedSecondOrder} The results in~\cref{ZeroTorsionNo2pNo1pComplicatedFirstOrder} repeated in the second-order formulation. Note the much larger quadratic expansion, but the consistent mass spectrum and unitarity conditions. All the quantities in this output are defined in~\cref{ZeroTorsionPalatini}.}
\end{figure*}

We show the general analysis of the theory in~\cref{ActionTLessMAG} in~\cref{ZeroTorsionGeneralFirstOrder,ZeroTorsionGeneralSecondOrder}, respectively for the first- and second-order formulations of the model. In~\cref{ZeroTorsionNo2pNo1pSimpleFirstOrder,ZeroTorsionNo2pNo1pSimpleSecondOrder,ZeroTorsionNo2pNo1pComplicatedFirstOrder,ZeroTorsionNo2pNo1pComplicatedSecondOrder} we consider tuned special-cases of the model in which the massive $1^-$ state is allowed to propagate.

Further theories considered in~\cref{SecTor}, whose matrices are too cumbersome for the appendices, are presented in~\cite{SupplementalMaterials}.

\section{Generic case with~\PSALTer{}}\label{SecGenApp}
\begin{figure*}[ht]
\includegraphics[width=\textwidth]{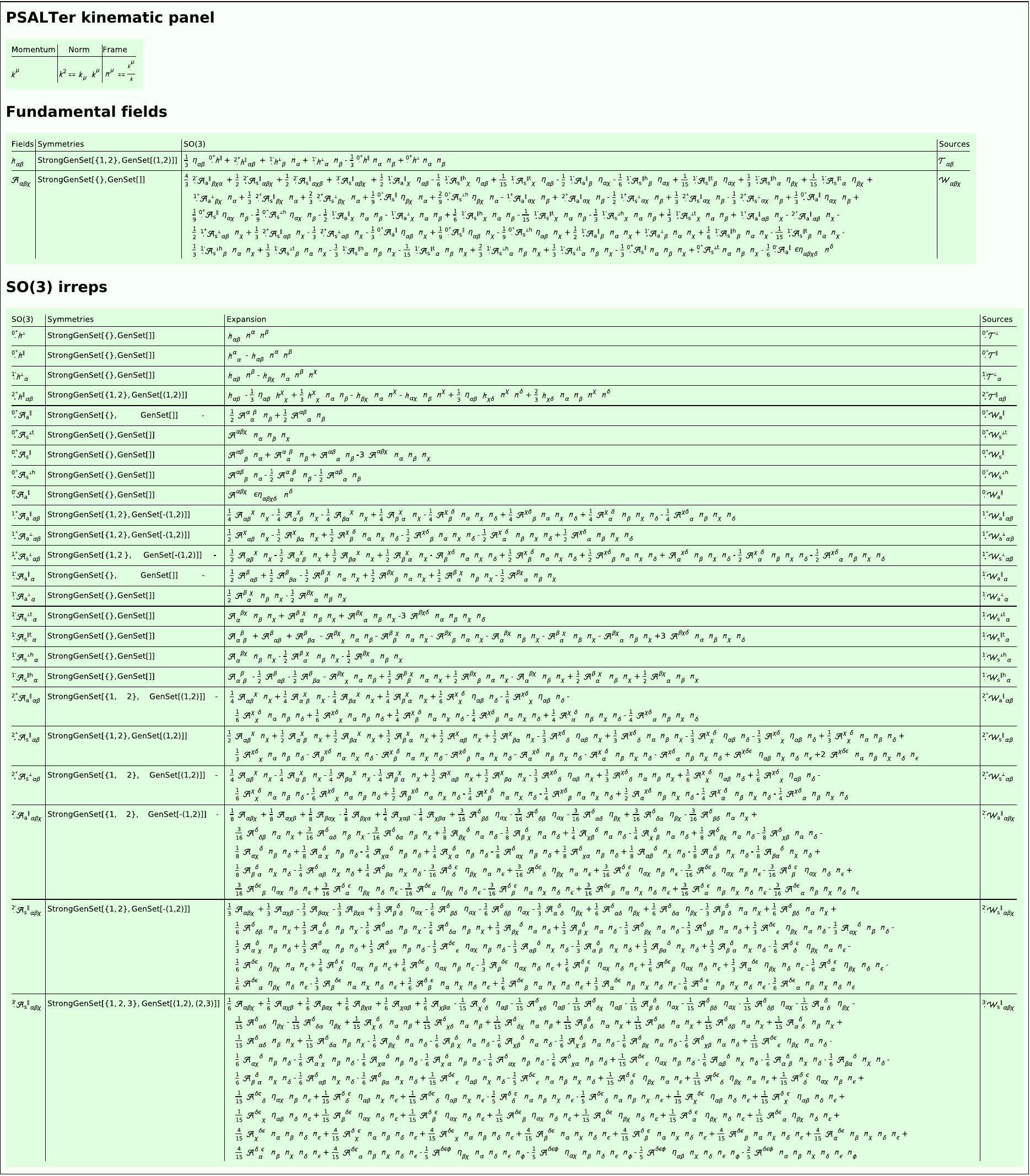}
	\caption{\label{MetricAffineGravity} Kinematic structure of the unrestricted MAG, as studied throughout~\cref{SecGen}. The $\mathrm{SO}(3)$ irreps precisely correspond to those in~\cref{ParticleDec1,ParticleDec2}, though the labelling of duplicate $J^P$ states is different from that in~\cite{Percacci:2020ddy}. As with~\cref{PoincareGaugeTheory,ZeroTorsionPalatini}, the $2^-$ and $3^-$ states have extra cyclic symmetries which are hidden. These definitions are used in~\cref{UnrestrictedGeneralFirstOrder,UnrestrictedGeneralSecondOrder}.}
\end{figure*}

As with~\cref{SecTorApp}, our \PSALTer{} analysis corresponding to~\cref{SecGen} is fully grounded in the MAG formulation. The generic MAG kinematic module is displayed in~\cref{MetricAffineGravity}. The generic first-order MAG kinematic module is of course larger than the zero-torsion counterpart in~\cref{ZeroTorsionPalatini}. By comparing~\cref{ParticleDecT} with~\cref{ParticleDec1} we see that the spin-zero and spin-one matrices will have extra rows and columns. As with~\cref{SecTorApp}, the transition to the second-order formulation is made using~\cref{Reparam}. The \PSALTer{} labelling of duplicate $J^P$ states is again different from that shown in~\cref{ParticleDec1,ParticleDec2}, and a new label ($\mathsf{a}$) is introduced.

\begin{figure*}[ht]
\includegraphics[width=\textwidth]{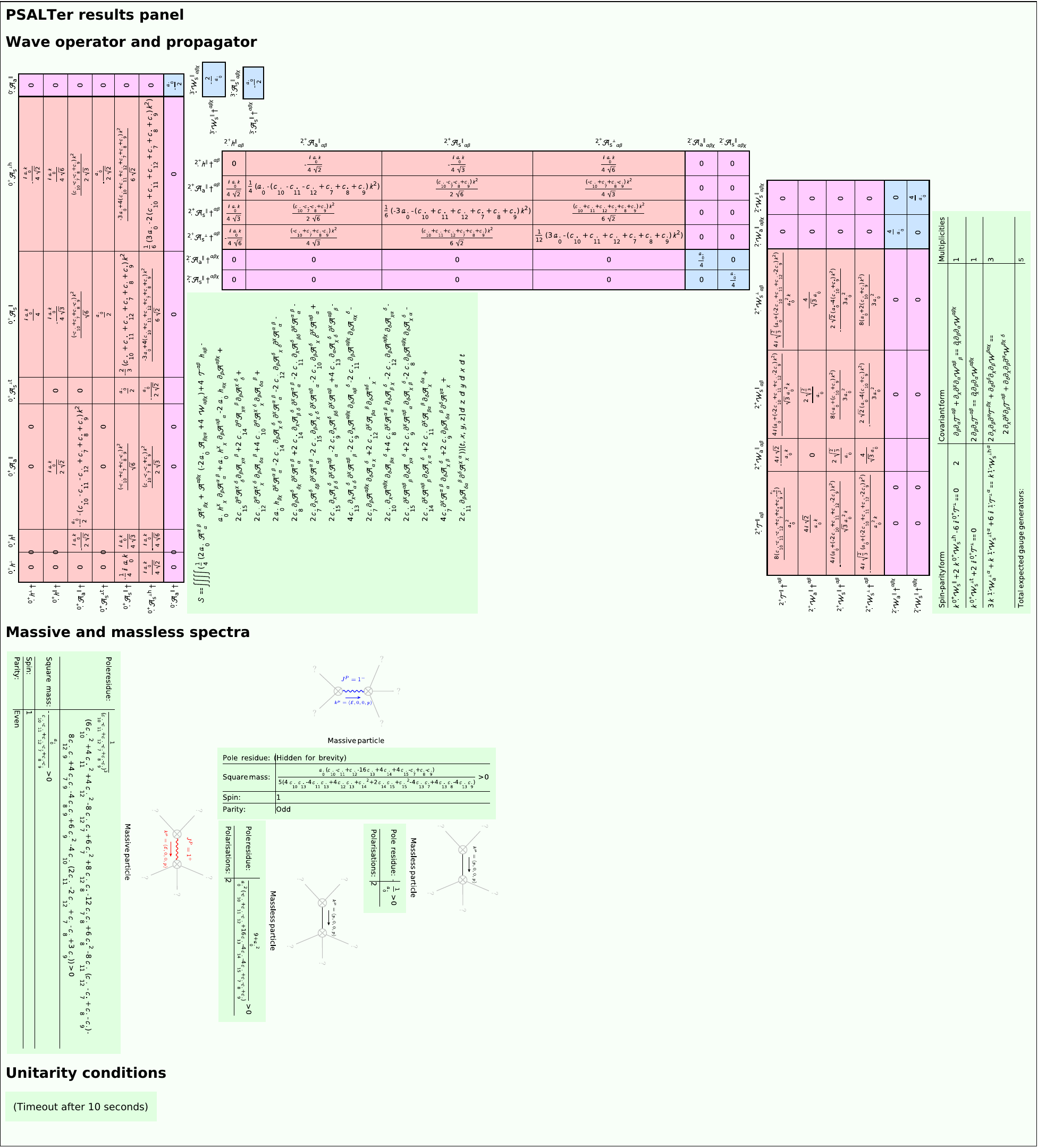}
	\caption{\label{UnrestrictedGeneralFirstOrder} The full spectrum of the general theory in~\cref{ActionMAG} using the first-order formulation. As with~\cref{ZeroTorsionGeneralFirstOrder,ZeroTorsionGeneralSecondOrder} completely general unitarity conditions cannot be found automatically within one minute, but we investigate the conditions in~\cref{SecGen} based on the masses and residues in these results. All the quantities in this output are defined in~\cref{MetricAffineGravity}.}
\end{figure*}

\begin{figure*}[ht]
\includegraphics[width=\textwidth]{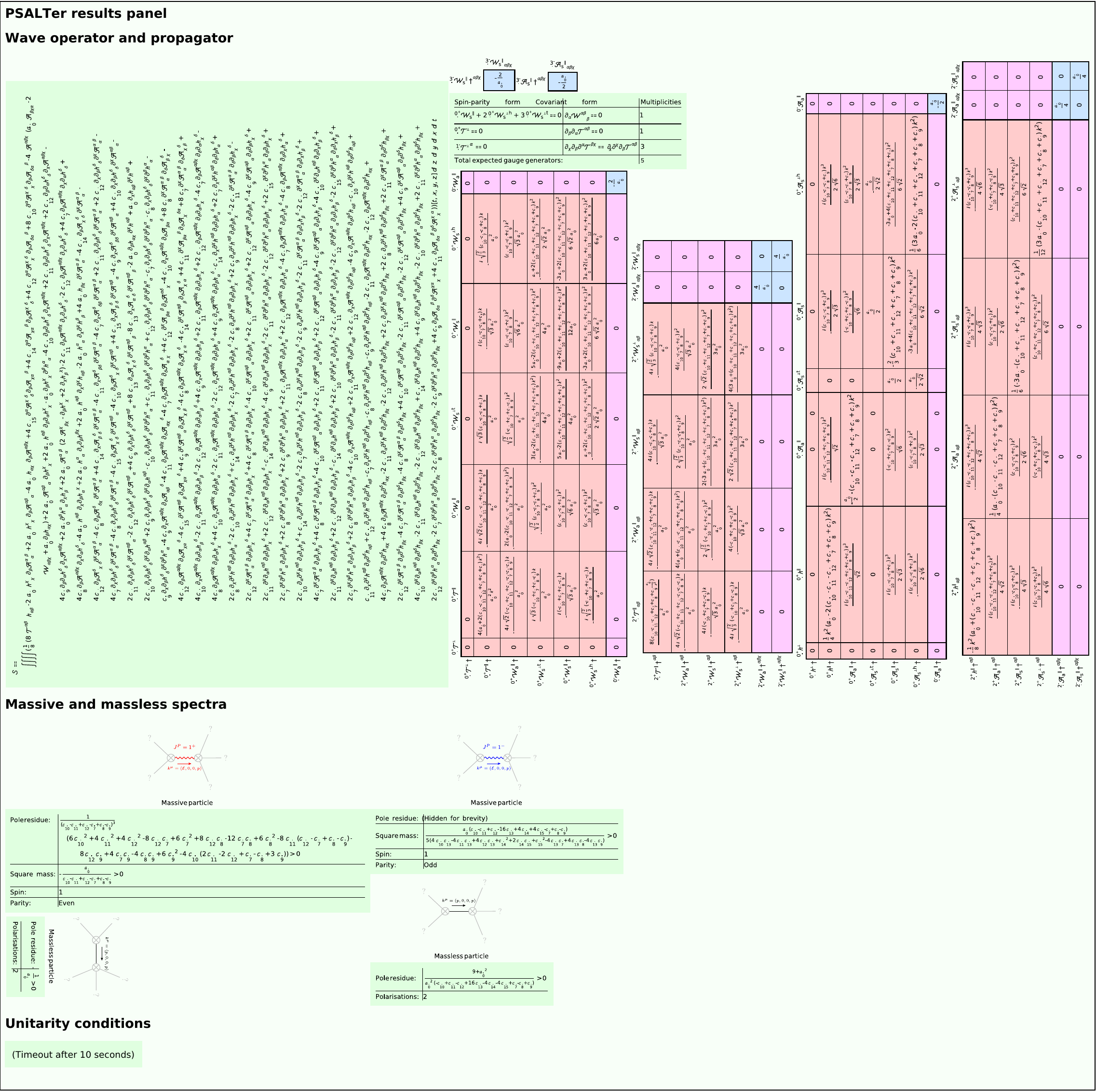}
	\caption{\label{UnrestrictedGeneralSecondOrder} The full spectrum of the general theory in~\cref{ActionMAG} using the second-order formulation. Note that the matrix elements differ from~\cref{UnrestrictedGeneralFirstOrder}, and the form of the gauge symmetries and pole residues seem to differ, but the mass spectrum is the same. Moreover, the number of gauge generators is the same, and the pole residues are expected to have the same implications for the unitarity. Note also the much larger quadratic expansion of the action. All the quantities in this output are defined in~\cref{MetricAffineGravity}.}
\end{figure*}

The full spectrum of the general theory in~\cref{ActionMAG} using the first-order formulation is given in~\cref{UnrestrictedGeneralFirstOrder}. The equivalent result in the second-order formulation is given in~\cref{UnrestrictedGeneralSecondOrder}.

Further theories considered in~\cref{SecGen}, whose matrices are too cumbersome for the appendices, are presented in~\cite{SupplementalMaterials}.

\end{document}